\pdfoutput=1
\documentclass[11pt]{article}
\usepackage{jheppub}

\usepackage{customprelude}
\usepackage{upgreek}

\newcommand{\nequiv}{\mathrel{\not\equiv}}
\newcommand{\tmop}[1]{\ensuremath{\operatorname{#1}}}

\newtheorem*{conjecture*}{Conjecture}

\newcommand{\noeq}{\mathrel{\phantom{=}}}
\newcommand{\e}{\mathrm{e}}

\renewcommand{\L}{\mathsf{L}}
\DeclareMathOperator{\Wil}{\mathsf{W}}
\DeclareMathOperator{\Span}{Span}
\newcommand{\Ss}{Ss}
\newcommand{\Sc}{Sc}
\newcommand{\cK}{\mathcal{K}}
\newcommand{\PD}{\mathrm{PD}}
\newcommand{\Kthree}{\ensuremath{\mathrm{K3}}}
\newcommand{\secref}[1]{\S\ref{#1}}

\title{\centering IIB flux non-commutativity\\and the global structure of field theories}

\author[\spadesuit]{I\~{n}aki Garc\'ia Etxebarria,}
\author[\clubsuit]{Ben Heidenreich}
\author{and Diego Regalado}

\affiliation[\spadesuit]{Department of Mathematical Sciences, Durham University,\\Durham, DH1 3LE, United Kingdom}
\affiliation[\clubsuit]{Department of Physics, University of Massachusetts,\\ Amherst, MA 01003 USA}

\emailAdd{inaki.garcia-etxebarria@durham.ac.uk}
\emailAdd{bheidenreich@umass.edu}
\emailAdd{diego.rgm@gmail.com}

\abstract{We discuss the origin of the choice of global structure for
  six dimensional $(2,0)$ theories and their compactifications in terms of their realization from IIB string theory on ALE spaces. We find that the ambiguity in
  the choice of global structure on the field theory side can be traced
  back to a subtle effect that needs to be taken into account when
  specifying boundary conditions at infinity in the IIB orbifold,
  namely the known non-commutativity of RR fluxes in spaces with
  torsion. As an example, we show how the classification of $\cN=4$ theories by Aharony, Seiberg and Tachikawa can
  be understood in terms of choices of boundary conditions for RR
  fields in IIB.  Along the way we encounter a formula for the
  fractional instanton number of $\cN=4$ ADE theories in
  terms of the torsional linking pairing for rational homology spheres.
  We also consider six-dimensional $(1,0)$
  theories, clarifying the rules for determining
  commutators of flux operators for discrete 2-form
  symmetries. Finally, we analyze the issue of global structure for
  four dimensional theories in the presence of duality defects.}

\begin{document}


\maketitle

\newpage

\section{Introduction}

One of the fundamental observables that we can use to characterize
Quantum Field Theories is their partition function on arbitrary
manifolds $\cM$. The partition function depends both on intrinsic data
$\cT$ defining the theory --- which we can provide without reference
to the underlying manifold --- and on background data on $\cM$, such
as a metric $g_{\mu \nu}$, a Spin connection $\omega^{ab}_{\mu}$, and
backgrounds $A_\mu$ for the global symmetries of the theory, which
might be continuous or discrete.\footnote{The separation into
  background and intrinsic data is sometimes arbitrary: if we restrict
  ourselves to four-dimensional Yang-Mills theories with constant
  coupling $\tau$ we could view $\tau$ as part of the data defining
  $\cT$. However, if we wish to allow for the possibility that $\tau$ varies across $\cM$
  then we must include it as part of the background data
  to be specified for each manifold. The second interpretation will
  be more natural from the point of view in this paper, and such
  configurations will play an interesting role below.} If $\cM$ is
non-compact, we need to specify boundary conditions for the theory,
which we denote as $\ket{\psi}$, for reasons that will become apparent
momentarily. For a theory $\cT$ on a manifold $\cM$ with this
structure specified, we can thus write
\begin{equation}
  Z_\cT[\cM(g,\omega,A,\ket{\psi})]
\end{equation}
for the partition function.

Our main interest in this paper will be the case in which $\cT$ is a
six-dimensional SCFT preserving $\cN=(2,0)$ supersymmetry, which we
construct as follows. Consider IIB string theory on a
manifold\footnote{In this paper we will take $\cM_6$ to be closed,
  Spin and orientable, and furthermore we will assume that the
  cohomology groups of $\cM_6$ are freely generated, so there is no
  torsion.}  $\cM_6\times \bC^2/\Gamma$, with $\Gamma\subset
SU(2)$. By the McKay correspondence \cite{McKay}, the relevant
discrete groups $\Gamma$ are in a one-to-one correspondence with the
simple Lie algebras of ADE type. Given such an algebra $\fg_\Gamma$,
we denote by $G_\Gamma$ the simply connected Lie group with algebra
$\fg_\Gamma$. It is a well supported conjecture that this system has a
non-trivial interacting fixed point at low energies, given by an
interacting six-dimensional $\cN=(2,0)$ SCFT \cite{Witten:1995zh},
known as the $(2,0)$ theory of type $\fg_\Gamma$. In fact, all known
interacting $(2,0)$ SCFTs that do not factorize into decoupled SCFTs
at the level of local operators can be obtained from this
construction.\footnote{The free, or ``abelian'', $(2,0)$ theory can be
  obtained by replacing $\bC^2/\Gamma$ by a single-centered Taub-NUT
  space.}

The $(2,0)$ theory of type $\fg_\Gamma$ has a number of remarkable
properties, one of the most exotic ones being that on generic $\cM_6$
there is no canonical choice for the background connection for its
global symmetries. More concretely, the $(2,0)$ theory of type
$\fg_\Gamma$ is believed to possess a discrete global 2-form symmetry
\cite{Gaiotto:2014kfa} given by the center $Z(G_\Gamma)$ of
$G_\Gamma$. The generators of this symmetry do not all commute with
each other, so quantum mechanically there is no way of setting all
background fields for the 2-form symmetry to zero. The following
consequence of this fact might be more familiar: upon compactification
on $T^2$ the $(2,0)$ theory becomes $\cN=4$ SYM with gauge algebra
$\fg_\Gamma$, and the 2-form symmetry gives rise to the 1-form
symmetries measuring the number of Wilson and 't Hooft lines. It is a
familiar fact that the associated symmetry generators do not commute
\cite{THOOFT19781,THOOFT1979141}.

Since the symmetry generators do not commute, they are not
simultaneous observables. The best we can do is to select a maximal
commuting subset of these operators and decompose the Hilbert space in
their simultaneous eigenbasis. By selecting an eigenvector from this
basis, an associated subset of the background fields for the 2-form
symmetry can all be set to zero, or to any definite value. However,
choosing a maximal commuting set of fluxes to fix requires explicit
reference to the structure of $H^3(\cM_6; Z(G_\Gamma))$, and
generically such a choice will not be invariant under large
diffeomorphisms of $\cM_6$. This can be naturally interpreted as an
anomaly (see \cite{Monnier:2019ytc} for an introduction), but the fact
that the ambiguity in the partition function is not just a phase makes
the situation exotic. This state of affairs is often described by
saying that the $(2,0)$ theory has a partition vector (of ``conformal
blocks'', in analogy with the situation for chiral theories in two
dimensions) as opposed to having a partition function, or sometimes,
more concisely, by saying that the $(2,0)$ theory is a ``metatheory''.

\bigskip

At this point we reach a puzzle, which this paper aims to clarify: we
have explained that generally there is no canonical choice of
partition function for the six-dimensional $(2,0)$ theory, due to the
non-commutativity of the operators generating the 2-form symmetry. But
on the other hand, we started our discussion by saying that the
$(2,0)$ theory of type $\fg_\Gamma$ can be constructed by considering
a low-energy limit of IIB string theory on $\bC^2/\Gamma\times
\cM_6$. The fact that there is no canonical choice of partition
function for the $(2,0)$ theory should then imply that there is no
canonical choice for the partition function of IIB string theory on
$\bC^2/\Gamma\times \cM_6$. We will argue that this is indeed the
case.

Briefly, in order to have a well defined partition function of the IIB
theory on $\bC^2/\Gamma\times \cM_6$ one needs to specify boundary
conditions for the RR fluxes, and in the presence of torsion this is a
fairly subtle affair due to the self-dual nature of RR fields in
string theory \cite{Freed:2006ya,Freed:2006yc}. We will show that
there is indeed no choice of boundary conditions in which all RR
fluxes are set to zero at infinity, and in fact the set of choices for
boundary conditions for IIB on $\bC^2/\Gamma\times \cM_6$ is in
one-to-one correspondence with the set of choices one makes in
choosing a partition function for the $(2,0)$ theory of type
$\fg_\Gamma$ on $\cM_6$.

This result removes a fair bit of mystery from the usual statement
that the $(2,0)$ theory has no well-defined partition function, since
the standard construction of such theories in string theory
\emph{requires} one to provide the missing data in the form of
boundary values for the RR fluxes. Remarkably, all possible choices
for the $(2,0)$ theory can be accommodated in the IIB construction. In
terms of symmetries our viewpoint provides a reinterpretation of the
2-form symmetry of the $(2,0)$ theory in terms of transformations of
the boundary conditions on IIB.

\bigskip

This whole discussion might come as a bit of a surprise to the reader
familiar with the proof in \cite{Witten:1996hc,Witten:1999vg} that
there \emph{is} a canonical partition function of IIB on a
ten-manifold $\cM_{10}$. The key assumption in the argument in
\cite{Witten:1996hc,Witten:1999vg} that does not hold for the
geometries analyzed in this paper is that $\cM_{10}$ has an
intersection form with unit determinant. This is always the case for
compact manifolds, but generically it is not the case for
$\cM_{10}=\cM_6\times \bC^2/\Gamma$ (except for the case associated
with $E_8$). Similarly, the statement that string theory always gives
rise to modular invariant theories (see for example
\cite{Seiberg:2011dr}) is true under the assumption that we have a
compact transverse space, so that the six-dimensional effective theory
of interest is coupled to six-dimensional gravity. But this does not
hold for the configurations that we study in this paper, in which the
metric is just a background field in the six-dimensional theory. The
effective six-dimensional theories that one finds in the case
$\cM_{10}=\cM_6\times\bC^2/\Gamma$ are not modular invariant, but
since six-dimensional gravity is non-dynamical there is no
contradiction. (The ten-dimensional gravity theory is dynamical, but
again there is no contradiction because as we will describe the lack
of modular invariance of the six-dimensional theory ultimately comes
from the lack of modular invariance of the choice of boundary
conditions on $\cM_6\times S^3/\Gamma$, and we do not sum over these
when doing the gravitational path integral.)

We emphasize that our viewpoint here, focusing purely on a careful
analysis of the original construction of $(2,0)$ theories in ten
dimensional type IIB string theory, is complementary to existing
viewpoints on the partition function of $(2,0)$ theories. One such
viewpoint is that of \emph{relative} QFTs articulated by Freed and
Teleman in \cite{Freed:2012bs}, where one views the $(2,0)$ theories
as furnishing the boundary degrees of freedom for certain
non-invertible seven dimensional TQFTs
\cite{Monnier:2014txa,Monnier:2016jlo,Monnier:2017klz}. For the $A_N$
cases one can also study the question using holography
\cite{Witten:1998wy}. We find that all three approaches give the same
results whenever they are simultaneously applicable.

\bigskip

We have organized this paper as follows. We start in
\S\ref{sec:quantization} by explaining how to choose boundary
condition for RR fields in IIB string theory on
$\cM_6\times \bC^2/\Gamma$. In \S\ref{sec:comparisonall} we compare
the results of \S\ref{sec:quantization} to the known results for the
behaviour of the $(2,0)$ partition function, and extend the results to
the $(1,0)$ case, refining a previous proposal in
\cite{DelZotto:2015isa}. We then show how one
can rederive the known classification of four dimensional $\cN=4$
theories \cite{Aharony:2013hda} (of ADE type) from the IIB
perspective. Along the way we encounter a simple geometric
reinterpretation of the fractional instanton number in $\cN=4$
theories with simply-connected gauge group, which we expect to
generalize to less supersymmetric cases. In \S\ref{sec:polarizations}
we explore these ideas in less familiar backgrounds: we will discuss
global aspects of 4d theories in the presence of duality defects (as
studied in
\cite{Harvey:2007ab,Cvetic:2011gp,Martucci:2014ema,Gadde:2014wma,Assel:2016wcr,Choi:2017kxf,Lawrie:2018jut,Lawrie:2016axq},
for instance) and subtleties having to do with modular invariance in
the context of 4d/2d dualities that arise when the four dimensional
manifold has two-cycles. We point out an interesting relation between
the Vafa-Witten partition function of self-dual $\fsu(p)$ theories on
\Kthree{} and Hecke operators acting on the partition function of chiral
bosons, and briefly discuss a (speculative, but suggestive) connection
between these partition functions and the $j$ invariant. In~\secref{sec:conclusions}, we conclude and list a
number of directions for further research. Appendix~\ref{app:homology-sphere-K-theory} contains
technical results on the complex K-theory groups of rational homology
spheres used in the main text, and
appendix~\ref{app:K3-partition-functions} discusses the Vafa-Witten partition
functions~\cite{Vafa:1994tf}  of $\cN=4$ theories with algebra $\fsu(N)$ on \Kthree{} for different choices of the
global form of the gauge group, and how their behavior under dualities
agrees with expectations.

\section{Quantization of type IIB string theory on \texorpdfstring{$\cM_6\times \bC^2/\Gamma$}{M6 x C2/Gamma}}

\label{sec:quantization}

We begin with a short informal outline of the main argument in
this section, without going into the technical details. Most of the
work in the rest of the section will be in making these arguments fully
precise.

Consider type IIB string theory compactified on
$\cM_6 \times \mathbb C^2/\mathbb Z_N$, which is believed to yield the
$A_{N-1}$ $(2,0)$ theory on $\cM_6$ at low energies.
Without changing the behaviour at infinity, we could
instead consider a resolution of the $\bC^2/\mathbb Z_N$ orbifold,
so that the spacetime curvature is arbitrarily small
and the string coupling is small and
constant. Thus, the subtlety in specifying boundary conditions cannot be due to any particular property of string theory in
singular spaces. Instead, it is due to the presence of the self-dual RR field $F_5 = \ast F_5$ in type IIB supergravity. As pointed out in beautiful work
by Freed, Moore and Segal \cite{Freed:2006ya,Freed:2006yc} (building
on
\cite{Gukov:1998kn,Moore:2004jv,Belov:2005ze,Witten:1998wy,Belov:2004ht,Burrington:2006uu,Burrington:2006aw,Burrington:2006pu}),
quantization of self-dual fields in spaces with torsion needs to be
done with care, even at arbitrarily weak coupling.

In more detail, in order to characterize the IIB background we should
specify boundary conditions for all the supergravity fields, including $F_5$. Classically, we would 
specify the background value for $F_5$ at infinity, which we could
simply set to zero if desired. Quantum mechanically, the story is
far more subtle. We describe it in detail below, but the
main point is that for each class
$\sigma\in \Tor(H^5(\cM_6\times S^3/\bZ_N;\bZ))$ there is a unitary flux
operator $\Phi_\sigma$, which measures the torsional part of the flux
on the homology class Poincaré dual to $\sigma$. The boundary
conditions are encoded in the expectation values of these operators,
and naively we could simply choose a state with
$\vev{\Phi_\sigma}=1$ for every $\sigma$, corresponding to a
background with no flux at infinity. Surprisingly, this is not possible, as the torsion flux operators for
self-dual forms on different cycles do not always commute
\cite{Freed:2006ya,Freed:2006yc}:\footnote{In general, electric and
  magnetic fluxes for $p$-form theories on spaces with torsion do not
  commute \cite{Freed:2006ya,Freed:2006yc}. The basic observation is
  that the action of the electric flux operator is to shift the
  connection by a closed form in $H^\bullet(X;U(1))$, while the
  magnetic flux operator measures the topological class of the bundle
  associated to the connection. This implies that whenever
  topologically non-trivial closed forms in $H^\bullet(X;U(1))$ exist
  (that is, in the presence of torsion, see
  footnote~\ref{fn:u(1)-tor-ses} below), electric and magnetic
  operators do not necessarily commute. It was argued in
  \cite{Freed:2006ya,Freed:2006yc} that analogously, fluxes for
  self-dual forms do not necessarily commute with each other whenever
  the spacetime has torsion.}
\begin{equation}\label{simplecom}
  \Phi_\sigma\Phi_{\sigma'} = e^{2\pi i\,\L(\sigma,\sigma')} \Phi_{\sigma'}\Phi_\sigma\, .
\end{equation}
Here $\L(\sigma,\sigma')$ is the linking pairing for the torsion
5-forms $\sigma,\, \sigma'$, taking values in $\bQ/\bZ$. Most of the
technical details in this section deal with the careful computation of
this linking pairing.

The nonvanishing commutator~\eqref{simplecom} implies that one
cannot specify the value of all fluxes simultaneously, and in
particular one cannot simply set the $F_5$ flux to
zero at infinity. Instead, the best we can do is to choose a maximal set of commuting
flux operators and set the corresponding fluxes to zero (or to another fixed value). Given such a choice we can in principle compute the partition function for type IIB on that
background, which also determines the partition function for the
$A_{N-1}$ $(2,0)$ theory on $\cM_6$. However, there is no canonical
choice for the maximal subset of commuting operators to set to zero
and, in fact, large diffeomorphisms on the boundary typically relate
different choices. In light of this, one might expect that the
collection of boundary conditions for type IIB in this background,
with the subtleties due to non-vanishing commutators properly taken
into account, is precisely the vector space of partition functions of
the $(2,0)$ theory on $\cM_6$. In the coming sections we will argue
that this expectation is indeed correct.

Note that the RR fields in IIB string theory are more properly
described in terms of differential K-theory (see
\cite{Hopkins:2002rd,Freed:2000ta,Freed:2006yc} for an introduction). This not only accounts for the local data of the $C_4$ connection (the
``differential'' qualifier), but also the fact that the flux quantization conditions
are better described by K-theory \cite{Moore:1999gb}. However, to understand the commutation relations it is
sufficient to restrict to ordinary K-theory, since the
commutators depend only on the K-theory class, and more specifically
its torsional component. Related to this, the class $\sigma$
really lives in $H^4(\cM_6\times S^3/\bZ_n;U(1))$ (or rather, its
generalization in differential K-theory) rather than
$\Tor(H^5(\cM_6\times S^3/\bZ_N;\bZ))$, but again to understand the flux commutation relations it will be
sufficient to restrict ourselves to torsion classes.\footnote{\label{fn:u(1)-tor-ses}The two groups are
  related by the short exact sequence
\begin{equation*}
  0 \to \Wil^4 \to H^4(\cM_6\times S^3/\bZ_n;U(1)) \to \Tor(H^5(\cM_6\times S^3/\bZ_N;\bZ)) \to 0 \,,
\end{equation*}
with $\Wil^4$ the group of topologically trivial $C_4$ Wilson
lines on $\cM_6\times S^3/\bZ_n$.}

\subsection{Flux operators and the Hilbert space \texorpdfstring{$\cH^{[RR]}(\cN_9)$}{HRR(N9)}}

\label{sec:construction-of-H}

Starting again from the beginning, we aim to specify the
boundary conditions for euclidean IIB string theory on a
ten-dimensional manifold $X_{10}=\cM_6\times \bC^2/\Gamma$, where
$\cM_6$ is closed, oriented, Spin, and without torsion. To
understand how to choose boundary conditions properly, we first
take a slight detour and review some basic aspects of quantum
field theory (see, e.g.,  \cite{HartmanLectures} for a less telegraphic
exposition).

In general, a $d$-dimensional quantum field theory associates a Hilbert space
$\cH(\cN_{d-1})$ to each $(d-1)$-dimensional manifold
$\cN_{d-1}$.
This Hilbert space is the one associated with
quantization of the original theory on $\cN_{d-1}\times\bR$, where
$\bR$ denotes the time direction. We stress that we are not yet
specifying the value of the fields on $\cN_{d-1}$, the Hilbert space
only depends on $\cN_{d-1}$ itself. Indeed, in the quantum theory a
choice of field configuration on $\cN_{d-1}$ corresponds to choosing a
state $\ket{\psi}\in \cH(\cN_{d-1})$.

Now consider the quantum field theory on a manifold $X_d$ with boundary $\cN_{d-1}=\partial X_d$. Then the path integral on
$X_d$, without specifying the boundary conditions, can be
understood as a dual vector $\bra{Z}\in \cH^*(\cN_{d-1})$, so the
value of the path integral \emph{with} boundary conditions specified
by $\ket{\psi} \in \cH(\cN_{d-1})$ is $\braket{Z}{\psi}\in\bC$.

Type IIB string theory in ten dimensions is most certainly not an
ordinary ten-dimensional quantum field theory, but a version of the
above is believed to hold whenever the ten-dimensional manifold is
non-compact, with $X_{10}$ asymptotically of the form
$\cN_{9}\times\bR$. Classically, we would specify the boundary
conditions on $\cN_{9}$ by giving boundary conditions at infinity for
the IIB supergravity fields. We focus on the RR fields, setting $B=0$,
which are classified by K-theory \cite{Moore:1999gb}. For the purposes
of studying the Heisenberg group of fluxes it is enough to consider
the topological class $K^1(\cN_{9})$ of the RR fields at the boundary
\cite{Freed:2006yc}.\footnote{Although it is not true in general, we
  will show that for the spaces discussed in this paper one
  has:\label{fn:cohomologysum}
  \begin{equation}
    K^1(\cN_{9}) = \bigoplus_{i\in2\bZ+1} H^i(\cN_{9})
  \end{equation}
  so the reader unfamiliar with K-theory can think instead of the
  formal sum of cohomology groups of odd degree. Note that whenever we
  write $H^i(Y)$ or $H_i(Y)$, without explicit mention of the
  coefficient ring, we are always referring to singular (co)homology
  theory with coefficients in $\bZ$.}

In analogy with the situation on QFT described above, we will assume
that there is a Hilbert space $\cH(\cN_{9})$
associated to quantum boundary conditions, and that a specific choice of boundary
conditions furnishes a vector in this Hilbert space.\footnote{If we
  specify the IIB geometry without choosing boundary conditions for
  the fields, then what we have is a dual vector of partition
  functions $\bra{Z}\in\cH^*(\cN_{9})$, which in the case of
  $\cM_d=\bC^2/\Gamma\times\cM_6$ will induce a partition vector on
  the $\fg_\Gamma$ $(2,0)$ theory on $\cM_6$.} (This prescription has
been used before, for instance in the case of AdS/CFT boundary
conditions \cite{Witten:1998wy}.)
In particular, if $X_{10}=\bC^2/\Gamma\times \cM_6$ with $\cM_6$ compact then $\cN_9=S^3/\Gamma\times \cM_6$. 

We will focus on the subsector of the Hilbert space $\cH(\cN_9)$
describing the topological class of the RR fields at the boundary,
which we will denote $\cH^{[RR]}(\cN_9)$. If the classical picture
were not modified quantum mechanically, then the answer would be that
$\cH^{[RR]}(\cN_{9})$ is graded by classes in $K^1(\cN_{9})$, or in
other words that the boundary conditions are determined topologically
by the K-theory class of the flux on the boundary. That this is not
the case was shown in \cite{Freed:2006ya,Freed:2006yc}. We refer the
reader to these papers for the derivation, and here just state the
result of the analysis as it applies to our case. Recall that the
K-theory group $K^1(\cN_{9})$ is an abelian group which might (and, in
our examples, will) contain a torsional subgroup
\begin{equation}
  \Tor(K^1(\cN_{9})) = \left\{x\in K^1(\cN_{9})\,\,\bigl|\,\,  nx=0 \text{ for some } n \in \bZ\right\}\, .
\end{equation}
We can also construct the group of fluxes modulo torsion
\begin{equation}
  \ov K^1(\cN_{9}) = \frac{K^1(\cN_{9})}{\Tor(K^1(\cN_{9}))}\, .
\end{equation}
Freed, Moore and Segal
\cite{Freed:2006ya,Freed:2006yc} showed that there is a grading of
$\cH^{[RR]}(\cN_{9})$ by $\ov K^1(\cN_{9})$; in other words the
non-torsional part of the flux can be specified without subtleties,
and the associated flux operators commute. Remarkably, they also
showed that this commutativity does not hold for the torsional
part.

To quantify this, we postulate a set of unitary operators $\Phi_x$, one for each K-theory class
$x\in \Tor K^1(\cN_9)$. The precise relation between these operators and the background RR fluxes will become clear shortly, but we remark for the present that they are essentially the integrals ``$\exp(i \int A_x \wedge F_{RR})$'' where $F_{RR}$ is the background flux and $A_x$ is a flat connection associated to the torsion class $x$.
As shown by Freed, Moore and Segal~\cite{Freed:2006ya,Freed:2006yc}, these operators do not commute. Instead,
\begin{equation}
\Phi_x \Phi_y = s(x, y) \Phi_y \Phi_x\,, \label{eq:flux-commutator}
\end{equation}
where $s(x_1,x_2)$ is a perfect pairing
\begin{equation}
  s\colon \Tor(K^1(\cN_{9})) \times \Tor(K^1(\cN_{9})) \to U(1)
\end{equation}
that we will discuss extensively below. 
Some useful properties of
$s(x,y)$ are that it is skew ($s(x,y)=s(y,x)^{-1}$), alternating
($s(x,x)=1$) and bimultiplicative ($s(x+y,z)=s(x,z)s(y,z)$ and
$s(x,y+z)=s(x,y)s(x,z)$). We say that a pairing
$A\times A\to U(1)$ is perfect if the induced map $A\to \Hom(A, U(1))$
is an isomorphism. The fact that the pairing is perfect implies, in
particular, that no non-trivial torsion flux commutes with all other
fluxes.

Note that it is not in general true that $\Phi_x \Phi_y =
\Phi_{x+y}$. Indeed, this would be incompatible
with~(\ref{eq:flux-commutator}). However, we will assume
that\footnote{This is actually not true when the order of
  $\Tor(K^1(\cN_9))$ is even. In this case, we believe that the
  correct statement is
  $s(x,y)=1 \implies\Phi_x\Phi_y = \pm\Phi_{x+y}$, with both signs
  being realized. Nonetheless, within any isotropic subspace of
  $\Tor(K^1(\cN_9))$ the flux operators can be redefined to satisfy
  $\Phi_x\Phi_y = \Phi_{x+y}$. In the following discussion, this is
  done implicitly, and this subtlety will have no effect on our
  subsequent analysis. It appears that there is some connection
  between this sign and the fractional instanton number discussed
  in~\secref{sec:ninst}, but we defer further consideration of this to
  future work.}
\begin{equation}
s(x,y) = 1 \quad\Longrightarrow\quad \Phi_x \Phi_y = \Phi_{x+y} \,.
\end{equation}
More generally, $\Phi_x \Phi_y$ and $\Phi_{x+y}$ will differ by a phase.

Since the flux operators do not commute, we cannot specify the
asymptotic values for all fluxes simultaneously. Instead, the
asymptotic values define a state in the Hilbert space
$\cH^{[RR]}(\cN_9)$, and this Hilbert space is a representation of the
Heisenberg group generated by the flux operators, defined
below.\footnote{See \cite{mumford2006tata} for background material on
  Heisenberg groups.} To construct this representation, we diagonalize
a maximal commuting subset of the flux operators, as follows.
(See \cite{Witten:1998wy,Tachikawa:2013hya} for previous discussions of this construction in related contexts.)

Consider a subgroup $L \subset \Tor(K^1(\cN_{9}))$. Define
\begin{equation}
  L^\perp \coloneqq \{x \in \Tor(K^1(\cN_{9})) \mid \forall y\in L, s(x,y)=1\}\, ,
\end{equation}
where $L^\perp$ is itself a subgroup of $\Tor(K^1(\cN_{9}))$.
We say that $L$ is \emph{isotropic} if $L\subseteq L^\perp$, and that
$L$ is a \emph{maximal isotropic} subspace of $\Tor(K^1(\cN_{9}))$ if there is no isotropic
subspace $L'$ such that $L\subset L'$, or equivalently, if $L=L^\perp$.

Clearly, $L$ is isotropic if and only if the group generated by the flux operators $\{\Phi_x | x \in L\}$ is abelian, hence choosing maximal isotropic $L$ corresponds to picking a maximal set of commuting observables.
Given maximal isotropic $L$, there is a unique state in the Hilbert space $\cH^{[RR]}(\cN_9)$ such that
\begin{equation}
  \Phi_x \ket{0;L} = \ket{0;L} \qquad \forall x \in L .
\end{equation}
As a unit eigenvector of the flux operators in $L$, this state is naturally thought of as a state of ``zero flux''. To see what fluxes we have turned off (and to turn them on with definite, non-zero, values), we consider the quotient:
\begin{equation}
  F_L \df \frac{\Tor(K^1(\cN_9))}{L}\, . \label{eq:flux-coset}
\end{equation}
Choosing a representative $f$ of each coset in $F_L$, we obtain a basis for $\cH^{[RR]}(\cN_9)$:
\begin{equation}
\ket{f;L} = \Phi_f \ket{0;L} \,,
\end{equation}
where the choice of representative only affects the overall phase of each basis element. The flux operators $\{\Phi_x | x \in L\}$ are diagonal in this basis: $\Phi_x |f; L\rangle = s(x,f) |f; L\rangle$ for all $x \in L$.

We conclude that in this basis the background RR flux belongs to a definite coset $f \in F_L$, whereas the flux operators $\Phi_x$, $x \in L$, are diagonalized with eigenvalues $s(x,f)$. Each maximal isotropic subspace $L \subset \Tor K^1(\cN_9)$ gives a different basis $|f; L\rangle$ for the same Hilbert space $\cH^{[RR]}(\cN_9)$, with different fluxes specified in different bases.

We reiterate at this point that it is only once we have specified
$\ket{\psi}\in\cH^{[RR]}(\cN_9)$ that have we completely fixed the IIB
background, and only in this case we expect to have a uniquely
determined partition function. How do we choose $\ket{\psi}$? In
ordinary quantum mechanics we would write
\begin{equation}
  \ket{\psi} = \sum_j a_j \ket{j}
\end{equation}
and we would choose the $a_i$ freely, giving rise to arbitrary
superpositions of basis states. In the current context we are dealing
with boundary conditions at infinity, so we expect the Hilbert space
to split into superselection sectors. Given that fluxes do not commute, the most conservative proposal (essentially the same choices studied in \cite{Witten:1998wy,Tachikawa:2013hya})) is to first specify a maximal isotropic subspace $L \subset \Tor K^1(\cN_9)$,
which will select the generators of
the discrete 2-form symmetries present in the $(2,0)$ theory. We then choose $\ket{\psi}=\ket{f; L}$ for arbitrary $f \in F$, specifying a background flux $f \in F_L$ for these 2-form symmetries.

As we discuss more extensively in \S\ref{sec:comparison-4d}, in the
particular case that $\cM_6=\cM_4\times T^2$ the different choices of
$L$ reproduce the choices of global form for the associated $\cN=4$
theory in four dimensions. More precisely, the state $\ket{0;L}$ is
associated with the $\cN=4$ theory with 1-form symmetries determined
by $L$ (and thus, with a specific choice of global form for the gauge
group and discrete theta angles \cite{Aharony:2013hda}), and no
background fluxes. 

\subsection{The K-theory groups of \texorpdfstring{$\cM_6\times S^3/\Gamma$}{M6 x S3/Gamma}} \label{subsec:Ktheory}

In the case of interest to us we have that
$\cN_{9}=\cM_6\times S^3/\Gamma$, so our task is to compute the
$K^1$ group of this space. Since $\cN_{9}$ is a product, we can make
use of the K\"unneth exact sequence for K-theory \cite{ATIYAH1962245}
\begin{equation} \label{eqn:Kunnethformula}
  0 \to \bigoplus_{i+j=m} K^i(X)\otimes K^j(Y) \to K^m(X\times Y) \to \bigoplus_{i+j=m+1} \Tor_\bZ(K^i(X), K^j(Y)) \to 0
\end{equation}
with all indices taken modulo 2. In this equation $\Tor_\bZ(A,B)$ is
the `$\Tor$' functor between $A$ and $B$ (see for instance
\cite{Hatcher:478079} for a definition), which has the property of
vanishing whenever $A$ or $B$ are free. Since we are assuming in our
case that the cohomology of $\cM_6$ has no torsion, we find
\begin{equation} 
  K^1(\cM_6\times S^3/\Gamma) = (K^0(\cM_6)\otimes K^1(S^3/\Gamma)) \oplus (K^1(\cM_6)\otimes K^0(S^3/\Gamma))\, .
\end{equation}

We will compute these K-theory groups by making use of some basic
properties of K-theory. Consider first a manifold $X$ without torsion,
such as $\cM_6$. The existence of the Chern isomorphism
\begin{equation}
  \label{eq:Chern-isomorphism}
  K^i(X)\otimes_\bZ \bQ \cong \bigoplus_{n \equiv i \text{ mod } 2} H^n(X; \bQ)
\end{equation}
immediately implies that
\begin{equation} \label{eqn:oddDegree}
  K^i(X) \cong \bigoplus_{n \equiv i \text{ mod } 2} H^n(X; \bZ)\, .
\end{equation}

The computation of the K-theory groups for $S^3/\Gamma$ is slightly more involved, since
this space has non-vanishing torsion. Remarkably, the end result is that~(\ref{eqn:oddDegree}) still applies. In particular, the cohomology groups of $S^3/\Gamma$ are
\begin{equation}
H^{\bullet}(S^3/\Gamma) = \{\bZ,0,\Gamma^{\text{ab}}, \bZ \} \,,
\end{equation}
where $\Gamma^{\text{ab}}\coloneqq \Gamma/[\Gamma, \Gamma]$ is the
abelianization of $\Gamma$, discussed further below, and we used $\pi_1(S^3/\Gamma) = \Gamma$ (since $S^3$ is the universal cover of $S^3/\Gamma$), along with
$H^2(S^3/\Gamma) = H_1(S^3/\Gamma) = \pi_1(S^3/\Gamma)^{\text{ab}} $
by Poincare duality and the Hurewicz theorem. Thus, (\ref{eqn:oddDegree}) would give
\begin{equation}
  K^0(S^3/\Gamma) = \bZ\oplus \Gamma^{\text{ab}} \,, \qquad K^1(S^3/\Gamma) = \bZ \,.
\end{equation}
That these are indeed the K-theory groups of $S^3/\Gamma$ is shown to be the case in appendix~\ref{app:homology-sphere-K-theory}.

Applying the K-theory K\"unneth formula~(\ref{eqn:Kunnethformula}) and comparing with the K\"unneth formula for cohomology, we see that likewise
\begin{equation}
  K^i(\cM_6 \times S^3/\Gamma) \cong \bigoplus_{n \equiv i \bmod  2} H^n(\cM_6 \times S^3/\Gamma; \bZ)\, ,
\end{equation}
so in this case K-theory reduces to cohomology. In particular,
\begin{equation}
 \Tor H^n(\cM_6 \times S^3/\Gamma) = H^{n-2}(\cM_6) \otimes \Gamma^{\text{ab}} \,,
\end{equation}
and so
\begin{equation} \label{eqn:K1N9}
  \Tor K^1(\cM_6 \times S^3/\Gamma) \cong \bigoplus_{n=1,3,5,7,9} \Tor H^n(\cM_6 \times S^3/\Gamma) = K^1(\cM_6)\otimes \Gamma^{\text{ab}}\, ,
\end{equation}
with potentially non-vanishing contributions in degrees 3, 5 and 7 arising from the degree 1, 3 and 5 components of $K^1(\cM_6) =  \bigoplus_{n=1,3,5} H^n(\cM_6) $, respectively.

\subsection{The defect group and the linking pairing} \label{subsec:linking}

The group $\Gamma^{\text{ab}}$ is easy to determine:\footnote{To
  avoid confusion, we refer to the binary dihedral group of $4n$
  elements as $\mathrm{Dic}_n$ (for \emph{dicyclic}, another name for
  the same family of discrete groups).}
\begin{equation}
  \label{eq:augmented-McKay}
  \begin{array}{c|c|c}
    \Gamma\subset SU(2) & \quad \fg_\Gamma \quad & \quad \Gamma^{\text{ab}} \\
    \hline
    \bZ_N & A_{N-1} & \bZ_N\\
    \text{Binary dihedral } \mathrm{Dic}_{(2k-2)} & D_{2k} & \bZ_2\oplus \bZ_2\\
    \text{Binary dihedral } \mathrm{Dic}_{(2k-1)} & D_{2k+1} & \bZ_4\\
    \text{Binary tetrahedral } 2T & E_6 & \bZ_3\\
    \text{Binary octahedral } 2O & E_7 & \bZ_2 \\
    \text{Binary icosahedral } 2I & E_8 & 1
  \end{array}
\end{equation}
The $\Gamma=\bZ_N$ case is clear, and that of $\Gamma = \mathrm{Dic}_n$ can be worked
out without much effort as follows. A presentation of $\mathrm{Dic}_n$ is
\begin{equation}
  \left\langle a,x \mid a^{2n}=1, x^2=a^n, x^{-1}ax = a^{-1}\right\rangle\, .
\end{equation}
We obtain the abelianization by adding the relation $ax=xa$, which
after some straightforward simplifications leads to
\begin{equation}
  \left\langle a,x \mid x^2=a^n, a^2=1\right\rangle
\end{equation}
which is $\bZ_2\oplus \bZ_2$ for $n$ even and $\bZ_4$ for $n$ odd. Similarly, one can verify the exceptional cases by adding the
relation $st=ts$ to the following presentations for the exceptional
groups
\begin{equation}
  \arraycolsep=4.4pt\def\arraystretch{1.2}
  \begin{array}{c|c}
    \qquad \Gamma \qquad & \text{Presentation} \\
    \hline
    2T & \left\langle s,t \mid (st)^2 = s^3 = t^3 \right\rangle \\
    2O & \left\langle s,t \mid (st)^2 = s^3 = t^4 \right\rangle \\
    2I & \left\langle s,t \mid (st)^2 = s^3 = t^5 \right\rangle
  \end{array}
\end{equation}

Notice that~\eqref{eq:augmented-McKay} follows a simple pattern: let $G_\Gamma$ be
the simply connected Lie group with algebra $\fg_\Gamma$, and
$Z(G_\Gamma)$ its center, then (as already pointed out in
\cite{Acharya:2001hq,DelZotto:2015isa})
\begin{equation}
  \label{eq:global-McKay}
  \Gamma^{\text{ab}} = Z(G_\Gamma)\, .
\end{equation}
This relation will play a key role below when we compare our IIB
analysis with the results of previous analyses of the global structure
of the $(2,0)$ theory. It is not hard to prove that this relation is
not accidental. Since $H_1(S^3/\Gamma) = H^2(S^3/\Gamma) =\Gamma^{\text{ab}}$ as previously remarked, it is sufficient to show that $H_1(S^3/\Gamma)=Z(G_\Gamma)$.

To do so, we first provide an alternate description of $H_1(S^3/\Gamma)$. Recall that whenever
we have a pair of spaces $(X,A)$ such that $A\subset X$ there is a
long exact sequence in homology of the form
\cite{Hatcher:478079}
\begin{equation}
  \ldots \to H_n(A) \to H_n(X) \to H_n(X,A) \to H_{n-1}(A) \to \ldots
\end{equation}
where $H_n(X,A)$ denotes the singular homology of $X$ relative to
$A$. We take $A$ to be $S^3/\Gamma$, and $X_\Gamma$ to be a smooth, simply-connected space
such that $\partial X_\Gamma = S^3/\Gamma$. More concretely,
$X_\Gamma$ can be taken to be a sufficiently large neighbourhood of
the origin of a resolved $\bC^2/\Gamma$. Since $H_1(X_\Gamma)=0$ and
$H_2(S^3/\Gamma)=0$, we have the short exact sequence
\begin{equation}
  \label{eq:exact-homology}
  0 \to H_2(X_\Gamma) \to H_2(X_\Gamma, S^3/\Gamma) \xrightarrow{\partial} H_1(S^3/\Gamma) \to 0\, .
\end{equation}
Geometrically, this exact sequence encodes the fact that one-cycles in
$S^3/\Gamma$ can be constructed by intersecting a non-compact 2-cycle
in $X_\Gamma$ with the $S^3/\Gamma$. Clearly, adding compact 2-cycles
has no effect on this description, hence the exact sequence.

More physically, we can understand the quotient
\begin{equation}
  \label{eq:defect-group}
  \cC\coloneqq H_1(S^3/\Gamma) = \frac{H_2(X_\Gamma, S^3/\Gamma)}{H_2(X_\Gamma)}
\end{equation}
as a ``defect group''~\cite{DelZotto:2015isa, Gukov:2018iiq}
describing the screening of surface operators, in analogy with the
field theory analysis in \cite{THOOFT19781,THOOFT1979141}. In brief,
$H_2(X_\Gamma, S^3/\Gamma)$ is expected to parametrize the surface
operators in the six-dimensional SCFT living at the singular point,
while $H_2(X_\Gamma)$ parametrizes the ``charge carriers'' of the
theory, and so $\cC$ measures how much of the charge of the surface
operators remains unscreened in the 6d SCFT. We refer the reader to
\cite{DelZotto:2015isa} for a more detailed discussion of $\cC$ from
this viewpoint.

Recall that we can identify $H_2(X_\Gamma)$ with the root lattice
$\Lambda^r_\Gamma$ of $\fg_\Gamma$.
 Because $\fg_\Gamma$ is simply laced, $\Lambda^r_\Gamma$ is also the coroot
lattice, whose dual is the weight lattice $\Lambda^w(G_\Gamma)$ of the universal
cover $G_\Gamma$. On the other hand, geometrically we have that
\begin{equation} \label{eqn:H2rel}
  H_2(X_\Gamma, S^3/\Gamma) = H^2(X_\Gamma)=\Hom(H_2(X_\Gamma), \bZ)
\end{equation}
where the first equality is Lefschetz duality and in the second
 we have used the universal coefficient theorem together with
$H_1(X_\Gamma)=0$. We are thus led to identify
$H_2(X_\Gamma,S^3/\Gamma)$ with $\Lambda^w(G_\Gamma)$. Therefore, we can
rephrase~\eqref{eq:defect-group} in group theory terms as
\begin{equation}
  H_1(S^3/\Gamma) = \frac{\Lambda^w(G_\Gamma)}{\Lambda^r_\Gamma} \,.
\end{equation}
It is well known that this quotient
is $Z(G_\Gamma)$, see for instance theorem 23.2 of
\cite{bump2004lie}.

\bigskip

We now come back to the perfect pairing $s(x,y)$ introduced
in~\eqref{eq:flux-commutator}. A key ingredient in constructing this pairing is the linking (or \emph{torsion}) pairing
$\mathsf{L}(x,y)$, which is a perfect pairing of the form
\begin{equation}
  \mathsf{L} \colon \Tor H_{p-1}(\cN_{n-1})\times \Tor H_{n-p-1}(\cN_{n-1}) \to \bQ/\bZ \,,
\end{equation}
describing the linking of torsion homology classes on a $(n-1)$-dimensional manifold $\cN_{n-1}$. To define this pairing, consider a
torsion homology class $[a] \in \Tor H_{p-1}(\cN_{n-1})$ of order $k_a$, so that $k_a [a] = 0$. Thus, given a representative $a_{p-1}$ of the class $[a]$, there is a chain $A_{p}$ such that $k_a a_{p-1} = \partial A_{p}$. We define
\begin{equation}
  \label{eq:linking-homology}
  \L(a,b) \equiv \frac{1}{k_a} (A_p \circ b_{n-p-1}) \pmod 1 \,,
\end{equation}
where $x \circ y$ denotes the signed intersection number between
transversely intersecting chains $x$, $y$ on $\cN_{n-1}$. This
definition is independent of the choice of $A_p$ for fixed $a_{p-1}$,
as the intersection number of $[b]$ (a torsion cycle) with any closed
cycle vanishes. Likewise, it does not depend on the choice of
representative $a_{p-1}$ within the torsion class $[a]$, as
$a_{p-1} \to a_{p-1} + \partial \lambda_p$ shifts $\L(a,b)$ by an
integer $\lambda_p \circ b_{n-p-1}$. Finally, noting that
\begin{equation}
\Sigma_p \circ \partial \Sigma_{n-p} = (-1)^{p(n-p)} \Sigma_{n-p} \circ \partial \Sigma_p \,,
\end{equation}
we find $\L(b,a) = (-1)^{p(n-p)} \L(a,b)$, implying that $\L(a,b)$ is also independent of the choice of representative $b_{n-p-1}$ of the torsion class $[b] \in \Tor H_{n-p-1}(\cN_{n-1})$.

By Poincar\'e duality, the linking pairing can also be framed in cohomology:
\begin{equation}
  \L\colon \Tor H^{n-p}(\cN_{n-1})\times \Tor H^p(\cN_{n-1}) \to \bQ/\bZ\,.
\end{equation}
To define it in cohomological terms, consider the short exact sequence
\begin{equation}
  0 \to \bZ \to \bQ \to \bQ/\bZ \to 0 \,,
\end{equation}
which induces a long exact sequence in cohomology of the form
\begin{equation}
  \ldots \to H^k(\cN_{n-1};\bZ) \xrightarrow{\pi} H^k(\cN_{n-1};\bQ) \to H^k(\cN_{n-1};\bQ/\bZ) \xrightarrow{\delta} H^{k+1}(\cN_{n-1};\bZ) \to \ldots \,,
\end{equation}
where $\delta$, (induced by) the coboundary operator, is sometimes called the Bockstein homomorphism. Given $x \in \Tor H^{n-p}(\cN_{n+1}; \bZ)$, $\pi(x) = 0$, and thus exactness of the above sequence implies that $X \in H^{n-p-1}(\cN_{n-1};\bQ/\bZ)$ exists such that $\delta X = x$. The linking pairing is then
\begin{equation}
  \L(x,y) = \int_{\cN_{n-1}} X \smile y \,,
\end{equation}
which is valued in $\bQ / \bZ$. Writing $y = \delta Y$ for $Y \in H^{p-1}(\cN_{n-1};\bQ/\bZ)$ as well, this becomes $\int_{\cN_{n-1}} X \smile \delta Y$ (schematically ``$\int X \wedge d Y$''), in which form the properties discussed in the preceding paragraph are readily established.

We now consider Maxwell theory for a $(p-1)$-form gauge potential, following Freed, Moore and Segal~\cite{Freed:2006yc}.
Given electric and magnetic torsion classes, $x \in H^{d-p}(\cN_{d-1})$ and $y \in H^{p}(\cN_{d-1})$ respectively, the corresponding flux operators $\Phi_x$ and $\Phi_y$ do not commute,
\begin{equation}
\Phi_x \Phi_y = \e^{2 \pi i\,\L(x,y)} \Phi_y \Phi_x \,,
\end{equation}
where $\L(x,y)$ is the linking pairing we have just discussed. The situation is slightly different for self-dual gauge fields, for which electric and magnetic fluxes are one and the same. In this case, the commutator is
\begin{equation}
\Phi_x \Phi_y = \e^{2 \pi i\, \L(x,y)} S(x,y) \Phi_y \Phi_x \,,
\end{equation}
where $S(x,y)$ is a correction of the form
\begin{equation} \label{eqn:Sxy}
 S(x,y) = \frac{1+S(x)+S(y)-S(x)S(y)}{2} \,, \qquad S(x) = (-1)^{\int_{\cN_{d-1}} x\smile \nu_{2k}} \,,
\end{equation}
with $k = \frac{d-2}{4}$ and $\nu_{2k}$ the Wu class of degree
$2k$.\footnote{We assume that $d/2$ is odd in the self-dual case. For a more general discussion, see~\cite{Freed:2006ya}.} Note that $S(x), S(y) = \pm 1$, with $S(x,y) = -1$ if $S(x) = S(y) = -1$ and $S(x,y) = +1$ otherwise. This correction is needed because, e.g., the linking pairing is not alternating on $H^{d/2}(\cN_{d-1})$~\cite{Freed:2006yc}.

In this paper, our primary interest is in type IIB string theory on $\cN_9 = \cM_6\times S^3/\Gamma$. Associated to the self-dual RR field $C_4$, there are flux operators labelled by torsion classes
\begin{equation}
  x=c_3\otimes \ell_2\in \Tor H^5(\cN_9) = H^3(\cM_6) \otimes H^2(S^3/\Gamma) \,.
\end{equation}
For such classes, $\int_{\cN_{9}} x\smile \nu_{4} \ne 0$ requires $\nu_4$ to have components of the form
$p_3\otimes q_1\in H^3(\cM_6) \otimes H^1(S^3/\Gamma)$. Since
$H^1(S^3/\Gamma)=0$, we conclude that $S(x)=1$ for all $x \in H^5(\cN_9)$, and so the $S(x,y)$ correction factor can be dropped.

More generally, the RR fluxes are described by K-theory rather than cohomology. However, we have shown that the K-theory groups of $\cN_9 = \cM_6\times S^3/\Gamma$ reduce to cohomology groups, and so it is natural to guess that the flux commutators likewise reduce to the cohomological ones discussed above, and in particular that the perfect pairing $s(x,y)$ introduced in~\eqref{eq:flux-commutator} is given by
\begin{equation} \label{eqn:Ktheorypairing}
s(x,y) = \e^{2 \pi i \L(x,y)} \,, \qquad x,y \in \Tor K^1(\cN_9) = \bigoplus_{i = 2k+1} \Tor H^i(\cN_9) \,,
\end{equation}
where $\L(x,y) = 0$ when the degrees of $x$ and $y$ do not add to $d=10$, and the correction factor $S(x,y)$ is absent per the above discussion. Indeed, (\ref{eqn:Ktheorypairing}) follows from the K-theory pairing found by Freed, Moore and Segal~\cite{Freed:2006yc}, validating this guess.\footnote{To see this, one can use the result in Klonoff's
  thesis \cite{klonoff2008index} to express the integral over
  differential K-theory classes in terms of the $\eta$-invariant. As
  shown by Atiyah, Patodi and Singer in \cite{APS-III} the $\eta$
  invariant on $\cM_6\times S^3/\Gamma$ will factor into the index on
  $\cM_6$ times $\eta$ on $S^3/\Gamma$. For odd forms the first term
  will be simply the intersection pairing on $\cM_6$, and since
  $\Omega_3^\Spin(\pt)=0$ the last quantity will be equal (mod 1) to
  the Chern-Simons invariant of the torsional class on $S^3/\Gamma$,
  reproducing the expression in cohomology.}

\bigskip

We now compute the linking pairing $\L(x,y)$ for $\cN_9 = \cM_6\times S^3/\Gamma$. It is convenient to work in homology.
Since torsion
comes from the $S^3/\Gamma$ component, we have
\begin{equation}
  \L(a\otimes \ell_1, b\otimes \ell_2) = (a\circ b) \L_\Gamma(\ell_1, \ell_2) \,,
\end{equation}
with $a,b\in H_*(\cM_6)$ and $\ell_1,\ell_2\in H_1(S^3/\Gamma)$. Thus, it is sufficient to compute the linking pairing $\L_\Gamma: H_1(S^3/\Gamma) \times H_1(S^3/\Gamma) \to \bQ / \bZ$, along with the intersection form on $\cM_6$.

To write down the linking pairing $\L_\Gamma$, it is convenient to use a construction of $H_1(S^3/\Gamma)$ that emphasizes
the intersection form on $X_\Gamma$. Using~(\ref{eqn:H2rel}), we can
rewrite~\eqref{eq:exact-homology} as
\begin{equation} \label{eqn:H1seq}
  0 \to H_2(X_\Gamma) \xrightarrow{Q} \Hom(H_2(X_\Gamma), \bZ) \xrightarrow{\partial} H_1(S^3/\Gamma) \to 0 \,,
\end{equation}
where $Q$ is the homomorphism
\begin{equation}
  \label{eq:Q-Hom}
  \begin{aligned}
  Q \colon H_2(X_\Gamma) & \to \Hom(H_2(X_\Gamma), \bZ) \\
  x & \mapsto q(x,\cdot)
  \end{aligned}
\end{equation}
with $q$ the intersection form on $H_2(X_\Gamma)$. 
Therefore,
\begin{equation}
  \label{eq:torsion-from-defect-group}
  H_1(S^3/\Gamma) = \frac{\Hom(H_2(X_\Gamma), \bZ)}{Q(H_2(X_\Gamma))}\, .
\end{equation}
The linking pairing on $S^3/\Gamma$ can be constructed from this short
exact sequence and the intersection form $q$, as follows (see also
\cite{Gukov:2018iiq}). Given $\sigma_1, \sigma_2 \in H_1(S^3/\Gamma)$,
we pick $\xi_i \in \partial^{-1}(\sigma_i)$. Then, since
$H_1(S^3/\Gamma)$ is pure torsion, there exists $n_i \ne 0$ such that
$\partial (n_i \xi_i) = n_i \sigma_i = 0$, and therefore we can pick
$\Sigma_i \in H_2(X_\Gamma)$ such that $n_i \xi_i = Q(\Sigma_i)$. The
linking pairing is then\footnote{There is some ambiguity in the
  literature regarding the overall sign of the linking number. We
  follow the conventions in \cite{ConwayFriedlHerrmann}.}
\begin{equation}
  \L_\Gamma(\sigma_1, \sigma_2) \equiv \frac{1}{n_1n_2}q(\Sigma_1,\Sigma_2) \equiv \frac{1}{n_2} \xi_1(\Sigma_2) \equiv \frac{1}{n_1} \xi_2(\Sigma_1)  \pmod 1 \,.
\end{equation}
Equivalently, this can be written as
\begin{equation}
  \L_\Gamma(\sigma_1, \sigma_2) \equiv q^{-1}(\xi_1, \xi_2) \pmod 1 \,, \label{eq:linking-from-intersection}
\end{equation}
with $q^{-1}: \Hom(H_2(X_\Gamma), \bZ) \times \Hom(H_2(X_\Gamma), \bZ) \to \bQ$ defined precisely by the above procedure.\footnote{Note that $q^{-1}$ need not be integral; since $q$ is integral, $q^{-1}$ is integral iff $\det q = \pm 1$.}

\begin{figure}
  \centering
  \includegraphics[width=0.4\textwidth]{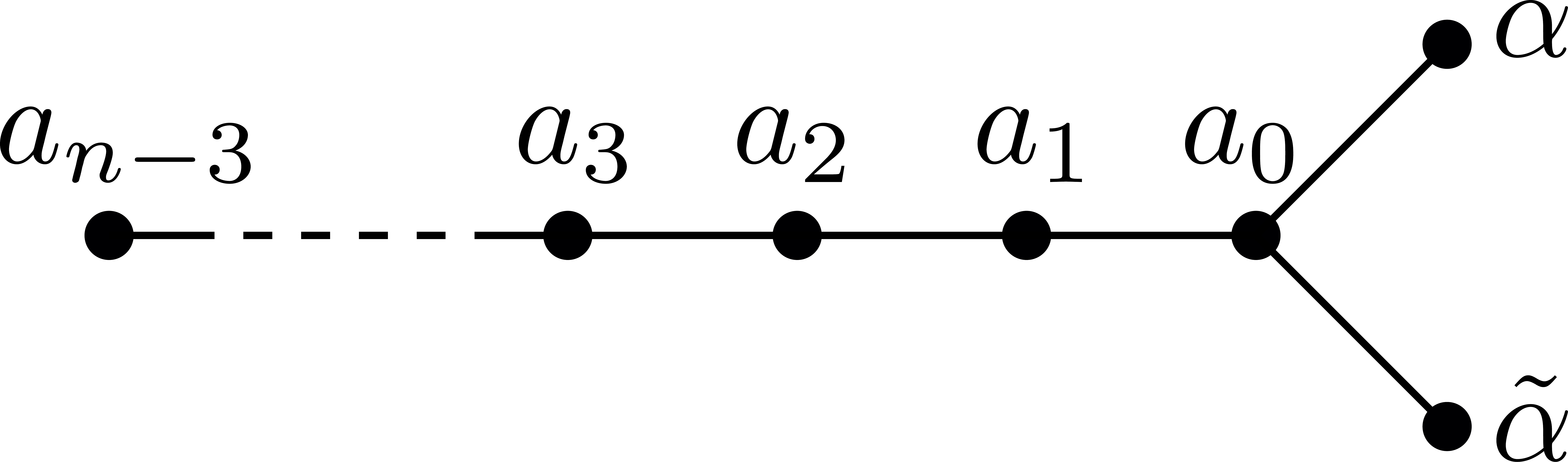}
  \caption{Dynkin diagram for $D_n$.}
  \label{fig:Dn-Dynkin}
\end{figure}

We now discuss examples, starting with the $D_{n}$ case. The
structure of $H_2(X_{D_{n}})$ together with its intersection form is
encoded in the Dynkin diagram shown in figure~\ref{fig:Dn-Dynkin},
where each dot represents a generator of $H_2(X_{D_{n}})$ and each
link between nodes indicates that the given homology classes intersect
once. Ordering the homology basis elements as
$\{\alpha,\tilde{\alpha},a_0,a_1,\ldots,a_{n-3}\}$ we have the intersection matrix
\begin{equation}
  q = \begin{pmatrix}
    -2 & 0 & 1 &  &    \\
    0 & -2 & 1 &  &    \\
    1 & 1 & -2 & 1 &     \\
    &  & 1 &\ddots & 1     \\
     &  &  & 1 & -2     \\
  \end{pmatrix}\, .
\end{equation}
We introduce a dual basis of $\Hom(H_2(X_{D_{n}}), \bZ)$ given by
$\{\alpha^*, \tilde{\alpha}^*, a_0^*, a_1^*, \ldots, a_{n-3}^*\}$, with the property
that $a^*_i(a_j)=\delta_{ij}$, and similarly for $\alpha$ and
$\tilde{\alpha}$. The relations introduced by $Q$ on
$\Hom(H_2(X_{D_{n}}), \bZ)$ are then
\begin{equation}
\begin{alignedat}{4}
  Q(\alpha) & = -2\alpha^* + a_0^* & &= 0\,, \hspace{1cm} & Q(a_1) & = -2a_1^* + a_0^* + a_2^* & &= 0\,, \\
    \MTFlushSpaceAbove
  &&&&&\vdotswithin{=}
  \MTFlushSpaceBelow
  Q(\tilde{\alpha}) & = -2\tilde{\alpha}^* + a_0^* & &= 0\,, &  Q(a_i) & = -2a_i^* + a_{i-1}^* + a_{i+1}^* & &=0\,,\\
  \MTFlushSpaceAbove
  &&&&&\vdotswithin{=}
  \MTFlushSpaceBelow
  Q(a_0) & = -2 a_0^* + a_1^* + \alpha^* + \tilde{\alpha}^* & &= 0\,, & Q(a_{n-3}) & = -2a_{n-3}^* + a_{n-4}^* & &= 0\,.
\end{alignedat}
\end{equation}
A little bit of algebra shows that these relations imply that
$a_k^*=(n-2-k)a_{n-3}^*$, so we can take
$a_{n-3}^*,\alpha^*,\tilde{\alpha}^*$ as generators of the
quotient~\eqref{eq:torsion-from-defect-group}, subject to the
remaining relations
  \begin{align}
    \alpha^* + \tilde{\alpha}^* & = (n-1) a_{n-3}^*\,, &
    2\alpha^* & = 2\tilde{\alpha}^* = (n-2) a_{n-3}^*\,, 
  \end{align}
which implies $2 a_{n-3}^*=0$. We now distinguish whether $n$ is even or
odd. For $n$ even we have
\begin{align}
  2\alpha^* & = 2\tilde{\alpha}^* = 0\,, & a_{n-3}^* &= \alpha^* + \tilde{\alpha}^*\, .
\end{align}
This is a $\bZ_2\oplus \bZ_2$ group, in agreement with
\eqref{eq:augmented-McKay}. We can choose $\alpha^*$ and $\tilde{\alpha}^*$ as
generators. Furthermore, it is easy to verify that when restricted to
$\alpha^*$ and $\tilde{\alpha}^*$ we have, using
\eqref{eq:linking-from-intersection}
\begin{equation}
  \label{eq:Leven-Lodd}
  \L_\Gamma = q^{-1} \bmod 1 = \begin{cases}
    \L_{\text{even}} \coloneqq \begin{pmatrix}
      0 & \frac{1}{2}\\
      \frac{1}{2} & 0
    \end{pmatrix} & \text{ for } n\in 4\bZ\, ,\\
    \L_{\text{odd}} \coloneqq \begin{pmatrix}
      \frac{1}{2} & 0\\
      0 & \frac{1}{2}
    \end{pmatrix} & \text{ for } n\in 4\bZ+2\, .
  \end{cases}
\end{equation}

If we instead choose $n$ to be odd, we obtain the equations
\begin{align}
  \alpha^* + \tilde{\alpha}^* &= 0\,, &  a_{n-3}^* &= 2\alpha^* = 2\tilde{\alpha}^* \, ,
\end{align}
which gives a presentation of a $\bZ_4$ group generated by
$\alpha^*$. From the inverse intersection form we obtain
\begin{equation}
  q^{-1}(\alpha^*,\alpha^*) = -\frac{n}{4} \,.
\end{equation}
Taking into account that $n$ is odd, we obtain a linking form
\begin{equation}
  \L_\Gamma(\partial \alpha^*, \partial \alpha^*) = \frac{(-1)^{\frac{n+1}{2}}}{4} \mod 1 \, .
\end{equation}
Other cases can be analyzed similarly; we will present the results
below.

The technology that we developed above is not restricted to ALE cases,
and applies equally well to any IIB background such that the horizon
manifold is smooth.\footnote{In some cases the IIB axio-dilaton might
  have non-trivial behaviour at infinity, so K-theory is not
  necessarily the right framework for classifying fluxes. (We refer
  the reader to \cite{Evslin:2006cj} for a review of some of the
  difficulties in trying to extend the K-theory classification to
  situations in which $SL(2,\bZ)$ dualities are important.) Our
  discussion below deals with $F_5$ only, which is invariant under
  $SL(2,\bZ)$ transformations, and we are in a context where K-theory
  reduces to cohomology, so we expect our results to survive in a more
  careful treatment.} We will
determine the linking pairing (and thus operator commutation relations
in the six dimensional theory) geometrically in a number of cases,
including those where more than one possibility exists at the level of
the algebra. In particular, we can apply this method to geometrically
engineered $(1,0)$ theories in six dimensions, as studied in
\cite{DelZotto:2015isa}.

\begin{figure}
  \centering
  \begin{subfigure}[t]{0.3\textwidth}
    \centering
    \includegraphics[height=1cm]{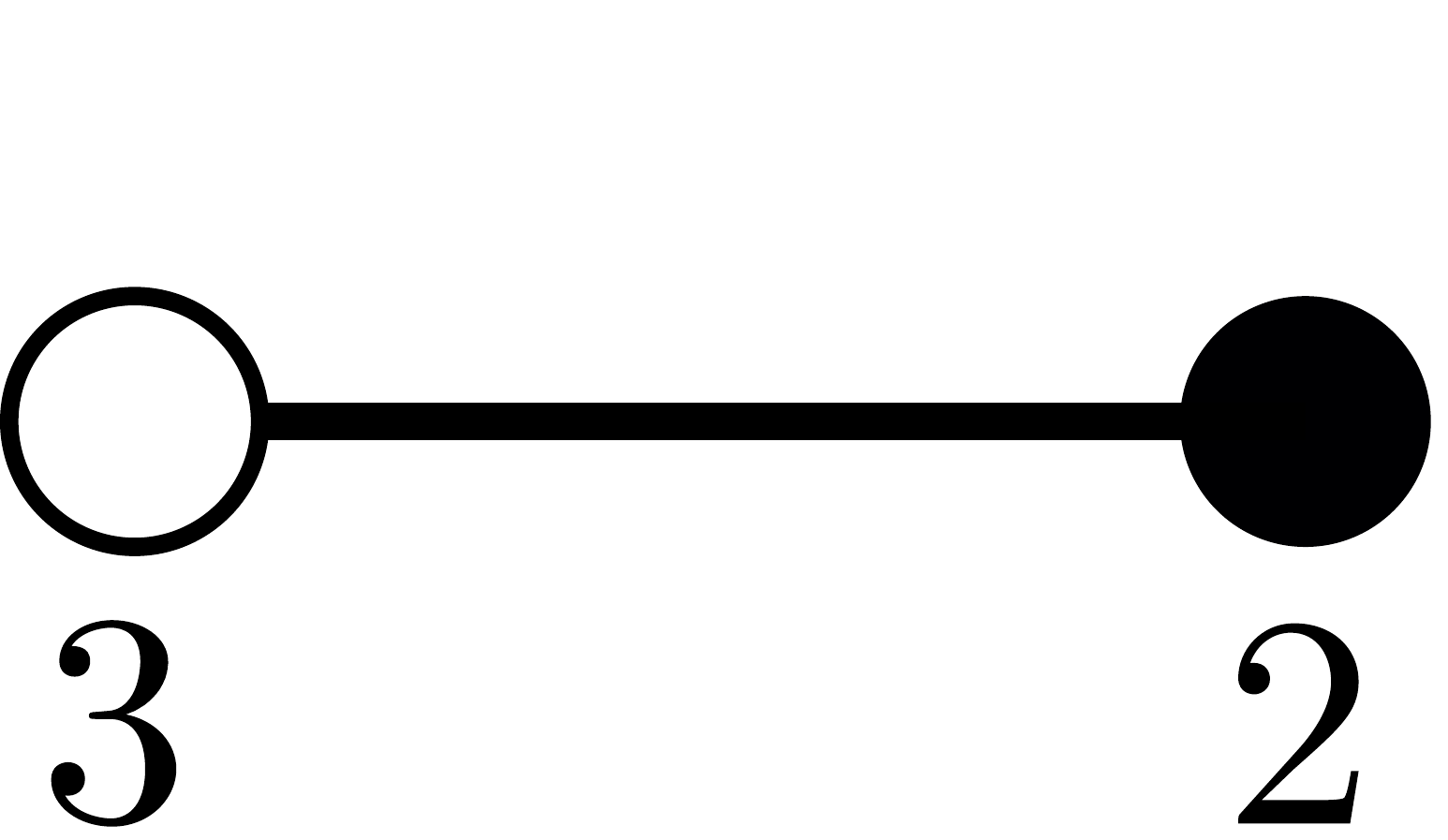}
    \caption{$A_{5,2}$}
    \label{sfig:32-Dynkin}
  \end{subfigure}
  \hfill
  \begin{subfigure}[t]{0.3\textwidth}
    \centering
    \includegraphics[height=2cm]{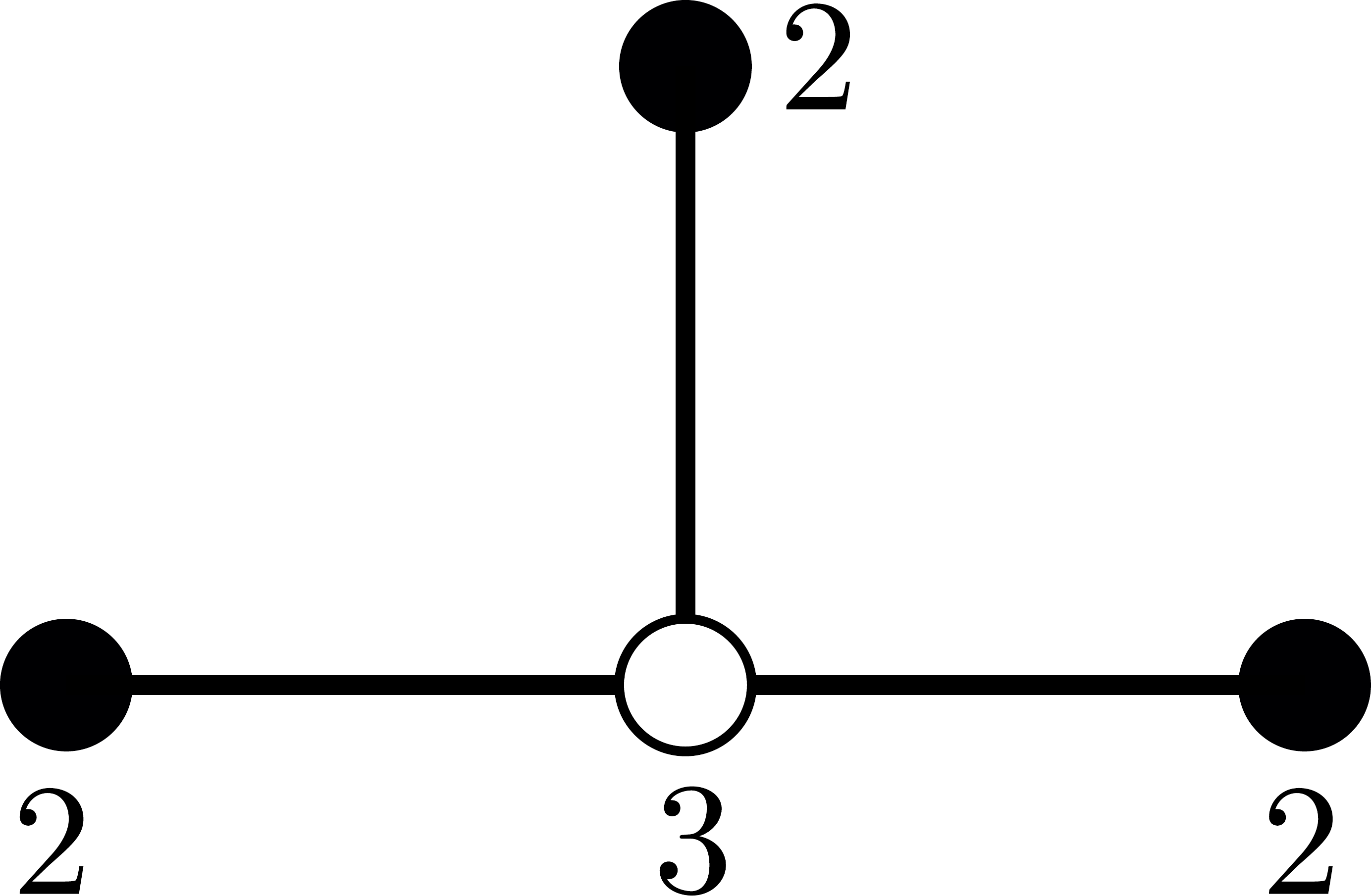}
    \caption{$D_{5,2}$}
    \label{sfig:D_{5,2}}
  \end{subfigure}
  \hfill
  \begin{subfigure}[t]{0.3\textwidth}
    \centering
    \includegraphics[height=2cm]{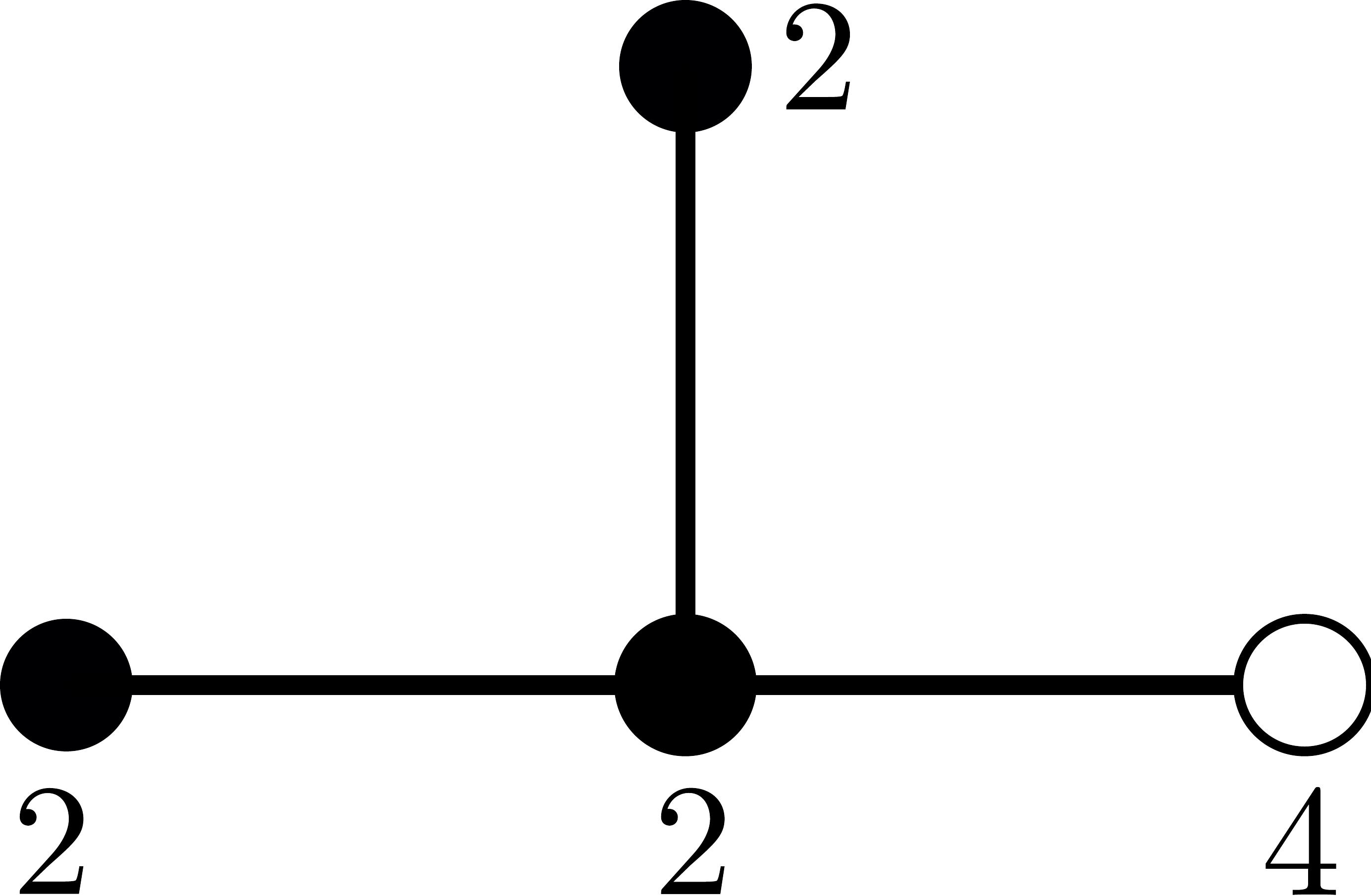}
    \caption{$D_{7,4}$}
    \label{sfig:D_{7,4}}
  \end{subfigure}
  \caption{The generalized geometries considered in the text. Nodes
    denote two-cycles, a line connecting two nodes indicates that the
    cycles intersect each other transversely, and a number next to the
    node denotes (minus) its self-intersection.}
\end{figure}

Consider for instance the case in which the small resolution of
$X_\Gamma$ has two curves $a_1$ and $a_2$, of self-intersection $-3$
and $-2$ respectively. The two curves intersect at a point. The
resulting intersection diagram is shown in
figure~\ref{sfig:32-Dynkin}. This geometry is one of the ``generalized
$A$-type'' configurations studied in
\cite{Heckman:2013pva,DelZotto:2015isa}, to which we refer the reader
interested in further details. The point of greatest interest to us is
that $X_\Gamma$ can be understood as a desingularization of
$\bC^2/\Gamma$, with $\Gamma$ a $\bZ_5$ subgroup of $U(2)$ acting as
\begin{equation}
  (z_1,z_2) \to (\omega z_1, \omega^2 z_2) \,,
\end{equation}
with $\omega=\exp(2\pi i/5)$. The intersection matrix for this
geometry, in the $a_i$ basis, is
\begin{equation}
  q = \begin{pmatrix}
    -3 & 1\\
    1 & -2
  \end{pmatrix} \,,
\end{equation}
leading to the relations
\begin{equation}
  3a_1^* = a_2^* \,, \qquad 2a_2^* = a_1^*\, .
\end{equation}
From here we learn that $H_1(S^3/\Gamma)=\bZ_5$, as expected. Given
that $2^{-1}=3$ in $\bZ_5$ we can take either $a_1^*$ or $a_2^*$ as
generators, let us take $a_1^*$ for convenience. We have
\begin{equation}
  \L_\Gamma(\partial a_1^*, \partial a_1^*) \equiv q^{-1}(a_1^*, a_1^*) \equiv \frac{3}{5} \pmod 1\, .
\end{equation}
Note that 3 is not a quadratic residue in $\bZ_5$, so this linking
form is inequivalent to the one with value $\frac{1}{5}$ for the
linking number of the generator with itself.

As another illustration, consider the $d=6$, $\cN=(1,0)$
compactifications classified in \cite{Heckman:2013pva,Heckman:2015bfa,Bhardwaj:2015xxa}. The associated defect
group was discussed in
\cite{DelZotto:2015isa}, where it was shown that for the
``generalized $D$-type'' singularities, obtained by taking the
quotient $\bC^2/D_{p+q,q}$,\footnote{$D_{p+q,q}$ is a certain subgroup of
$U(2)$ acting freely on the $S^3$ at infinity. For details see
\cite{DelZotto:2015isa}.} one has
\begin{equation}
  H_1(S^3/D_{p+q,q}) = \frac{\Hom(H_2(X_{D_{p+q,q}}), \bZ)}{Q(H_2(X_{D_{p+q,q}}))} = \bZ_2 \oplus \bZ_{2p} \,,
\end{equation}
whenever $q$ is even. Consider for example the case $p=3$. In this
case $\bZ_{2p}\cong \bZ_2\oplus\bZ_3$. Up to a sign that we specify
below, there is a unique linking form for the $\bZ_3$ factor, but for
the remaining $\bZ_2\oplus\bZ_2$ factor we have two possibilities,
given by $\L_{\text{even}}$ and $\L_{\text{odd}}$
in~\eqref{eq:Leven-Lodd}. And indeed both possibilities appear: a
straightforward application of the techniques above shows that the
$(p,q)=(3,2)$ case (in figure~\ref{sfig:D_{5,2}}) has intersection
form $\L_{\text{even}}$, while the $(p,q)=(3,4)$ case (in
figure~\ref{sfig:D_{7,4}}) has a linking form given by
$\L_{\text{odd}}$.

Finally, let us consider the ``generalized $D_N$'' theory of type
$D_{p+q,q}$ with $(p,q)=(2,7)$. It was shown in
\cite{DelZotto:2015isa} that the defect group in this case is
$\bZ_8$. A computation along the lines described above shows that the
linking form is
\begin{equation}
  \L_\Gamma(\ell,\ell) = \frac{5}{8} \mod 1
\end{equation}
with $\ell$ the generator of $H_1(S^3/D_{9,7})$. Note that 5 is not a
residue modulo 8, so this linking form is inequivalent to the naive
pairing $\L(\ell,\ell)=\frac{1}{8}$ mod 1. Another inequivalent
pairing
\begin{equation}
   \L_\Gamma(\ell, \ell) = \frac{3}{8} \mod 1
\end{equation}
is also realized, for instance by choosing $(p,q)=(2,9)$. More
generally, one finds that $S^3/D_{p+q,q}$ for $(p,q)=(2,2k+1)$ has pairing
\begin{equation}
   \L_\Gamma(\ell,\ell) = \frac{3q}{8} \mod 1
\end{equation}
despite the defect group always being $\bZ_8$, so all possible
pairings are realized.

Other generalized $D_N$ theories can be analyzed similarly, we will
briefly state the results without proof. For instance, consider the
family of theories $D_{p+q,q}$ with $(p,q)=(3,3k+1)$. One finds that
the defect group is $\bZ_3\oplus \bZ_2\oplus \bZ_2$ for $q$ even (or
equivalently, $k$ odd), and $\bZ_{12}$ for $q$ odd
\cite{DelZotto:2015isa}. In this last case we find the linking form
\begin{equation}
  \L_\Gamma(\ell, \ell) = -\frac{q}{12} + \frac{1}{2} \mod 1
\end{equation}
with $\ell$ the generator of $\bZ_{12}$. For $q$ even, we find instead
\begin{equation}
  \L_\Gamma(\ell_3, \ell_3) = \frac{2}{3} \mod 1
\end{equation}
with $\ell_3$ a generator of the $\bZ_3$ factor, and
\begin{equation}
  \L_\Gamma|_{\bZ_2\oplus\bZ_2} = \begin{cases}
    \L_{\text{odd}} & \text{when } q\in 4\bZ\\
    \L_{\text{even}} & \text{when } q\in 4\bZ+2
  \end{cases}
\end{equation}
for the restriction of the linking form to the $\bZ_2\oplus\bZ_2$
factor. One can also see that the $(p,q)=(3, 3k-1)$ case leads to
precisely the same results as the ones we have just given.

\section{Comparison with known results in four and six dimensions} \label{sec:comparisonall}

Let us summarize the story so far. Quantizing type IIB string theory on a non-compact manifold $\cM_6 \times \bC^2 / \Gamma$ requires a choice of flux boundary conditions on $\cN_9 = \cM_6 \times S^3 / \Gamma$. Because electric and magnetic flux operators do not commute, there is no canonical ``zero flux'' boundary condition that we can choose. Instead, the possible boundary conditions for the RR fluxes are states in a Hilbert space acted on by the flux operators $\Phi_x$, $x \in \Tor K^1(\cN_9)$, with commutation relations
\begin{equation}
\Phi_x \Phi_y = s(x,y) \Phi_y \Phi_x \,,
\end{equation}
where $s(x,y)$ is a perfect pairing. Maximal commuting subsets of the flux operators are in direct correspondence with maximal isotropic subspaces $L \subset \Tor K^1(\cN_9)$ with respect to the perfect pairing $s(x,y)$. Given maximal isotropic $L$, there is a basis of eigenstates $|f;L\rangle$ labeled by cosets $f \in F_L = \Tor K^1(\cN_9) / L$ with
\begin{equation}
\forall x \in L, \qquad \Phi_x |f; L\rangle = s(x,f) |f; L\rangle \,.
\end{equation}
These states have boundary flux in a definite coset $f \in F_L$, the strongest condition that we can consistently impose. In particular, $f\cong 0$ (restricting the flux to lie along $L$) is the closest we can come to a ``zero flux'' boundary condition. The resulting quantization depends on the choice of maximal isotropic subspace $L \subset \Tor K^1(\cN_9)$.

Note that the flux operators $\Phi_x$ generate a Heisenberg group, summarized by the short exact sequence,\footnote{To be precise, this sequence is exact if we take $\cW$ to be generated by the flux operators and arbitrary $\U(1)$ phase factors. If we take $\cW$ to be generated by the flux operators alone, then $\U(1)$ must be replaced by $\bZ_N$ in the exact sequence, where $N$ is the order of the largest cyclic subgroup of $\Tor K^1(\cN_9)$.}
\begin{equation}
  0 \to U(1) \to \cW \xrightarrow{\pi} \Tor K^1(\cN_9) \to 0\, ,
\end{equation}
where $\pi(\Phi_x) = x$. The Hilbert space of flux boundary conditions discussed above is the unique irreducible representation of $\cW$, and so the Heisenberg group $\cW$ is a convenient avatar for the choice of boundary conditions.

For $\cN_9 = \cM_6 \times S^3 / \Gamma$ with $\cM_6$ torsion-free, $K^1(\cN_9)$ is the sum of cohomology groups of odd degree and the perfect pairing is
\begin{equation} \label{eqn:N9linking}
  s(a_1\otimes \ell_1, a_2\otimes \ell_2) = \exp\biggl(2\pi i \,  \L_\Gamma(\ell_1, \ell_2)\, \int_{\cM_6} a_1 \smile a_2 \biggr) \,,
\end{equation}
where $a_i\in H^{1,3,5}(\cM_6)$, $\ell_i\in H^2(S^3/\Gamma) = \Gamma^{\text{ab}}$, and $\L_\Gamma$ is the linking pairing for $S^3 /\Gamma$, which can be computed using the methods described in the previous section. For instance, for $\Gamma \subset \SU(2)$ --- leading to the $(2,0)$ theories --- we find (see also \cite{Hikami:2009ze})
\begin{equation}
  \label{eq:ADE-linking-forms}
  \arraycolsep=4.4pt\def\arraystretch{1.2}
  \begin{array}{cccc}
    \Gamma & G_\Gamma & \Gamma^{\text{ab}} &  \L_\Gamma \\
    \hline
    \bZ_N & SU(N) & \bZ_N & \frac{1}{N} \\
    \mathrm{Dic}_{(4N-2)} & \Spin(8N) & \bZ_2\oplus \bZ_2 & \L_{\text{even}} \\
    \mathrm{Dic}_{(4N-1)} & \Spin(8N+2) & \bZ_4 & \frac{3}{4} \\
    \mathrm{Dic}_{(4N)} & \Spin(8N+4) & \bZ_2\oplus \bZ_2 & \L_{\text{odd}} \\
    \mathrm{Dic}_{(4N+1)} & \Spin(8N+6) & \bZ_4 & \frac{1}{4} \\
    2T & E_6 & \bZ_3 & \frac{2}{3} \\
    2O & E_7 & \bZ_2 & \frac{1}{2} \\
    2I & E_8 & 0 & 0
  \end{array} 
\end{equation}
When the defect group $\Gamma^{\text{ab}}$ is cyclic, we list $\L_\Gamma(a,a)$ for the generator $a$, whereas for $\Gamma^{\text{ab}} = \bZ_2\oplus\bZ_2$, we refer to the two cases in~(\ref{eq:Leven-Lodd}).

We emphasize that the correct linking pairing is in general \emph{not}
determined by the defect group. For instance, the defect groups for
$\Spin(8N)$ and $\Spin(8N+4)$ are both $\bZ_2 \oplus \bZ_2$, but the
linking pairings are distinct. This has physical consequences, e.g.,
for S-duality in 4d compactifications of these theories, and we will
see that the linking pairings in~(\ref{eq:ADE-linking-forms})
correctly reproduce known results from the literature. This is a
sensitive test of our methods.

\subsection{\texorpdfstring{$(2,0)$}{(2,0)} theories}

\label{sec:comparison-6d}

This solves the problem of specifying the RR flux boundary conditions for type
IIB string theory compactified on $\cM_6 \times \bC^2/\Gamma$. We now compare our results with known results about the global structure of 6d $(2,0)$ theories with simple Lie algebras.
To do so, we use the universal
coefficient theorem, which is the short exact sequence (see theorem~2.33 in
\cite{DavisKirk})
\begin{equation}
  0 \to H^n(X)\otimes A \to H^n(X; A) \to \Tor(H^{n+1}(X), A) \to 0 \,.
\end{equation}
Applying~(\ref{eqn:K1N9}) along with the assumption that $\cM_6$ is torsion-free, we find
\begin{equation}
  \Tor K^1(\cM_6 \times S^3/\Gamma) = H^1(\cM_6; \Gamma^{\text{ab}})\oplus H^3(\cM_6; \Gamma^{\text{ab}}) \oplus H^5(\cM_6; \Gamma^{\text{ab}})\, .
\end{equation}
Thus, the Heisenberg group can be presented as
\begin{equation}
  0 \to U(1) \to \cW \to H^1(\cM_6; \Gamma^{\text{ab}}) \oplus H^3(\cM_6; \Gamma^{\text{ab}}) \oplus H^5(\cM_6; \Gamma^{\text{ab}}) \to 0 \,, 
\end{equation}
as is typically done in the $(2,0)$ literature. Note, however, that the cohomology theory of $\cM_6$ with coefficients in $\Gamma^{\text{ab}}$ does not in itself define the perfect pairing $s(x,y)$. Instead, this depends on the topology of $S^3/\Gamma$, as we have seen.

Since (\ref{eqn:N9linking}) involves the cup product on $\cM_6$, the Heisenberg group splits naturally into a direct sum $\cW=\cW_{1,5}\oplus \cW_3$, where
\begin{align}
  0 \to U(1) &\to \cW_{1,5} \to H^1(\cM_6; \Gamma^{\text{ab}}) \oplus H^5(\cM_6; \Gamma^{\text{ab}}) \to 0 \,,  \label{eq:Heisenberg-W15} \\
  0 \to U(1) &\to \cW_3 \to H^3(\cM_6; \Gamma^{\text{ab}}) \to 0\, . \label{eq:Heisenberg-W3}
\end{align}
The $\cW_{1,5}$ factor is associated with D1 and D5 branes wrapping
torsional cycles in $S^3/\Gamma$, and stretching from infinity to the
singularity, giving rise to point and codimension two operators in the
six dimensional theory. We also expect to have operators related to
these by $SL(2,\bZ)$ transformations of the IIB background, that is
$(p,q)$ 5-branes and $(p,q)$ 1-branes. It would be interesting
to understand these operators more fully from the field theoretic
viewpoint, but we will not do so here, simply noting that a choice
of maximal isotropic subspace within $H^1(\cM_6; \Gamma^{\text{ab}}) \oplus H^5(\cM_6; \Gamma^{\text{ab}})$ can be done
canonically, without reference to the details of $\cM_6$. For example, we can choose $L=H^1(\cM_6;\Gamma^{\text{ab}})$, or with equal validity $L=H^5(\cM_6;\Gamma^{\text{ab}})$.

Likewise, $\cW_3$ is associated with D3 branes wrapping torsion cycles in $S^3/\Gamma$, giving rise to 2-surface operators in the six dimensional theory. However, unlike before, there is no $\cM_6$-independent choice of boundary
conditions (except in some special cases, see~(\ref{eq:ADE-self-dual}) below).
This differs from the situation at the classical level, where all background fluxes can be set to zero if desired.
Due to the non-commutativity of fluxes in the presence of torsion, this canonical
choice ceases to exist in the quantum theory: trying to set
all fluxes to zero would be akin to trying to fix both the position
and momentum of a particle in ordinary quantum mechanics.

These IIB results have clear implications for 6d $(2,0)$ theories. In
order to fully specify the partition function of a six-dimensional
$(2,0)$ theory on a manifold $\cM_6$ we need to specify the background
fields for the global 2-form symmetries of the theory. These background
fields are inherited from the asymptotic boundary conditions for the $F_5$ flux
--- or somewhat more precisely, from the holonomies of $C_4$ on torsion
cycles, see \cite{Freed:2006ya,Freed:2006yc} and the remarks at the
beginning of \S\ref{sec:construction-of-H}. But we have just argued
that completely fluxless boundary conditions for $F_5$ are impossible. Thus, the background fields for the 2-form symmetries of the $(2,0)$ theory cannot all be set to zero.
Instead, only a subset can be fixed, the remainder being summed over. The choice of this subset is a choice of maximal abelian subgroup of the Heisenberg group $\cW_3$ in~\eqref{eq:Heisenberg-W3}, equivalently the choice of a maximal isotropic subspace of $H^3(\cM_6;\Gamma^{\text{ab}})$.

Indeed, precisely the same structure has been previously argued --- by
different means --- to describe the global structure of $(2,0)$
\cite{Witten:1998wy,Witten:2009at} and $(1,0)$ \cite{DelZotto:2015isa}
theories. The IIB viewpoint that we have developed here encompasses
all previously understood cases, and allows us to determine the
precise commutation relations for the 2-form flux operators, as
illustrated above for $\bC^2/D_{2n}$ and $\bC^2/D_{p+q,p}$. For
instance, the distinction between $\L_{\text{even}}$ and
$\L_{\text{odd}}$ for the case $p=3$ with $q$ even should lead to
distinct S-duality patterns after compactification on $T^2$; to
our knowledge this is not yet explored in the literature.

\subsection{Theories and metatheories} \label{subsec:metatheories}

As we have seen, the $(2,0)$ theories are generally ``metatheories'': they have a partition vector --- associated to a choice of maximal isotropic subspace $L \subset \Gamma^{\text{ab}} \otimes H^3(\cM_6)$ --- rather than a partition function. If we can devise a prescription for choosing $L$, independent of the details of $\cM_6$,\footnote{A precise way of stating this is the following: the 6d metatheory $D$ may be viewed as a choice of boundary condition for a 7d anomaly theory on a half-infinite line. To generate a genuine 6d theory, we place the anomaly theory on an interval, with $D$ on one boundary and gapped boundary conditions $T$ on the other. Distinct choices of $T$ lead to distinct genuine theories with the same spectrum of local operators (determined by $D$). If gapped boundary conditions are not possible then there are no genuine theories corresponding to $D$. (We thank Davide Gaiotto for discussions on this point.)}
then the partition vector becomes a partition function, and we obtain a ``genuine'' theory. In particular, this is true when
\begin{equation} \label{eqn:M6independent}
L = L_0 \otimes H^3(\cM_6) \,,
\end{equation}
where $L_0 \subset H^2(S^3/\Gamma) = \Gamma^{\text{ab}}$ is ``self-dual,'' i.e., equal to its orthogonal complement $L_0 = L_0^\perp$ with respect to the linking pairing $\L_\Gamma$.

Crucially, since the linking pairing $\L_\Gamma$ on $H^2(S^3/\Gamma)$ is \emph{symmetric} (unlike the linking pairing $\L$ on $H^5(\cN_9)$, which is antisymmetric), self-dual $L_0$ need not exist. In particular, one can show that $|L| |L^\perp| = |\Gamma^{\text{ab}}|$, and so the order of the defect group must be a perfect square.
Examining~(\ref{eq:ADE-linking-forms}), the possibilities corresponding to simple Lie algebras are easily classified:\footnote{The $\Ss(8k)$ and $\Sc(8k)$ cases are related by an outer automorphism of $\Spin(8k)$, and triality relates them to $\SO(8)$ for $k=1$. The case $\SU(4)/\bZ_2 = \SO(6)$ appears twice in the table due to an exceptional isomorphism.}
\begin{equation}
  \label{eq:ADE-self-dual}
  \arraycolsep=4.4pt\def\arraystretch{1.2}
  \begin{array}{cccc}
    \mathfrak{g} & \Gamma^{\text{ab}} & L_0 & G \\
    \hline
    A_{k^2-1} & \bZ_{k^2} & k & \SU(k^2) / \bZ_k \\
    D_{k} & \bZ_{2} \oplus \bZ_2 \text{\ or\ } \bZ_4 & (1,1) \text{\ or\ } 2 & \SO(2k) \\
    D_{4k} & \bZ_2\oplus \bZ_2 & (1,0) & \Ss(8k) \\
    D_{4k} & \bZ_2\oplus \bZ_2 & (0,1) & \Sc(8k) \\
    E_8 & 0 & 0 & E_8
  \end{array} 
\end{equation}
where in each case $L_0$ is cyclic and we indicate its generator.
Fixing these maximal isotropic subspaces, we obtain genuine $(2,0)$ theories (see,
e.g.,~\cite{Tachikawa:2013hya,Gaiotto:2014kfa}), where $G$ is the 5d
gauge group that results from compactification on $S^1$ (see below)
and $\Ss(4k) = \Spin(4k)/\bZ_2^{(L)}$ and
$\Sc(4k) = \Spin(4k)/\bZ_2^{(R)}$ are the semispin groups. Notice in
particular that the linking pairing $\L_{\text{even}}$ leads to
additional genuine $(2,0)$ theories that are not present for
$\L_{\text{odd}}$.

The distinction between metatheories and genuine theories is further illuminated by considering the behavior of extended operators. For instance, the $(2,0)$ theory with Lie algebra  $D_{4k}$ contains three types of 2-surface operators, corresponding to the three non-zero elements of the defect group $\bZ_2 \oplus \bZ_2$. These 2-surface operators are not ``mutually local'', in that correlation functions containing multiple types of 2-surface operators will have branch cuts when one type circles another. We can solve this problem by declaring only one type of 2-surface operator to be ``genuine''~\cite{Witten:1998wy,Kapustin:2014gua,Gaiotto:2014kfa}. The remaining ``non-genuine'' 2-surface operators are then interpreted as lying at the boundaries of 3-surface operators (the branch cuts), with the correlation functions only topologically dependent on the position of the 3-surfaces.

Indeed, depending on which 2-surface operator we designate as genuine, we obtain one of the genuine theories $\SO(8k)$, $\Ss(8k)$, or $\Sc(8k)$ listed in the table above, where the generator of the maximal isotropic subspace $L_0$ corresponds to the genuine 2-surface operator. The other genuine theories also correspond to choosing genuine 2-surface operators in the same manner, but with the added complication that some 2-surface operators fail to be ``self-local'', in that two operators of the same type can generate a branch cut upon circling each other.\footnote{For instance, the $A_1$ theory has one non-trivial 2-surface operator, which fails to be self-local. As a result, there is no maximal isotropic subspace of the kind~(\ref{eqn:M6independent}) for this Lie algebra.}

To see how these properties follow from the string theory picture discussed previously, it is convenient to consider first the conceptually simpler 6d $(1,1)$ theories.

\subsection{Wilson and 't Hooft operators}

To obtain 6d $\cN = (1,1)$ Yang-Mills theories with simple ADE Lie algebra $\fg_\Gamma$, we replace type IIB string theory with type IIA string theory in our discussion above, with 
$\Gamma\subset SU(2)$.\footnote{Note that F-theory is not available to restore $(1,0)$ supersymmetry in the
  $\Gamma\subset U(2)$ cases, unlike in IIB. It would be interesting to consider IIA
  backgrounds with varying dilaton and compare with a geometric analysis in M-theory, but we do not attempt this here.}
To define the partition function for this theory on a
manifold $\cM_6$ we again need to choose boundary conditions at infinity. The
main difference with the IIB case is that in IIA the RR fluxes live in
$K^0(X)$, instead of $K^1(X)$ \cite{Moore:1999gb}. Repeating the
analysis above, \textit{mutatis mutandis}, we obtain
\begin{equation}
  \Tor(K^0(\cM_6\times S^3/\Gamma)) = (H^0(\cM_6)\oplus H^2(\cM_6)\oplus H^4(\cM_6)\oplus H^6(\cM_6))\otimes \Gamma^{\text{ab}} \,,
\end{equation}
so that once more the Heisenberg group splits naturally into two components
\begin{align}
  0 \to U(1) &\to \cW_{0,6} \to H^0(\cM_6;\Gamma^{\text{ab}}) \oplus H^6(\cM_6;\Gamma^{\text{ab}}) \to 0 \,, \\
  0 \to U(1) &\to \cW_{2,4} \to H^2(\cM_6;\Gamma^{\text{ab}}) \oplus H^4(\cM_6;\Gamma^{\text{ab}}) \to 0\, ,
\end{align}
both with the commutation relations coming from the torsion pairing in
$S^3/\Gamma$ times the intersection number in $\cM_6$. As we will see, the Heisenberg algebra $\cW_{2,4}$ is associated to Wilson and 't Hooft operators, and correspondingly the maximal isotropic subspaces of $H^2(\cM_6;\Gamma^{\text{ab}}) \oplus H^4(\cM_6;\Gamma^{\text{ab}})$ are related to the global form of the gauge group.
The significance of $\cW_{0,6}$ is less clear, and we defer further consideration of it to a future work.\footnote{In the IIA description the associated operators come from
  D0 branes wrapping the torsion cycle (suggestive of fractional
  instanton effects in the field theory \cite{Brodie:1998bv}), and D6
  branes wrapping the torsion cycle and extending from infinity to the
  singularity, where they become $\Gamma^{\text{ab}}$-valued domain
  walls.}

Wrapping a D2 brane on a torsion one-cycle $\sigma_a$ of $S^3/\Gamma$ and extending it from the singularity off to infinity, we obtain a Wilson line operator in the 6d gauge theory. To determine whether the Wilson line operator is genuine, we move it around a closed path in $\cM_6$, tracing out a two-cycle $\Sigma_2$, and ask whether the correlation function has changed once it returns to its original position. If we initially deform the D2 brane only within a distance $r < r_0$ of the singularity, then the net result of the deformation is to add a D2 brane wrapped on $\sigma_a \times \Sigma_2$ at radius $r=r_0$. Extending the deformation outward ($r_0 \to \infty$) corresponds to moving the wrapped D2 brane far away from the singularity. The Chern-Simons coupling $\oint_{\sigma_a \times \Sigma_2} C_3$ of the wrapped D2 brane contributes a phase to the path integral unless the holonomy of $C_3$ on $\sigma_a \times \Sigma_2$ vanishes. Explicitly, pulling back to $S^3/\Gamma \times \Sigma_2$, the phase is $\exp(2\pi i \L_{\Gamma}(\PD[\sigma_a], f))$ where $\PD$ denotes the Poincar\'e dual within $S^3/\Gamma$ and $f \in \Tor H^4(S^3/\Gamma\times \Sigma_2) \cong H^2(S^3/\Gamma) \otimes H^2(\Sigma_2) \cong H^2(S^3/\Gamma)$ is the torsion component of the $F_4$ flux along $S^3/\Gamma\times \Sigma_2$. Thus, the correlation function has a branch cut unless the linking pairing $L_\Gamma(\PD[\sigma_a], f)$ vanishes.

Likewise, a D4 brane wrapped on $\sigma_b$ and extended from the singularity to infinity yields a 't Hooft 3-surface operator in the gauge theory. Consider the link $\Sigma_2$ of the 3-surface wrapped by the 't Hooft operator within $\cM_6$. The presence of the D4 brane generates torsional flux $f = \PD[\sigma_b]$ within $\Tor H^4(S^3/\Gamma\times \Sigma_2) \cong H^2(S^3/\Gamma)$, and so deforming a Wilson line associated to the torsion cycle $\sigma_a$ along $\Sigma_2$ we pick up a phase $\exp(2 \pi i \L_{\Gamma}(\sigma_a, \sigma_b))$: the Wilson and 't Hooft operators are not mutually local.

Suppose that we wish to designate all Wilson lines as genuine. Per the above discussion, this requires a boundary condition where the torsion component of $[F_4]$, classified by $\Tor H^4(S^3/\Gamma \times \cM_6) \cong \Gamma^{\text{ab}} \otimes H^2(\cM_6)$, vanishes. The corresponding maximal isotropic subspace is $L=\Gamma^{\text{ab}} \otimes H^4(\cM_6)$. As this choice is independent of the details of $\cM_6$, it produces a genuine $(1,1)$ theory with Wilson lines classified by $\Gamma^{\text{ab}} = Z(G_\Gamma)$. In a gauge theory with gauge group $G$, we expect a Wilson line operator for each element of $Z(G)$ (see, e.g., \cite{Aharony:2013hda}), so we interpret this theory as the 6d $(1,1)$ theory with simply connected gauge group $G_\Gamma$.

More generally, for any subgroup $L_W \subseteq \Gamma^{\text{ab}}$, we can choose the maximal isotropic subspace
\begin{equation} \label{eqn:genuine11pol}
L = [L_W\otimes H^4(\cM_6)] \oplus [L_H\otimes H^2(\cM_6)] \,,\qquad L_H = L_W^\perp \,,
\end{equation}
for which the Wilson lines $L_W$ and 't Hooft lines $L_H = L_W^\perp$ are genuine. By the same reasoning as above, this is a genuine $(1,1)$ theory with gauge group $G_\Gamma / L_H$.\footnote{To make this statement precise, we need to specify a canonical map between $\frac{\Lambda^w(G)}{\Lambda^r}$ and $Z(G)$. In particular, we choose this map so that $g \in Z(G)$ gives a phase $\exp(2\pi i \L_\Gamma(g,r))$ to representations in the coset $r \in \frac{\Lambda^w(G)}{\Lambda^r}$. This is the natural choice, but has potentially unexpected consequences for the case $D_{4k}$, e.g., the left-handed spinor coset maps to the generator of $\bZ_2^{(R)}$.} In this way, the $\cM_6$-independent maximal isotropic subspaces reproduce the different global forms of the gauge group.

\bigskip

This result can also be understood from the viewpoint of generalized
global symmetries \cite{Gaiotto:2014kfa}. Consider, as an example, the
six-dimensional $(1,1)$ theory with algebra $\fsu(N)$. The choice of a
global form of the gauge group can be understood as a choice of which
higher-form symmetries are present in the theory. For instance, if we choose
global form $\SU(N)$ then there is a $\bZ_N$
discrete 1-form symmetry counting Wilson lines (which are ``genuine'',
in this theory), while if we choose global form $SU(N)/\bZ_N$ there is instead a $\bZ_N$ 3-form symmetry counting 't Hooft 3-surface
operators. In the former case, we can couple the theory to a background 2-form $\bZ_N$ gauge field, with the non-trivial gauge bundles classified by $H^2(\cM_6;\bZ_N)$. These gauge bundles for the background 2-form should correspond to the background flux $f \in F_L$, where $F_L$ is given by~\eqref{eq:flux-coset}. Thus, the global form $\SU(N)$ corresponds to the maximal isotropic subspace $L = H^4(\cM_6;\bZ_N)$ (for which $F_L \cong H^2(\cM_6;\bZ_N)$), in agreement with the above analysis. The case $\SU(N) / \bZ_N$ is analyzed similarly.

\subsection{\texorpdfstring{$\cN=4$}{N=4} theories of ADE type}

\label{sec:comparison-4d}

The above discussion is readily generalized to the $(2,0)$ theories, with corresponding changes in the dimensions of branes/operators and the ranks of fluxes. However, as many $(2,0)$ theories do not admit an $\cM_6$-independent maximal isotropic subspace, see~\S\ref{subsec:metatheories}, it is particularly interesting in this case to consider Lagrangian subspaces that depend on $\cM_6$. The discussion of the previous section can be summarized as follows: switching to homology using Poincar\'e duality, the Lagrangian subspace $L \subset H_1(S^3/\Gamma) \otimes H_3(\cM_6) \cong H_3(\cM_6; \Gamma^{\text{ab}})$ is the space of cycles on which the holonomy of $C_4$ is asymptotically fixed to zero by the boundary conditions. As such, these are the cycles around which we can deform the 2-surface operators without encountering a branch cut. When $L$ is $\cM_6$ dependent, this means that some but not all branch cuts are eliminated, and in general no 2-surface operators are genuine when deformed around an arbitrary three-cycle.

Having understood the behavior of 6d $(2,0)$ theories in terms of boundary conditions in type IIB string theory, we can apply the same ideas to compactifications of the $(2,0)$ theory. Our goal in the remainder of this section is to demonstrate that the classification of 4d $\cN=4$ theories given by \cite{Aharony:2013hda} (see also \cite{Gaiotto:2010be}) is reproduced in this framework. To do so, we consider $T^2\times \cM_4$ compactifications of the $(2,0)$ theories, following a similar approach to Tachikawa~\cite{Tachikawa:2013hya} but using Heisenberg group commutators computed directly in the type IIB picture discussed above, rather than inferred from four dimensional reasoning~\cite{Gaiotto:2010be,Aharony:2013hda}. The Lie algebras $D_{4k}$ and $D_{4k+2}$ (not analyzed in~\cite{Tachikawa:2013hya}) provide a particular sensitive test of our reasoning, as the different linking pairings $\L_{\text{even}}$ and $\L_{\text{odd}}$ for these two cases lead to different patterns of 4d S-duality, in agreement with \cite{Aharony:2013hda}.

First note that, in the absence of torsion on $\cM_4$, we have
by the K\"unneth formula
\begin{equation} \label{eqn:M4T2Kunneth}
  H^3(\cM_4\times T^2) = H^3(\cM_4) \oplus [H^2(\cM_4)\otimes H^1(T^2)] \oplus H^1(\cM_4)\, .
\end{equation}
Again for degree reasons we have a natural splitting of the associated
Heisenberg group $\cW_3=\cW_{1,3}\oplus \cW_2$, with
\begin{align}
  0 \to U(1) &\to \cW_{1,3} \to H^1(\cM_4; \Gamma^{\text{ab}})\oplus H^3(\cM_4; \Gamma^{\text{ab}}) \to 0 \,,  \label{eq:Heisenberg-W13} \\
  0 \to U(1) &\to \cW_2 \to H^2(\cM_4)\otimes H^1(T^2)\otimes \Gamma^{\text{ab}} \to 0\, .   \label{eq:Heisenberg-W2}
\end{align}
The Heisenberg group $\cW_{1,3}$ is associated to point and 2-surface operators in the 4d theory. Noting that, $\cM_4$ and $T^2$-independent choices of maximal isotropic subspace are always possible within this factor, such as $L_{1,3} = H^1(\cM_4; \Gamma^{\text{ab}})$ or $L_{1,3} = H^3(\cM_4; \Gamma^{\text{ab}})$, we ignore it for the time being, instead focusing on the factor $\cW_2$ describing line operators.

To obtain genuine 4d theories, we consider $\cM_4$-independent maximal isotropic subspaces of $H^2(\cM_4)\otimes H^1(T^2)\otimes \Gamma^{\text{ab}}$. These are of the form
\begin{equation}
  \label{eq:T^2-polarizations}
  L = H^2(\cM_4) \otimes L_{T^2}
\end{equation}
where $L_{T^2}$ is a maximal isotropic subspace of $H^1(T^2; \Gamma^{\text{ab}})$, corresponding to the Heisenberg algebra
\begin{equation}
  \label{eq:W-T^2}
  0 \to U(1) \to \cW_{T^2} \to H^1(T^2; \Gamma^{\text{ab}}) \to 0\, .
\end{equation}
Since a maximal isotropic subspace of $H^1(T^2; \Gamma^{\text{ab}})$ always exists there are genuine theories corresponding to every Lie algebra, unlike in six dimensions. Instead, the absence of a genuine six dimensional theory causes a ``modular anomaly'': tracing a closed path in the complex structure moduli space of the torus changes the partition function. 

In particular, the partition function is generally not a modular-invariant function of the holomorphic gauge coupling $\tau$. From the 6d perspective, $\tau$ is the complex structure
of the torus and modular transformations $\tau \to \frac{a \tau + b}{c \tau+ d}$ are large diffeomorphisms in the background metric. Thus, the failure of modular invariance (in the absence of additional background fields along the torus) is the result of a 6d anomaly in large diffeomorphisms. Depending on the 6d anomaly, a characteristic pattern of S-dualities is generated, as in, e.g.,~\cite{Aharony:2013hda}.

Note that if we view fixed $\tau$ as part of the defining data
of the theory then we would not consider the non-invariance of the partition function under $SL(2,\bZ)$ transformations of $\tau$ to be a 4d anomaly, but rather a
consequence of deforming along a fixed line from one theory to
another. 
On the other hand, when considering four-dimensional backgrounds with varying $\tau$, the anomaly viewpoint becomes more natural.
We revisit this point below in the context of theories with codimension-two duality defects.\footnote{See
  \cite{Seiberg:2018ntt,Hsieh:2019iba,Cordova:2019jnf,Cordova:2019uob}
  for recent work studying other aspects of anomalies on the space of
  coupling constants.}

Thus, for each maximal isotropic subspace of $L_{T^2} \subset H^1(T^2; \Gamma^{\text{ab}})$ there is a genuine 4d $\cN=4$ theory. Wrapping the 2-surface operators of the 6d $(2,0)$ theory on different cycles of the torus, we obtain different types of 4d line operators. For instance, reducing the $(2,0)$ theory first on the $A$ cycle of the torus, we obtain a five-dimensional gauge theory, with Wilson line and 't Hooft 2-surface operators. Reducing again on the $B$ cycle, the 't Hooft operators become lines. Thus, 2-surface operators wrapped around the $A$ and $B$ cycles are Wilson and 't Hooft lines, respectively, whereas those wrapped around a combination of the two are dyonic lines.

We can identify the genuine theory in question by specifying which of these line operators are genuine. In particular, by the same reasoning as in the previous section, there is a one-to-one correspondence between the elements of the maximal isotropic subspace $L_{T^2}$ and the genuine line operators in the 4d theory, and so the result can be directly compared with~\cite{Aharony:2013hda}.

Consider for example the $\cN = 4$ theories with Lie algebra $\fsu(N)$, corresponding to the $A_{N-1}$ $(2,0)$ theory on a torus. In general $H^1(T^2;\Gamma^{\text{ab}}) = H^1(T^2)\otimes \Gamma^{\text{ab}}=\Gamma^{\text{ab}} \oplus \Gamma^{\text{ab}}$ with the perfect pairing $s(a,b) = \exp(2 \pi i \L_{T^2}(a, b))$, where $\L_{T^2}$ is the linking pairing on $S^3/\Gamma \times T^2$
\begin{equation} \label{eqn:LT2}
\L_{T^2}(a, b) = \L_\Gamma(\pi_A(a),\pi_B(b)) - \L_\Gamma(\pi_B(a),\pi_A(b)) \, \qquad a,b \in \Gamma^{\text{ab}} \oplus \Gamma^{\text{ab}}\,,
\end{equation}
and $\pi_A: \Gamma^{\text{ab}} \oplus \Gamma^{\text{ab}} \to \Gamma^{\text{ab}}$ and $\pi_B: \Gamma^{\text{ab}} \oplus \Gamma^{\text{ab}} \to \Gamma^{\text{ab}}$ project onto the first and second summand, respectively. In the $\fsu(N)$ case, $\Gamma^{\text{ab}} = \Gamma = \bZ_N$, and we obtain the perfect pairing
\begin{equation} \label{eqn:ZNperfectpairing}
s(e_1 \mathbf{p} + m_1 \mathbf{q}, e_2 \mathbf{p} + m_2 \mathbf{q}) = \exp\biggl(\frac{2 \pi i}{N}(e_1 m_2 - e_2 m_1)\biggr)
\end{equation}
from~(\ref{eq:ADE-linking-forms}), where $\mathbf{p}$ and $\mathbf{q}$ denote the $A$ and $B$ cycles of the torus, respectively. Here $e_i$ and $m_i$ denote the Wilson and 't Hooft charges of the associated line operators, respectively. For instance, $L = \{m \mathbf{q} | 0\le m < N\}$ is a maximal isotropic subspace whose elements correspond to 't Hooft lines of every possible $\bZ_N$ charge; the associated gauge theory is therefore $(\SU(N)/\bZ_N)_0$ in the notation of~\cite{Aharony:2013hda}.

\begin{figure}
  \centering
  \begin{subfigure}{0.20\textwidth}
    \centering
    \includegraphics[width=\textwidth]{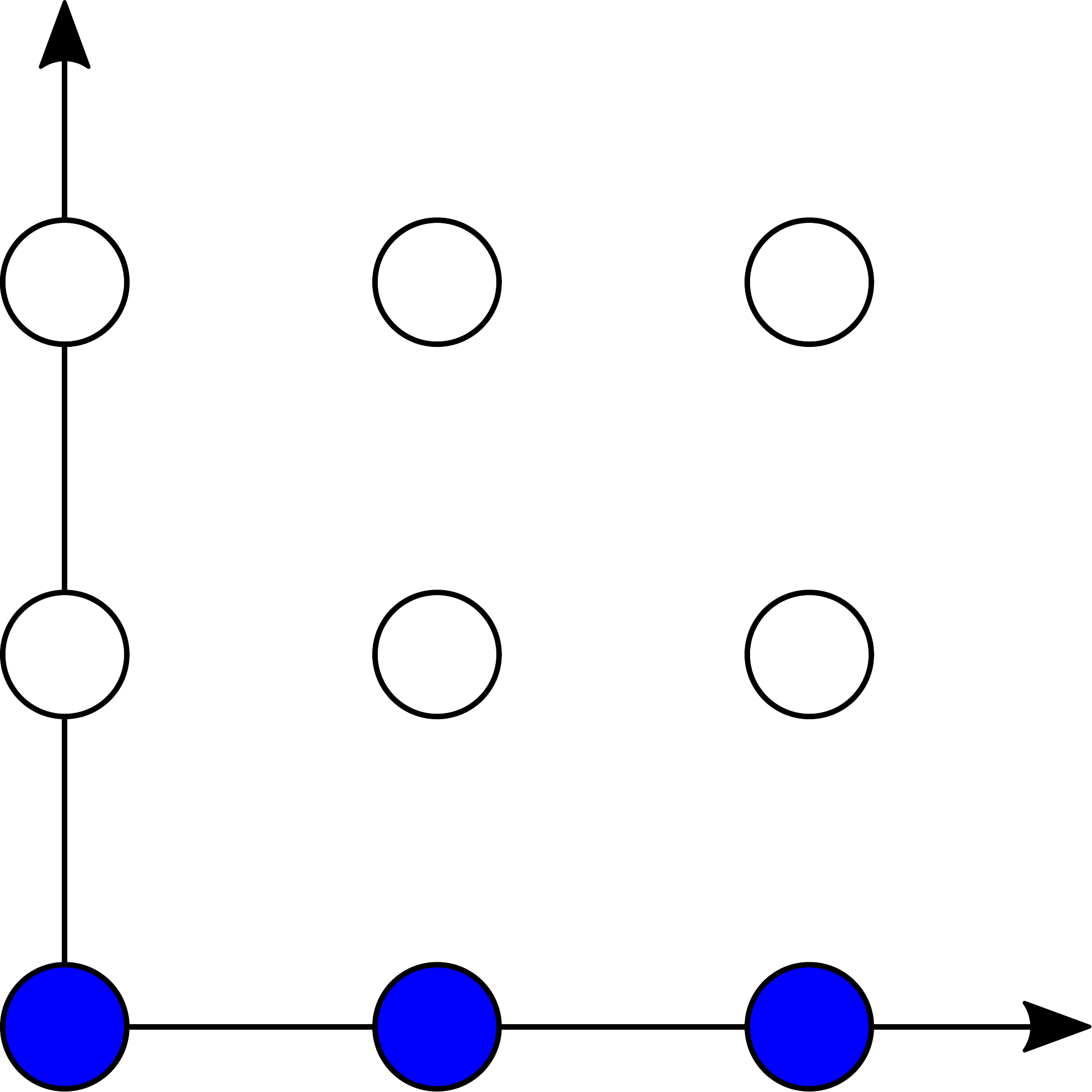}
    \caption{$SU(3)$}
  \end{subfigure}
  \hfill
  \begin{subfigure}{0.20\textwidth}
    \centering
    \includegraphics[width=\textwidth]{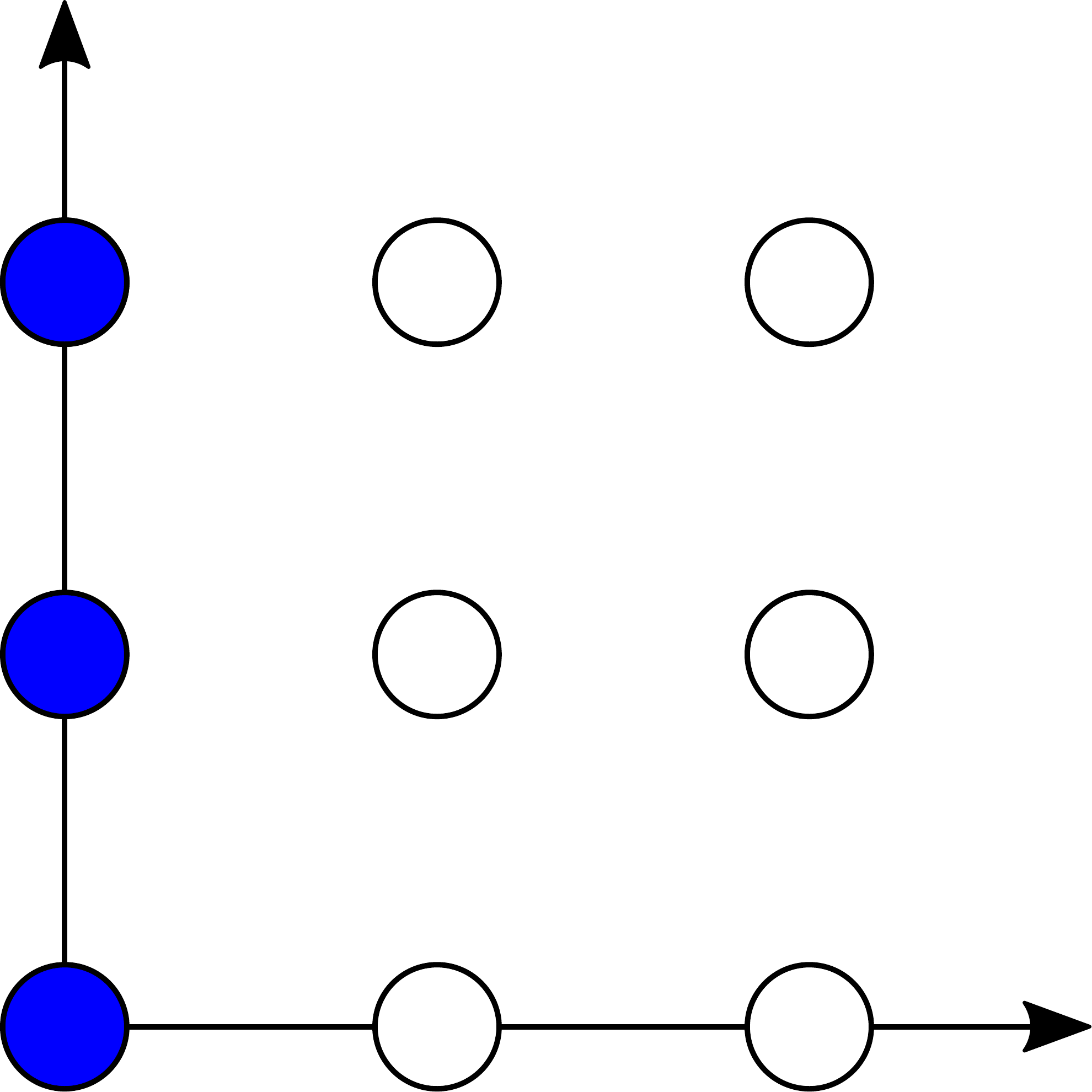}
    \caption{$(SU(3)/\bZ_3)_0$}
  \end{subfigure}
  \hfill
  \begin{subfigure}{0.20\textwidth}
    \centering
    \includegraphics[width=\textwidth]{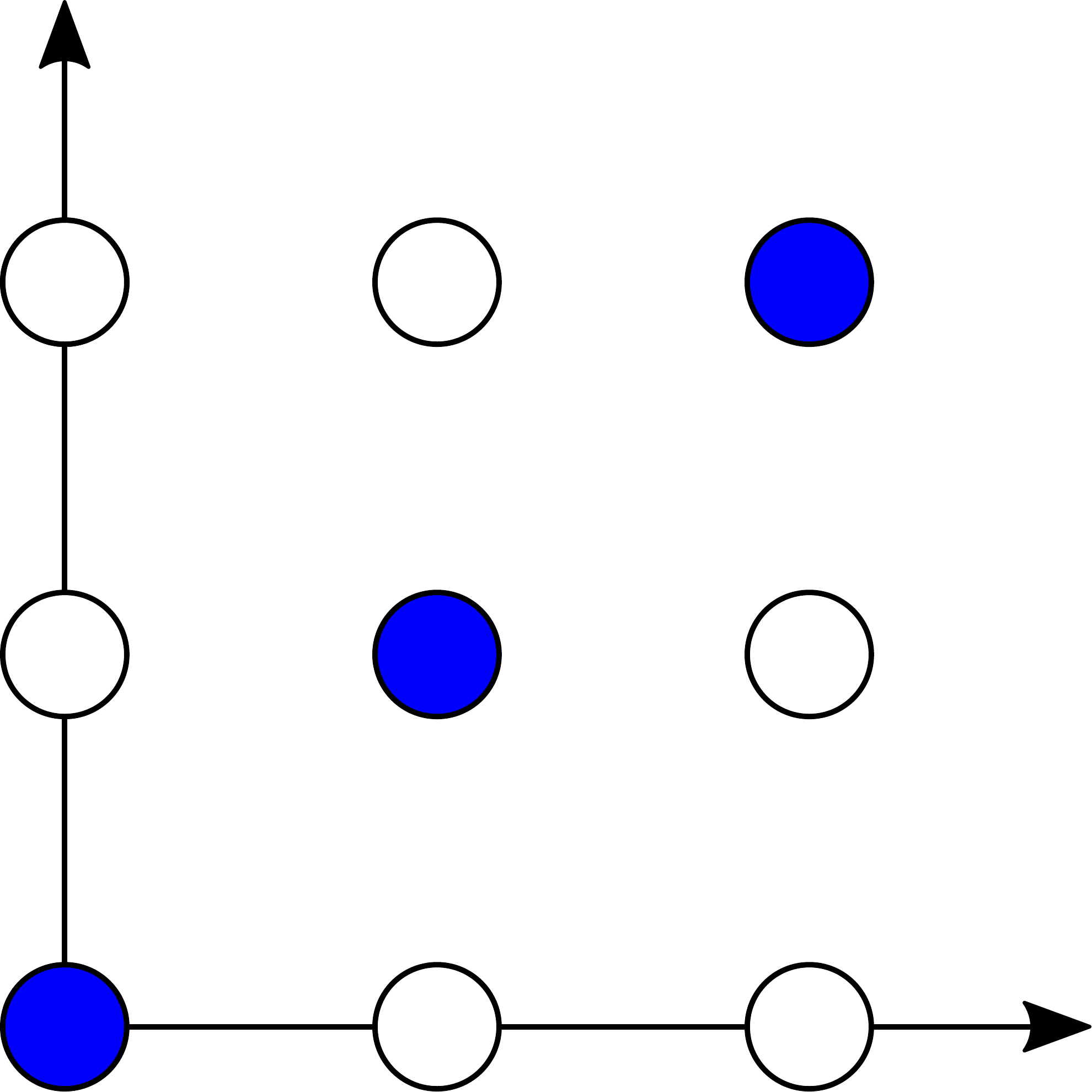}
    \caption{$(SU(3)/\bZ_3)_1$}
  \end{subfigure}
  \hfill
  \begin{subfigure}{0.20\textwidth}
    \centering
    \includegraphics[width=\textwidth]{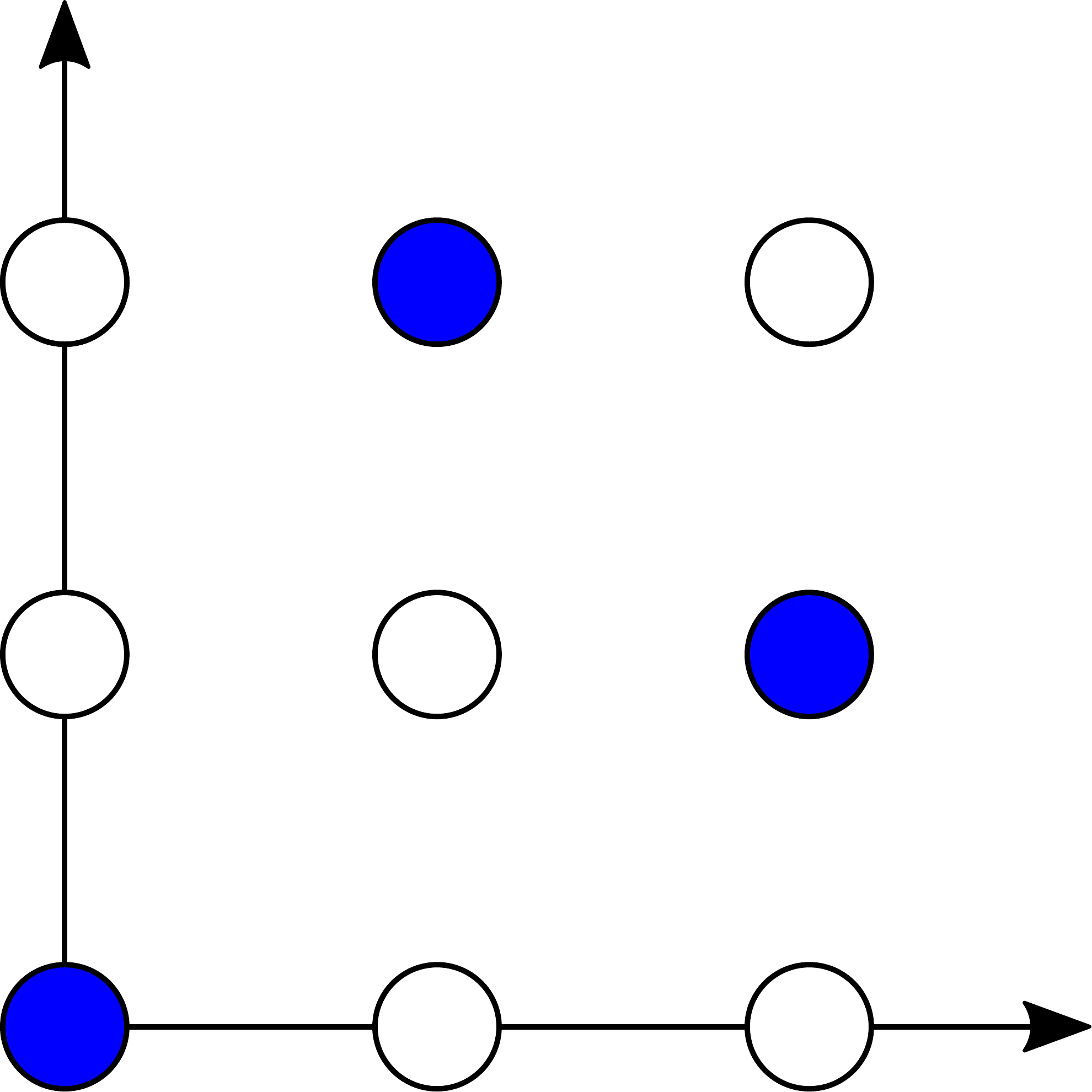}
    \caption{$(SU(3)/\bZ_3)_2$}
  \end{subfigure}
  \caption{Maximal isotropic subspace of
    $H^1(T^2;\bZ_3)=\bZ_3\oplus\bZ_3$ with respect to the perfect pairing~(\ref{eqn:ZNperfectpairing}). We have labelled the
    possibilities using the nomenclature of \cite{Aharony:2013hda}. Each filled dot corresponds to a genuine line operator.}
  \label{fig:su(3)-lattices}
\end{figure}

For any fixed $N$, it is a simple exercise to enumerate the maximal isotropic subspaces of $\bZ_N \oplus \bZ_N$ with respect to the perfect pairing~(\ref{eqn:ZNperfectpairing}). For instance, the case $N=3$ is shown in figure~\ref{fig:su(3)-lattices}, with results that are easily seen to agree
with \cite{Aharony:2013hda}. More generally, flux operators $\Phi_1$ and $\Phi_2$ commute if
\begin{equation}
e_1 m_2 - e_2 m_1 \equiv 0 \pmod N \,.
\end{equation}
This is the same as the mutual locality constraint found in~\cite{Aharony:2013hda}, so the results will agree in general.

The $\fso(4k+2)$ and $E_{6,7,8}$ theories are handled similarly. However, the case $\fso(4k)$ deserves special attention, as the defect group $\Gamma^{\text{ab}} = \bZ_2 \oplus \bZ_2$ is not cyclic, and there are multiple possible linking pairings, each with different consequences. For $\fso(8k)$, the linking pairing is $\L_\Gamma = \L_{\text{even}}$ per~\eqref{eq:ADE-linking-forms}, so we obtain the perfect pairing
\begin{equation}
s(e_1 \mathbf{p} + \tilde{e}_1 \tilde{\mathbf{p}}+ m_1 \mathbf{q}+ \tilde{m}_1 \tilde{\mathbf{q}}, e_2 \mathbf{p} + \tilde{e}_2 \tilde{\mathbf{p}}+ m_2 \mathbf{q}+ \tilde{m}_2 \tilde{\mathbf{q}})
= (-1)^{e_1 \tilde{m}_2 + \tilde{e}_1 m_2 + e_2 \tilde{m}_1 + \tilde{e}_2 m_1}
\end{equation}
using~\eqref{eqn:LT2}, where $\mathbf{p}$ and $\tilde{\mathbf{p}}$ denote the $A$ cycle of the torus tensored with the two generators of $\bZ_2 \oplus \bZ_2$ and likewise for $\mathbf{q}$ and $\tilde{\mathbf{q}}$. For $\fso(8k+4)$ the linking pairing is $\L_\Gamma = \L_{\text{odd}}$, so we obtain instead
\begin{equation}
s(e_1 \mathbf{p} + \tilde{e}_1 \tilde{\mathbf{p}}+ m_1 \mathbf{q}+ \tilde{m}_1 \tilde{\mathbf{q}}, e_2 \mathbf{p} + \tilde{e}_2 \tilde{\mathbf{p}}+ m_2 \mathbf{q}+ \tilde{m}_2 \tilde{\mathbf{q}})
= (-1)^{e_1 m_2 + \tilde{e}_1 \tilde{m}_2 + e_2 m_1 + \tilde{e}_2 \tilde{m}_1} \,.
\end{equation}
These agree with (5.4) of~\cite{Aharony:2013hda}, which is a sensitive check of our analysis.

The above analysis generalizes readily to compactifications of the $(2,0)$ theory on an arbitrary compact Riemann surface $\Sigma$; $\cM_4$-independent maximal isotropic subspaces are now of the form
$  L = H^2(\cM_4)\otimes L_{\Sigma}$,
associated to the Heisenberg group
 $ 0 \to U(1) \to \cW_{\Sigma} \to H^1(\Sigma)\otimes \Gamma^{\text{ab}} $,
with the perfect pairing
$ s(c_1\otimes \ell_1, c_2\otimes \ell_2) = \exp\Bigl(2\pi i\, (c_1 \circ c_2)\, \L_\Gamma(\ell_1,\ell_2)\Bigr)$,
where $\circ$ is the intersection form on $\Sigma$. It would be interesting to understand how adding punctures on $\Sigma$---as in class $\cS$ constructions---changes this story.

\subsection{Fractional instanton numbers and the linking form}

\label{sec:ninst}

Although this is somewhat outside the main line of development of our
paper, we point out that in the $\cN=4$ cases one can give a simple
expression for the fractional instanton numbers for
$G_\Gamma/\Gamma^{\text{ab}}$ bundles, as computed in
\cite{Witten:2000nv} (see also \cite{Aharony:2013hda}) in terms of the
linking pairing discussed above. Let us assume that $\cM_4$ has no
torsion and also that it is a Spin manifold. Consider the class
$w_2\in H^2(\cM_4; \Gamma^{\text{ab}})$ measuring the obstruction to
lifting the given $G_\Gamma/Z(G_\Gamma) = G_\Gamma/\Gamma^{\text{ab}}$
bundle to $G_\Gamma$. Since
$\Tor H^4(\cM_4\times S^3/\Gamma) \cong H^2(\cM_4;
\Gamma^{\text{ab}})$ along the same lines as above, we can rewrite
this as a class $\widehat{w}_2\in \Tor H^4(\cM_4\times
S^3/\Gamma)$. Denoting by $\hat{\L}$ the linking form in
$\cM_4\times S^3/\Gamma$, one can check that the fractional instanton
number can be expressed as\footnote{Recall that we are taking $\cM_4$
  to be a Spin manifold, so
  $\frac{1}{2}\int_{\cM_4} w_2\smile w_2$ is an integer.}
\begin{equation}
  \label{eq:ninst-linking}
  n_{\text{inst}} \equiv \frac{1}{2} \hat{\L}(\widehat{w}_2, \widehat{w}_2) \pmod 1\, ,
\end{equation}
in the conventions where the minimal local $G_\Gamma$-instanton on
$\bR^4$ has instanton number 1.\footnote{In comparing with the results
  of \cite{Aharony:2013hda}, it might be useful to recall that in the
  case at hand one can define the Pontryagin square of
  $x\in H^2(\cM_4,\bZ_2)$ by $\cP(x)=\ov{x}^2$ mod 4, where
  $\ov{x}\in H^2(\cM_4)$ is an uplift of $x$.} This relation is less
surprising if we recall the fact that the fractional instanton number
$n_{\text{inst}}$ encodes the change in the partition function of
$\cN=4$ super-Yang-Mills under $\tau\to\tau+1$, up to a factor
$c(\cM_4)$ that depends on the topology of $\cM_4$ but not on $w_2$
\cite{Vafa:1994tf}:\footnote{That is, the fractional instanton number
  encodes an anomaly under $\theta\to\theta+2\pi$. See
  \cite{Cordova:2019uob} for recent work discussing this viewpoint in
  more detail.}
\begin{equation}
  \label{eq:Vafa-Witten}
  Z_{w_2}(\tau+1) = \exp\biggl(2\pi i (c(\cM_4) + n_{\text{inst}})\biggr) Z_{w_2}(\tau)\, .
\end{equation}
From the type IIB string theory perspective this is a change in
the phase of the partition function resulting from a large diffeomorphism of
the $T^2$ factor in the $\cM_4\times T^2\times \bC^2/\Gamma$ geometry,
in the presence of a RR 5-form flux given by $\widehat{w}_2\otimes x$,
with $x$ a generator of $H^1(T^2)$.

\begin{figure}
  \centering
  \begin{subfigure}[t]{0.35\textwidth}
    \centering
    \includegraphics[width=\textwidth]{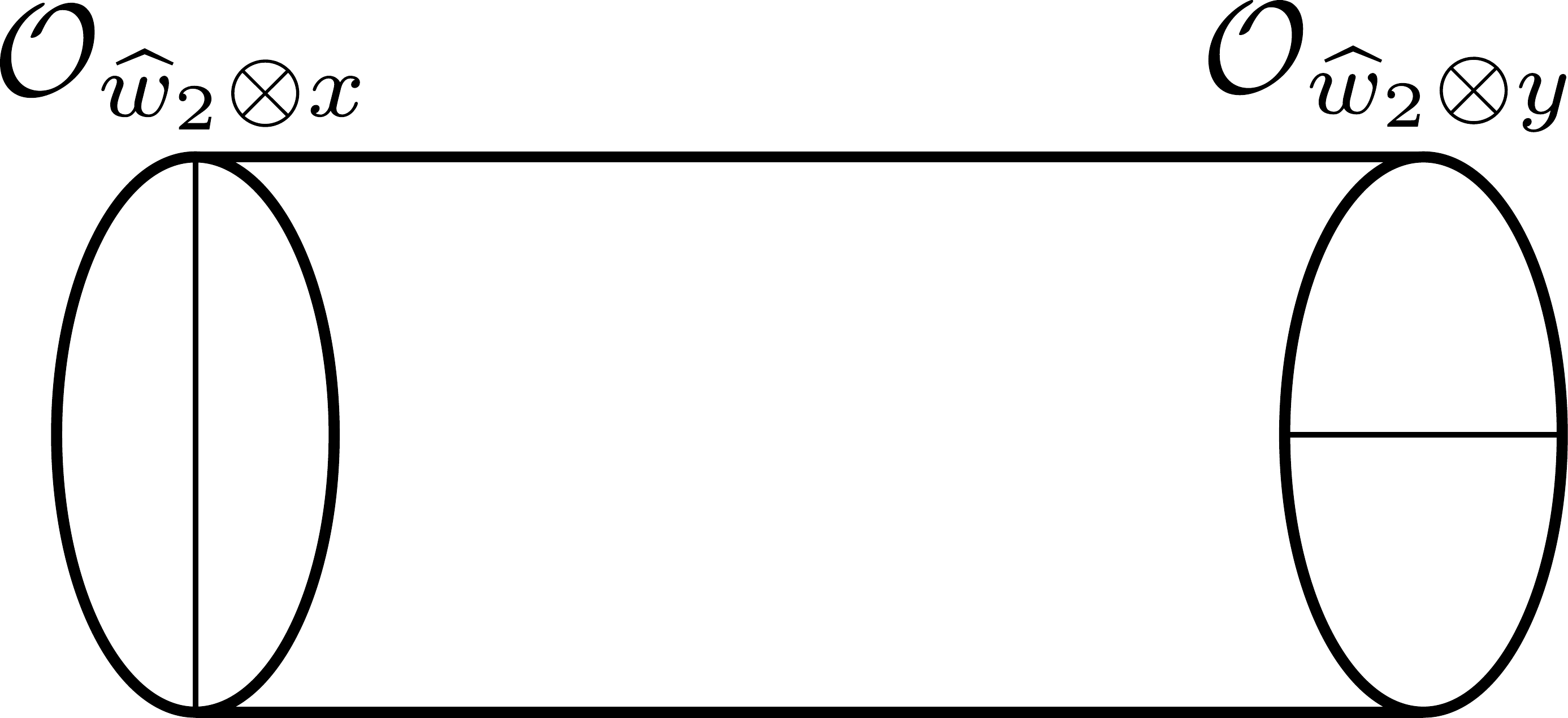}
    \caption{Action of the large diff on IIB.}
    \label{sfig:flux-diff-a}
  \end{subfigure}
  \hspace{0.05\textwidth}
  \begin{subfigure}[t]{0.55\textwidth}
    \centering
    \includegraphics[width=\textwidth]{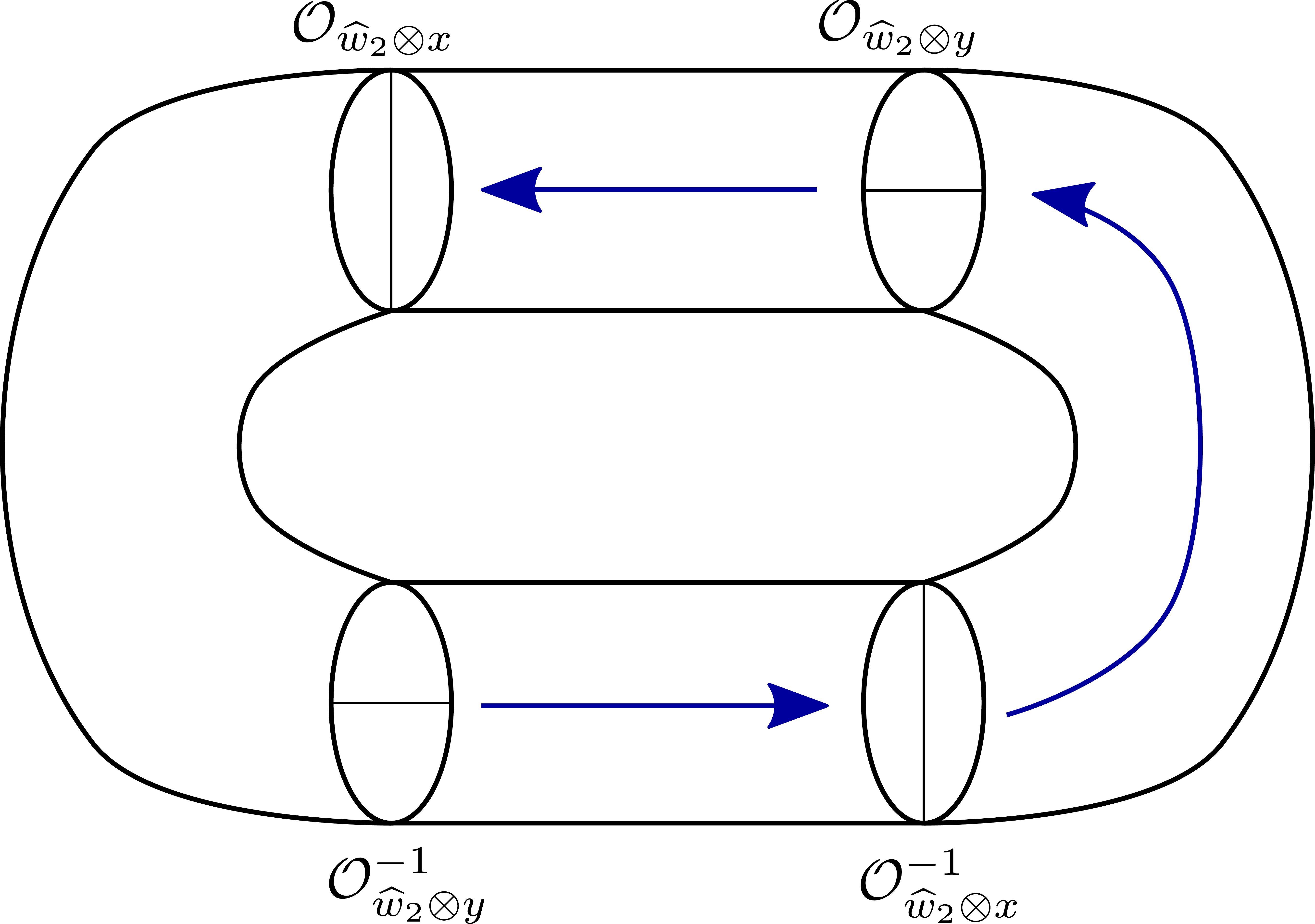}
    \caption{Building the commutator.}
    \label{sfig:flux-diff-b}
  \end{subfigure}
  \caption{\subref{sfig:flux-diff-a} The fractional instanton number
    can be viewed as the anomaly coming from a large diffeomorphism in
    the presence background torsion flux. At the level of the fluxes,
    this can be implemented by the insertion of suitable operators in
    the anomaly theory. \subref{sfig:flux-diff-b} Gluing two copies of
    the configuration giving the anomaly to two configurations without
    flux we obtain the anomaly theory with four operator
    insertions. Bringing the operators together we obtain the
    commutator, a c-number.}
\end{figure}

As such, a rough argument for~\eqref{eq:ninst-linking} is as
follows. Heuristically, we could express the path integral of IIB
string theory on a manifold $X_{10}$ with background flux $F$ as the
the partition function of the anomaly theory $\cA$ on a manifold
$Y_{11}$ with $\partial Y_{11}=X_{10}$, and an insertion in $Y_{11}$
of an appropriate flux operator $\Phi_F$ (see for instance
\cite{Witten:1998wy} for a similar construction in the context of
AdS/CFT). In our situation, depicted in figure~\ref{sfig:flux-diff-a},
we are interested in computing the partition function of $\cA$ on a
cylinder with flux $\widehat{w}_2\otimes x$ on one end, and (due to
the large diff on the $T^2$ factor) a flux $\widehat{w}_2\otimes(x+y)$
on the other end. In order to create these fluxes, we introduce
operators $\Phi_{\widehat{w}_2\otimes x}$ and
$\Phi_{\widehat{w}_2\otimes y}$ into the bulk of the anomaly
theory.

Now take two copies of the cylinder constructed above, and glue them
together, along with two trivial cylinders, into a torus, as in
figure~\ref{sfig:flux-diff-b}. Bringing the four insertions together
we obtain the commutator
$s(\widehat{w}_2\otimes x, \widehat{w}_2\otimes y)$ which is a
c-number, and can be taken out of the path integral. The $c(\cM_4)$
factor in~\eqref{eq:Vafa-Witten} is associated to the change in the
partition function with no flux, so it is natural to conjecture
that it is associated with the value of the partition function in the
absence of $\Phi_F$ insertions.\footnote{\label{footnote:phases-computation}It should in principle be
  possible to compute this change in the partition function of IIB
  string theory in terms of an eleven-dimensional anomaly theory $\cA$
  (see
  \cite{Belov:2004ht,Belov:2006jd,Belov:2006xj,Monnier:2011rk,Monnier:2013kna})
  on a cylinder with boundary the ten dimensional configurations
  related by the large diffeomorphism. A natural stepping stone
  towards the full eleven-dimensional computation would be to
  reproduce the anomalous phases of the partition function from the
  behaviour of the anomaly theory for the six-dimensional $(2,0)$
  theory \cite{Monnier:2017klz}. See \cite{Seiberg:2018ntt} for an
  analysis following this approach for the abelian case (or more
  generally, for six-dimensional theories with an invertible anomaly
  theory).} Removing this overall factor, the construction implies
that
\begin{equation}
  s(\widehat{w}_2\otimes x, \widehat{w}_2\otimes y) = \biggl[\exp(2\pi i \, n_{\text{inst}})\biggr]^2\, .
\end{equation}
Using the relations between $s$ and the linking form given above, this
implies~\eqref{eq:ninst-linking} up to a sign, which depends on
choices of orientation that we have not been careful about.

The above argument is somewhat heuristic.
It would be interesting to work it out in detail and determine
its implications beyond the $\cN=4$ case. It seems natural to
conjecture, for instance, that~\eqref{eq:ninst-linking} still holds if
we consider $(1,0)$ theories compactified on $T^2$. Even though the resulting theory may be non-Lagrangian,
\eqref{eq:ninst-linking} is a natural guess for the behavior
of the partition function in the presence of backgrounds for the 1-form symmetries.

\subsection{Product groups}

Consider the case of the $E_8$ theory in
six-dimensions, arising from IIB on $\bC^2/E_8$. Since\footnote{The
  space $S^3/E_8$ is known as the ``Poincar\'e homology sphere'', and
  is well known to have the same homology groups as $S^3$. As we have
  explained above, this statement is equivalent to the fact that the
  centre of $E_8$ is trivial.}  $H^2(S^3/E_8)=0$ the $(2,0)$ theory of
$\mathfrak{e}_8$ type is a genuine six-dimensional theory: no choice
of IIB boundary conditions at infinity is needed in order to define
the theory on any six-manifold. This implies, in particular, the
well-known fact that the $\cN=4$ theory with gauge group $E_8$ is
invariant under $SL(2,\bZ)$ dualities. The group $E_8$ has a maximal
subgroup $(E_6\times SU(3))/\bZ_3$, and one can check that the $\cN=4$
theory with this gauge group is also invariant under $SL(2,\bZ)$.

We can reproduce this result from our geometric perspective, by
showing that there is a genuine six-dimensional theory of type
$\mathfrak{e}_6\oplus \fsu(3)$. Consider a local \Kthree\ with
singularities of type locally $\bC^2/E_6$ and $\bC^2/\bZ_3$. We link
the singularities by small rational homology spheres $S^3/E_6$ and
$S^3/\bZ_3$, with total manifold their disjoint union
$S_u\df S^3/E_6\sqcup S^3/\bZ_3$. From~\eqref{eq:ADE-linking-forms} we
obtain
\begin{equation}
  H^2(S_u) = H^2(S^3/E_6) \oplus H^2(S^3/\bZ_3) = \bZ_3\oplus\bZ_3
\end{equation}
with linking form
\begin{equation}
  \L_u = \L_{E_6}\oplus \L_{\bZ_3} = \begin{pmatrix}
    \frac{2}{3} & 0\\
    0 & \frac{1}{3}
  \end{pmatrix}\, .
\end{equation}
Let  $a$ and $b$ be the generators of $H^2(S_u)$ corresponding to the
$H^2(S^3/E_6)$ and $H^2(S^3/\bZ_3)$ factors, respectively. Because $\L_u(a+b,a+b)=0$, $H^2(S_u)$ has a self-dual subspace generated by $a+b$ (as well as one generated by $a-b$).
 Associated to this, there is a maximal isotropic subspace of $H^2(S_u)\otimes H^3(\cM_6)$ given by
\begin{equation}
  L_u = \Span(a+b)\otimes H^3(\cM_6)\, .
\end{equation}
Following the same reasoning as above, after reduction on $T^2$ we
obtain a 4d theory with line operators that carry equal charge under the
$\bZ_3$ 1-form symmetries associated to $\mathfrak{e}_6$ and
$\fsu(3)$, hence the global form of the gauge group is indeed
$(E_6\times SU(3))/\bZ_3$.

We emphasize that in this last example it was essential that
$\L_{E_6}=-\L_{\bZ_3}$, so this is another sensitive check of our
arguments. This condition can also be understood along the lines of the previous section: because of the change in sign of
the linking form, the induced fractional instanton numbers associated to the two factors are equal and opposite, so that the $\tau\to\tau+1$
transformation becomes anomaly-free.

Similar checks can be performed for the rest of the maximal
subgroups of $E_8$. For instance,
$(Spin(10)\times SU(4))/\bZ_4$ is another example where the precise signs
in~\eqref{eq:ADE-linking-forms} are crucial to get the right
results.\footnote{As in~\eqref{eq:ADE-self-dual}, we use the global form of the 5d gauge theory that results from circle compactification to label genuine $(2,0)$ theories.}
An interesting case is
$(SU(5)\times SU(5))/\bZ_5$, for which
$H^2(S^3/\bZ_5\sqcup S^3/\bZ_5)=\bZ_5\oplus\bZ_5$ with linking pairing
\begin{equation}
  \L = \begin{pmatrix}
    \frac{1}{5} & 0\\
    0 & \frac{1}{5}
  \end{pmatrix}\, .
\end{equation}
Since $2^2\equiv -1$ mod 5 and $\gcd(2,5)=1$, we can perform
an invertible change of basis $a'=2a$ of the first $\bZ_5$ factor so
that the linking pairing becomes
\begin{equation}
  \L = \begin{pmatrix}
    -\frac{1}{5} & 0\\
    0 & \frac{1}{5}
  \end{pmatrix}
\end{equation}
and we can proceed as above.

As pointed out in \cite{Vafa:1994tf}, there are
self-dual $\cN=4$ theories with gauge group
$(SU(N)\times SU(N))/\bZ_N$ for any $N$. At first, this poses a bit of a puzzle,
since in general there is no $n$ such that $n^2\equiv -1$ mod $N$. For
instance, for $N$ prime, the condition for such an $n$ to exist (i.e., for $-1$ to be a quadratic residue) is that $N\equiv 1$ mod
4. Choose $N=3$, for example. The linking pairing is
\begin{equation}
  \L = \begin{pmatrix}
    \frac{1}{3} & 0 \\
    0 & \frac{1}{3}
  \end{pmatrix}
\end{equation}
and it is easy to see that $H^2(S_u)$ has no self-dual subspaces.

The resolution of the puzzle is that the theories described in
\cite{Vafa:1994tf} are really of the form
$(\ov{SU(N)}\times SU(N))/\bZ_N$, meaning that in the IIB string
theory realization the 16 supercharges preserved by the first factor
are precisely the 16 supercharges \emph{broken} by the second factor,
as in brane-antibrane systems. In the deep infrared, the two $A_{N-1}$
theories decouple, and each is invariant under 16
supercharges. However, the preserved supercharges have opposite
chiralities ($(2,0)$ versus $(0,2)$), and the full theory is
non-supersymmetric at the massive level.

Geometrically, this is achieved by gluing
ALE spaces with opposite orientation to each other. The
change of orientation flips the overall sign of the intersection form on the ALE space, which likewise flips the sign of
the linking pairing on $S_3/\bZ_N$ by~\eqref{eq:linking-from-intersection}.
The correct linking pairing on $H^2(S_u)$ is therefore
\begin{equation}
  \L = \begin{pmatrix}
    -\frac{1}{N} & 0 \\
    0 & \frac{1}{N}
  \end{pmatrix}
\end{equation}
which admits self-dual subspaces, such as $\Span(a+b)$.

\section{Self-dual boundary conditions}

\label{sec:polarizations}

In the previous sections we have discussed how a careful treatment of
boundary conditions in IIB string theory in
$\cM_6 \times \bC^2/\Gamma$ allows us to reproduce the known global
structure of $(2,0)$ theories of type $\fg_\Gamma$ on $\cM_6$, giving
in particular a systematic way of understanding the set of discrete
2-form symmetries of the $(2,0)$ theory and their commutation
relations, as encoded in the Heisenberg group
\begin{equation}
  0 \to U(1) \to \cW_3 \to H^3(\cM_6, \Gamma^{\text{ab}}) \to 0\, .
\end{equation}
We have seen that $\cW_3$ can be naturally understood as the group of
asymptotic fluxes for the self-dual RR 5-form on
$\cM_6\times \bC^2/\Gamma$. The known classification of $\cN=4$
theories arises beautifully from this viewpoint.

Have understood these rather subtle properties of the
6d $(2,0)$ and 4d $\cN=4$ theories from the IIB viewpoint, the following question naturally arises. Say that we choose
$\cM_6=\cM_4\times\Sigma$, as above. In choosing boundary conditions
for type IIB on $\cM_4\times\Sigma\times \bC^2/\Gamma$ we generally
need to choose between breaking large diffeomorphisms on $\Sigma$ or
on $\cM_4$. What makes the IIB boundary conditions that are invariant
under the large diffeomorphisms of $\cM_4$ special from the IIB
viewpoint? The answer, clearly, is that there is nothing special about
them from the 10d perspective. As such, it is in principle an interesting question
to choose different boundary conditions and examine their consequences. In fact,
we will argue that in some contexts it is more natural to choose boundary conditions that are invariant under large diffeomorphisms of $\Sigma$.
Perhaps surprisingly, we show that these alternate ``self-dual'' boundary conditions are possible whenever $\cM_4$ satisfies a few basic assumptions, regardless of whether a genuine $(2,0)$ theory exists in six dimensions.

\subsection{On the global structure of \texorpdfstring{$\cN=4$}{N=4} theories with duality defects}

As a warm-up, and to provide additional motivation, we first describe a situation where it becomes impossible to choose boundary conditions that are invariant under large diffeomorphisms of $\cM_4$.

We consider the $(2,0)$ theory of type $\fg_\Gamma$ compactified on $\cM_6 = \Kthree \times \Sigma$, where $\Sigma$ is a
Riemann surface and the \Kthree{} is elliptically fibered. More
concretely, we construct \Kthree{} as a hypersurface $\{P=0\}$ of degree
$(12,6)$ in a toric space $Y$ described by the gauged linear sigma
model with charges
\begin{equation}
  \begin{array}{c|ccccc}
    & u_1 & u_2 & x & y & z\\
    \hline
    \bC^*_1 & 1 & 1 & 4 & 6 & 0\\
    \bC^* & 0 & 0 & 2 & 3 & 1
  \end{array}
\end{equation}
We can write the fibration in Weierstrass form
\begin{equation}
  P = -y^2 + x^3 + f(u_1,u_2)xz^4 + g(u_1,u_2)z^6
\end{equation}
where $f$ and $g$ are sections of the line bundles $\cO_{\bP^1}(8)$ and
$\cO_{\bP^1}(12)$, respectively. (That is, locally they are
homogeneous functions of $s_1,s_2$ of degrees 8 and 12, respectively.)

There is a fibration map $\pi\colon \Kthree \to \bP^1$ induced
by the ambient space fibration $\pi_a\colon Y\to \bP^1$
\begin{equation}
  \pi_a(u_1, u_2, x, y, z) = (u_1, u_2)\, .
\end{equation}
The generic fiber $\pi^{-1}(u_1,u_2)$ is $T^2$. The Calabi-Yau space \Kthree{} has a section, namely an embedding
$\bP^1\to X$ intersecting each fiber once, given by
$\{z=0\}\cap\{P=0\}$.

There are two interesting limits to consider. When the volume of $\Sigma$ is very small, we recover a 4d $\cN = 4$ theory on \Kthree{} along the same lines as we have already discussed. If instead the volume of the $T^2$ fiber of \Kthree{} is very small, we expect an effective local description in terms of 4d $\cN=4$ SYM on $\bP^1 \times \Sigma$ with algebra $\fg_\Gamma$. However, this description is qualitatively different from the previous case,
due to the presence of
\emph{duality defects}. Recall that the complexified gauge coupling of the
$\cN=4$ theory is given by the complex structure of the $T^2$
fiber. As the fibration is non-trivial in this case, the gauge coupling varies across $\cM_4 = \bP^1 \times \Sigma$, and is now better viewed as a background field, rather than a ``constant''.%
\footnote{See
\cite{Harvey:2007ab,Cvetic:2011gp,Martucci:2014ema,Gadde:2014wma,Assel:2016wcr,Choi:2017kxf,Lawrie:2018jut,Lawrie:2016axq}
for studies of such backgrounds.} There are
codimension two loci along the $\bP^1\times \Sigma$ base---the duality defects---located at
the vanishing points of the discriminant
\begin{equation}
  \Delta(u_1,u_2)=4f(u_1, u_2)^3+27g(u_1,u_2)^2\, ,
\end{equation}
around which the complexified gauge coupling has $SL(2,\bZ)$
monodromies. Notice that $\Delta$ is a section of $\cO_{\bP^1}(24)$,
so generically it vanishes at 24 points in the base $\bP^1$.

For a generic fibration, it is easy to see that the monodromy group for loops beginning and ending at any fixed base point is the entire $\SL(2,\bZ)$.
 As an explicit example, let us start with a
class of \Kthree{} manifolds introduced by Sen \cite{Sen:1996vd}, where
\begin{align}
  \label{eq:Sen-constant-coupling}
  f(s_1, s_2) &= \alpha Q(s_i)^2 \,, & g(s_1, s_2) &= Q(s_i)^3\,, &
  Q(s_i) &= \prod_{i=1}^4(s_1 - a_is_2) \,,
\end{align}
with $\alpha$ and $a_i$ arbitrary complex constants and $a_i \ne a_j$ for $i \ne j$. The monodromy around each zero of $Q$ is given by
\begin{equation}
  \sM = \begin{pmatrix}
    -1 & 0\\
    0 & -1
  \end{pmatrix}\, .
\end{equation}
These \Kthree{} manifolds have the peculiarity that the complex structure
of the torus is constant. In fact they have a familiar interpretation
in the context of F-theory \cite{Vafa:1996xn}, where each defect corresponds to four D7 branes
on top of an O7$^-$ plane \cite{Sen:1996vd}.\footnote{We refer the
  reader interested in reading more about F-theory to the excellent
  reviews
  \cite{Denef:2008wq,Weigand:2010wm,Maharana:2012tu,Weigand:2018rez}.}
In this configuration we have
\begin{equation}
  \Delta = (4\alpha^3 + 27)Q(s_i)^6\, .
\end{equation}
That is, there are six zeroes of $\Delta$ coalescing on each zero of
$Q(s_i)$. We now study what happens around each zero of $Q(s_i)$
when we perturb $f,g$ away from the special
form~\eqref{eq:Sen-constant-coupling}. The answer is well known in the
context of F-theory (and before that, from the analysis of the
Seiberg-Witten solution of $\cN=2$ $SU(2)$ with four flavours
\cite{Seiberg:1994rs,Seiberg:1994aj}): the six zeroes of $\Delta$
split into four mutually local degenerations (the D7 branes, in
F-theory) and two mutually non-local degenerations (the O7$^-$
plane).

\begin{figure}
  \centering
  \includegraphics[width=0.5\textwidth]{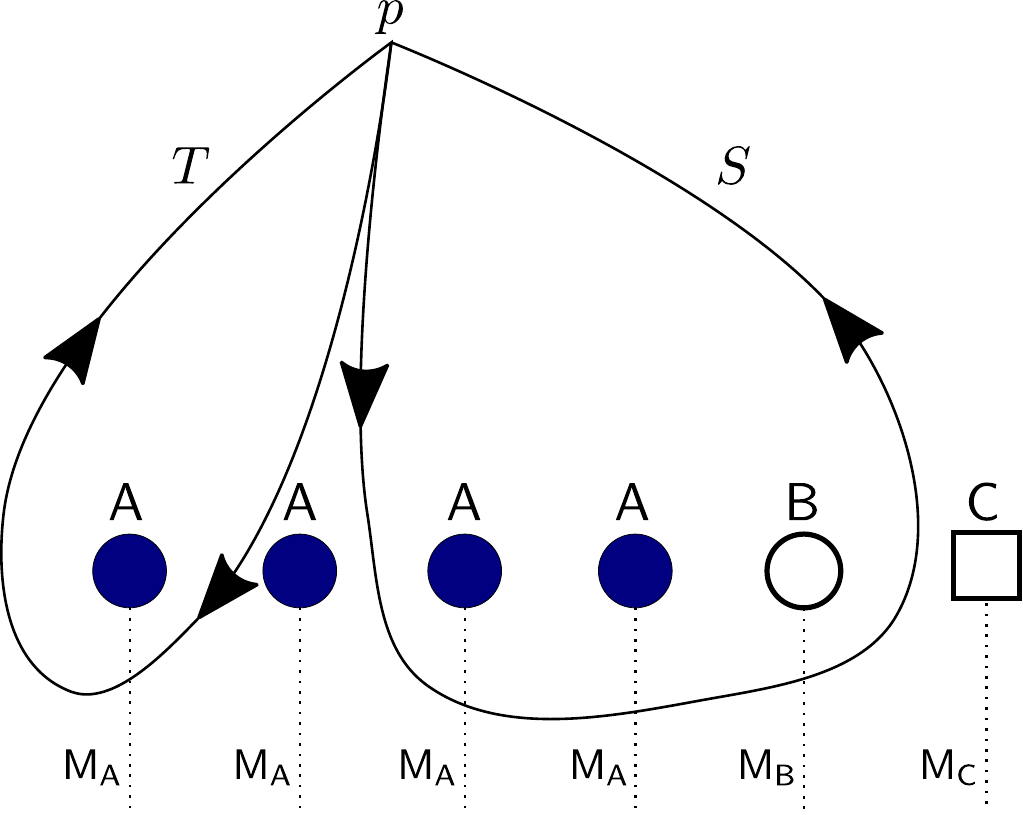}
  \caption{The paths that implement the monodromies $S$ and $T$ at the
    point $p$ in the presence of the degenerations of the elliptic
    fibration discussed in the text. The dotted lines indicate branch
    cuts. The monodromies indicated in the text are obtained by
    crossing the branch cut counterclockwise.}
  \label{fig:SO(8)-deformed}
\end{figure}

To show explicitly that the monodromy group is the full $SL(2,\bZ)$, we choose an explicit basis
for the geometry, following the conventions of
\cite{Gaberdiel:1998mv,DeWolfe:1998zf}. The defect described above splits into four
degenerations of type $\sA$, one of type $\sB$ and one of type
$\sC$. The $\sA$ degenerations are associated with degenerations the $(1,0)$
cycle of the $T^2$, the $\sB$ with degenerations of the $(1,-1)$
cycle, and $\sC$ with degenerations of the $(1,1)$ cycle (all defined
relative to a common canonical basepoint).  The $SL(2,\bZ)$
monodromies associated to these degenerations are
\begin{equation}
  \sM_\sA = \begin{pmatrix}
    1 & -1\\
    0 & 1
  \end{pmatrix} \quad ; \quad
  \sM_\sB = \begin{pmatrix}
    0 & -1\\
    1 & 2
  \end{pmatrix} \quad ; \quad
  \sM_\sC = \begin{pmatrix}
    2 & -1\\
    1 & 0
  \end{pmatrix}\, .
\end{equation}
One can obtain the two standard generators
\begin{equation}
  T = \begin{pmatrix}
    1 & 1\\
    0 & 1
  \end{pmatrix} \qquad ; \qquad
  S = \begin{pmatrix}
    0 & -1\\
    1 & 0
  \end{pmatrix}
\end{equation}
of $SL(2,\bZ)$ from here. Clearly $T=\sM_\sA^{-1}$, and one can also
see easily that $S=\sM_\sC\sM_\sA^2$. This situation is depicted in
figure~\ref{fig:SO(8)-deformed}.

\bigskip

\begin{figure}
  \centering
  \includegraphics[width=0.7\textwidth]{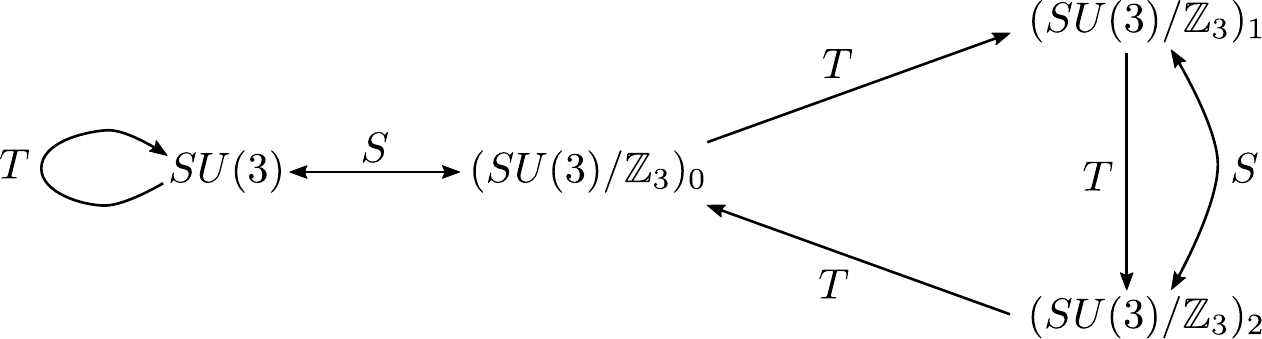}
  \caption{$SL(2,\bZ)$ duality orbits for the $\cN=4$ $\fsu(3)$
    theories, from \cite{Aharony:2013hda}.}
  \label{fig:SU(3)-dualities}
\end{figure}

We now specialize to the $A_2$ theory, for
concreteness, the generalization to other algebras being clear.\footnote{One subtlety in
  the $A_1$ case is that the 4d $\cN=4$ theory with algebra
  $\fsu(2)$ can also be understood as part of the $\fsp(2N)$ family.
  As the discussion in this paper does not cover such cases, we avoid using $A_1$ for the following argument.} Thus, we aim to describe an $\cN=4$ theory with algebra $\fsu(3)$ on
$\bP^1\times \Sigma$ in the presence of duality defects. What is the global form of the gauge group of this theory? Intriguingly, this question is not answerable, because \emph{none} of the genuine $\fsu(3)$ theories can be placed in a background with generic duality defects. This because these theories are not invariant under the $\SL(2,\bZ)$ monodromy group of a generic collection of defects, see figure~\ref{fig:SU(3)-dualities}.
More concretely, say that we declare that the gauge group on a $\bR^4$
neighbourhood of the point $p$ in figure~\ref{fig:SO(8)-deformed} is
of the form $SU(3)$. By taking the path with monodromy $\tau \to -1/\tau$, we end up with $(SU(3)/\bZ_3)_0$
instead, in contradiction with our initial assertion.

There are two kinds of solutions to this problem. More conservatively, we can restrict to some particular genuine theory, which will restrict to a particular class of duality defects leaving the choice of theory invariant. While this is well suited to certain problems, this restrictive viewpoint is not always satisfactory. For instance, the 6d viewpoint suggests the possibility of a 4d/4d correspondence between duality defects on $\bP^1 \times \Sigma$ and elliptically fibered \Kthree{} with constant gauge coupling. However, if we fix a choice of genuine theory in the former case, then the boundary conditions will not be invariant under large diffeomorphisms of \Kthree{} in the latter, whereas they \emph{will} be invariant under large diffeomorphisms of $\Sigma$.

This suggests a second solution, which is to view the 4d theory with varying $\tau$ as a kind of metatheory, just like the $(2,0)$ theory. In classifying general boundary conditions, we are led back to the Heisenberg group
\begin{equation}
  0 \to U(1) \to \cW_3 \to H^3(\Kthree\times\Sigma; \bZ_3) \to 0\, ,
\end{equation}
arising from the 6d perspective. However, in some cases it is interesting to focus on more restricted choices other than those arising from genuine 4d theories. For instance, as we have seen above, it is particularly natural to consider choices that are invariant under large diffeomorphisms of $\Sigma$. We discuss a further motivation for $\Sigma$-independent boundary conditions in~\S\ref{subsec:2d4d} below.

Does $\cW_3$ contain a maximal isotropic subspace that is
invariant under the large diffeomorphisms of $\Sigma$?  It is a non-trivial result, shown below,
that such a subspace does exist, not just for the $\fsu(3)$ theory on \Kthree{} but for any $(2,0)$ theory on any smooth, compact, Spin manifold $\cM_4$ without torsion. 

\subsection{Self-dual subspaces for smooth Spin four-manifolds}

If $\cM_4$ has no torsion, then (cf.~\eqref{eqn:M4T2Kunneth}):
\begin{equation}
  H^3(\cM_4\times \Sigma; \Gamma^{\text{ab}}) = H^3(\cM_4; \Gamma^{\text{ab}}) \oplus [H^2(\cM_4; \Gamma^{\text{ab}})\otimes H^1(\Sigma)] \oplus H^1(\cM_4; \Gamma^{\text{ab}})\, .
\end{equation}
Thus, a $\Sigma$-independent maximal isotropic subspace of $H^3(\cM_4 \times \Sigma; \Gamma^{\text{ab}})$ should take the form
\begin{equation} \label{eqn:L123}
L = L^3 \oplus (L^2 \otimes H^1(\Sigma)) \oplus L^1 \,,\qquad L^2 = (L^2)^\perp \,, \qquad L^3 = (L^1)^\perp \,,
\end{equation}
where $L^i \subseteq H^i(\cM_4; \Gamma^{\text{ab}}) = H^i(\cM_4) \otimes H^2(S^3/\Gamma)$ and $(\ldots)^\perp$ denotes the orthogonal complement with respect to the linking pairing
\begin{equation}
\L_{\cM_4}(a_1 \otimes \ell_1, a_2\otimes \ell_2) =  (a_1 \circ a_2) \L_\Gamma(\ell_1, \ell_2)\,,
\end{equation}
with $a_1 \circ a_2 = \int_{\cM_4} a_1 \smile a_2$ the intersection form on $\cM_4$.\footnote{This is a slight abuse of notation, since the intersection form is defined on homology; however, the difference is an implicit application of Poincar\'e duality, which we won't need to keep track of.}

Note that the linking pairing $\L_{\cM_4}$ is symmetric. Following the nomenclature of~\S\ref{subsec:metatheories}, we call a subspace $L$ of the form $L = L^\perp$ ``self-dual'', where in particular this differs from a maximal isotropic subspace (applicable to an antisymmetric linking pairing) in that a self-dual subspace is not guaranteed to exist. Because we can freely pick $L^1$, fixing $L^3 = (L^1)^\perp$ (or vice versa), the existence of a $\Sigma$-independent maximal isotropic subspace of $H^3(\cM_4 \times \Sigma; \Gamma^{\text{ab}})$ is equivalent to the existence of a self-dual subspace of $H^2(\cM_4; \Gamma^{\text{ab}})$. Below, we show that such a subspace exists for any smooth, compact Spin manifold $\cM_4$ without torsion and for any $\Gamma$.

\subsubsection{The \texorpdfstring{$\fsu(2)$}{su(2)} case} \label{subsec:su2}

We begin with the $A_1$ case, deferring a
general statement till later. As above, we assume that $\cM_4$ is compact, Spin, and without torsion.
 Additionally, in order to be able to apply some general
results of Donaldson, we require that $\cM_4$ is smooth.

Since $\cM_4$ is torsion-free, $H^i(\cM_4;\bZ_2)=H^i(\cM_4)\otimes
\bZ_2$. Denote by $\rho\colon H^i(\cM_4)\to H^i(\cM_4;\bZ_2)$ the
associated mod-2 reduction of cohomology classes. Explicitly, $\rho$ is constructed from the short exact sequence
\begin{equation}
  0 \to \bZ \xrightarrow{\times 2} \bZ \xrightarrow{\text{mod}\ 2} \bZ_2 \to 0\, ,
\end{equation}
which induces the long exact sequence in cohomology
\begin{equation}
  \label{eq:long-exact-torsion}
  \ldots \to H^2(\cM_4) \xrightarrow{\psi} H^2(\cM_4) \xrightarrow{\rho} H^2(\cM_4;\bZ_2) \xrightarrow{\phi} H^3(\cM_4) \to \cdots
\end{equation}
Since there is no torsion in $\cM_4$ the map $\phi=0$ in
\eqref{eq:long-exact-torsion} is necessarily vanishing, and then
$H^2(\cM_4;\bZ_2)\cong\coker(\psi)$. Since $\psi$ is simply
multiplication of elements in $H^2(\cM_4)$ by 2, we can write
\begin{equation}
  H^2(\cM_4;\bZ_2) = \frac{H^2(\cM_4)}{2H^2(\cM_4)}\, ,
\end{equation}
and think of elements of $H^2(\cM_4;\bZ_2)$ as the reduction modulo 2
of elements in $H^2(\cM_4)$.

For any $x\in H^2(\cM_4)$ we have
\begin{equation}
  x^2 = \rho(x)^2 = \Sq2(x) = \nu_2 \circ x \mod 2
\end{equation}
where $\Sq2(x)$ denotes the Steenrod square \cite{Hatcher:478079} and
$\nu_2$ is the second Wu class \cite{spanier}, which in terms of
Stiefel-Whitney classes can be written as $\nu_2=w_2+w_1^2$. A Spin
manifold has $w_1=w_2=0$, so we learn that the intersection form is
even. A theorem of Donaldson \cite{donaldson1983,donaldson1987} then
implies that the intersection form is necessarily of indefinite
signature.

Likewise, the intersection form is unimodular when $\cM_4$ is compact.
The classification of even unimodular forms of indefinite signature is
a classical result, known as the Hasse-Minkowski classification (see
\cite{MilnorHusemoller}), implying that one can choose a basis
for $H^2(\cM_4)$ such that the intersection form is given by a block
diagonal matrix of the form\footnote{Here we assume that the signature $\sigma(\cM_4) = -8p$ is non-positive. When the signature is positive, we replace $C(E_8) \to -C(E_8)$.}
\begin{equation}
  (-C(E_8))^{\oplus p} \oplus H^{\oplus q}\, .
\end{equation}
Here $p \ge 0$ and $q>0$ are integers depending on $\cM_4$, $H$ is the matrix
\begin{equation}
   H = \begin{pmatrix}
     0 & 1\\
     1 & 0
   \end{pmatrix} \,,
\end{equation}
and $-C(E_8)$ is the negative of the Cartan matrix of $E_8$, which in
our conventions will be written
\begin{equation}
  \label{eq:C(E_8)}
  C(E_8) = \begin{pmatrix}
    2 & 0 & -1 & 0 & 0 & 0 & 0 & 0 \\
    0 & 2 & 0 & -1 & 0 & 0 & 0 & 0 \\
    -1 & 0 & 2 & -1 & 0 & 0 & 0 & 0 \\
    0 & -1 & -1 & 2 & -1 & 0 & 0 & 0 \\
    0 & 0 & 0 & -1 & 2 & -1 & 0 & 0 \\
    0 & 0 & 0 & 0 & -1 & 2 & -1 & 0 \\
    0 & 0 & 0 & 0 & 0 & -1 & 2 & -1 \\
    0 & 0 & 0 & 0 & 0 & 0 & -1 & 2
  \end{pmatrix}\, .
\end{equation}

We would like to construct a $\bZ_2$-symplectic structure on
$\cM_4$, by which we mean a choice of basis for $H^2(\cM_4;\bZ_2)$
such that the $\bZ_2$ valued intersection form in this basis
has the form $H^{\oplus
  (4p+q)}$. Clearly, the problem reduces to finding a change of basis
for each $C(E_8)$ block. 
The $\bZ_2$-valued intersection form on
$H^2(\cM_4;\bZ_2)$ is defined in terms of that on $H^2(\cM_4)$ by
\begin{equation}
  a\cdot b \equiv \hat{a} \circ \hat{b} \pmod 2\, ,
\end{equation}
where for conciseness we have denoted $\hat{a}=\rho^{-1}(a)$ and
$\hat{b}=\rho^{-1}(b)$ for the uplifts to $H^2(\cM_4)$. Likewise, there is a quadratic refinement $(\cdot)^2/2: H^2(\cM_4;\bZ_2) \to \bZ_2$ (the Pontryagin square) given by
\begin{equation}
  a^2/2 \equiv (\hat{a}\circ \hat{a})/2 \pmod 2\, .
\end{equation}
As discussed above, the intersection form is even, so $a^2/2 \in \bZ_2$ as required. While we are primarily interested in the intersection form, the Pontryagin square shows up in certain calculations (see~\eqref{eq:ninst-linking} and~\cite{Aharony:2013hda}), so we will keep track of it as well. In particular, the symplectic basis can be chosen so that $a^2/2 = 0$ for each basis element $a$.

We now explicitly construct a $\bZ_2$-symplectic basis for the $C(E_8)$ intersection form. Denote by $\hat{e}_i$ the generators of $H^2(\cM_4)$ with
$\hat{e}_i\cdot \hat{e}_j = -C(E_8)_{ij}$ as above. They define an
associated basis $\{e_i\}$ of $H^2(\cM_4;\bZ_2)$, by taking
$e_i=\rho(\hat{e}_i)$. The desired symplectic basis is given by
$s_i=S_{ij}e_j$ with
\begin{equation}
  S = \begin{pmatrix}
    1 & 1 & 1 & 0 & 1 & 0 & 1 & 1 \\
    0 & 1 & 1 & 0 & 1 & 0 & 1 & 1 \\
    1 & 1 & 0 & 1 & 0 & 0 & 0 & 0 \\
    1 & 0 & 0 & 1 & 0 & 0 & 0 & 0 \\
    0 & 1 & 0 & 0 & 1 & 0 & 0 & 0 \\
    0 & 1 & 0 & 0 & 1 & 1 & 0 & 0 \\
    1 & 1 & 0 & 0 & 1 & 0 & 1 & 0 \\
    1 & 0 & 0 & 0 & 0 & 0 & 0 & 1
  \end{pmatrix}\, .
\end{equation}
Consider the lift $\hat{s}_i = S_{ij} \hat{e}_j$. We have
\begin{equation}
  \hat{s}_i\cdot \hat{s}_j = -(SC(E_8)S^t)_{ij} = -\begin{pmatrix}
    8 & 7 & 0 & -2 & 4 & 2 & 6 & 2 \\
    7 & 8 & -2 & -4 & 4 & 2 & 4 & 0 \\
    0 & -2 & 4 & 3 & 0 & 0 & 2 & 2 \\
    -2 & -4 & 3 & 4 & -2 & -2 & 0 & 2 \\
    4 & 4 & 0 & -2 & 4 & 3 & 4 & 0 \\
    2 & 2 & 0 & -2 & 3 & 4 & 2 & 0 \\
    6 & 4 & 2 & 0 & 4 & 2 & 8 & 1 \\
    2 & 0 & 2 & 2 & 0 & 0 & 1 & 4
  \end{pmatrix}_{ij}\, .
\end{equation}
Reducing modulo 2, we conclude that
\begin{equation}
  s_i \cdot s_j = (H^{\oplus 4})_{i j} \,.
\end{equation}
Since the diagonal elements are likewise multiples of four, $s_i^2/2 = 0$, and we are done.

\subsubsection{\texorpdfstring{$SL(2,\bZ)$}{An SL(2,Z)}-invariant partition function on \texorpdfstring{\Kthree{}}{K3}}
\label{sec:K3-partition-function}

The symplectic basis we have just constructed yields a self-dual subspace $\cI_0$ of
$H^2(\cM_4; \bZ_2)$ in a rather trivial manner: divide the generators $s_i$ into pairs $(e_i,\bar{e}_i)$,
with the property that $e_i\cdot e_j = \bar{e}_i\cdot \bar{e}_j=0$ and
$e_i\cdot \bar{e}_j=\delta_{ij}$. Then $\cI_0 =\Span\{e_1,\ldots,e_n\}$ is self-dual, where $n=4p+q$ and $|\cI_0| = 2^n$. There are many such subspaces because, e.g., we can exchange $e_1 \leftrightarrow \bar{e}_1$, and likewise for the other pairs. Note that $\cI_0$ is also ``null'', which we define to mean $x^2/2 = 0$ for all $x \in \cI_0$.\footnote{Thus, the
elements of $\cI_0$ are \emph{even}, in the classification of \cite{Vafa:1994tf}.} Self-dual does not imply null, since, e.g., $\Span\{e_1+\bar{e}_1,\ldots,e_n+\bar{e}_n\}$ is self-dual but not null. This distinction makes a difference for some calculations.

Letting $\cM_4$ be simply connected for simplicity (so as to be able to ignore $L^{1,3}$ in~\eqref{eqn:L123}), corresponding to the self-dual subspace $\cI_0$ there is a maximal isotropic subspace
\begin{equation} \label{eqn:selfdualL}
L_0 = \cI_0 \otimes H^1(\Sigma) \subset H^3(\cM_4 \times \Sigma)\,.
\end{equation}
The associated choice of boundary conditions is invariant under large diffeomorphisms of $\Sigma$, as desired, but not under large diffeomorphisms of $\cM_4$.

We now perform some checks on this
construction. Consider type IIB string theory on $\cM_4\times T^2\times \bC^2/\bZ_2$ with the
boundary condition $\ket{L_0}$ associated to the maximal isotropic subspace $L_0$ in~\eqref{eqn:selfdualL}. Since $L_0$ is invariant under large diffeomorphisms of
$T^2$, in the small $T^2$ limit we expect to obtain an effective 4d description on
$\cM_4$ that is $\SL(2,\bZ)$ invariant. For concreteness, choose $\cM_4=\Kthree$, which should give a four
dimensional $\cN=4$ theory with algebra $\fsu(2)$ on \Kthree{}, with a
peculiar choice of global structure that is not invariant under large
diffeomorphisms of the \Kthree{},\footnote{A peculiarity of the \Kthree{} case
  is that the theory can be topologically twisted without changing the
  partition function \cite{Vafa:1994tf}, which implies that the
  partition function itself will, in fact, be invariant under large
  diffeomorphisms of the \Kthree{}. But we have no reason to expect this to
  be true for general $\cM_4$.\label{fn:toptwist}} but is invariant under the $SL(2,\bZ)$
duality group of $\cN=4$ $\fsu(2)$ theory.\footnote{The existence of
  self-dual phases of the $\cN=4$ $\fsu(2)$ theory has been suggested
  by Argyres and Martone \cite{Argyres:2016yzz}. Here we have shown
  that something similar can be constructed in IIB, at the price of breaking
  invariance under large diffeomorphisms in four dimensions. We
  emphasize that there is another class of constructions one could
  consider in this context, the $N=1$ case in the $\fsp(2N)$ family,
  which we have not yet analysed.} We will refer to this choice of
global structure as $SO(3)_0$. We emphasize that in contrast to genuine 4d field theories, but
reflecting its origin in the $(2,0)$ theory of type $A_1$ (equivalently from the IIB $\bC^2/\bZ_2$ orbifold), the construction of this ``theory'' makes explicit reference to the
topology of $\cM_4$ in the form of a particular choice of self-dual $\cI$, which
is not invariant under large diffeomorphisms of $\cM_4$. Note that by the 4d/4d correspondence discussed above, this theory is related to one with 24 duality defects on $\bP^1\times T^2$ where invariance under large diffeomorphisms has been imposed along the $T^2$. The existence of self-dual $\cI$ is a non-trivial check (at the level of the partition function) that this is possible.

To proceed further, we write $\ket{L_0}$ in a particular basis, which will give an expansion of the partition function
$Z(\tau) = \braket{Z}{L_0}$ in terms of ``conformal blocks.'' A convenient choice is the one associated to
the ordinary $SU(2)$ theory, with basis elements $\ket{v}$ for each $v\in H^2(\cM_4;\bZ_2)$, each associated to a background flux $v$ for the $\bZ_2$ one-form symmetry
of the theory. Equivalently, as in~\cite{Vafa:1994tf}, we can think of the basis elements $\ket{v}$ as representing classes of $\SO(3)$ gauge bundles that are not $\SU(2)$ gauge bundles. Referring to~\secref{sec:construction-of-H}, \secref{sec:comparison-4d} we see that
\begin{equation}
\Phi_{u \otimes a} \ket{v} = (-1)^{u \cdot v} \ket{v} \,, \qquad \Phi_{u \otimes b} \ket{v} = \ket{u+v} \,,
\end{equation}
where $a, b$ denote the generators of the $A$ and $B$ cycles of $T^2$ and we fix the phases of the basis elements so that $\ket{v} = \Phi_{v \otimes b} \ket{0}$.

The associated conformal blocks $Z_v(\tau)=\braket{Z}{v}$ were computed by Vafa and Witten~\cite{Vafa:1994tf}. We will
not need the precise expressions, only their transformations under
$SL(2,\bZ)$, which are given below. By linearity, given
$Z_v(\tau)$ we can determine the partition function for the $\SO(3)_0$
theory on \Kthree{} once we decompose $\ket{L_0}$ in the
$\{\ket{v}\}$ basis. To do so, recall that the defining property of
$\ket{L_0}$ is that
\begin{equation}
  \label{eq:I0-eigen}
  \Phi_{x}\ket{L_0} = \ket{L_0}
\end{equation}
for all $x\in L_0 = \cI_0 \otimes H^1(T^2)$. Since $\{\ket{v}\}$ is a basis, we can write
\begin{equation}
  \ket{L_0} = \sum_{v} c_v \ket{v}
\end{equation}
for some $c_v$ to be determined. We have $L_0 = L_0^{(a)} \oplus L_0^{(b)}$ where $L_0^{(a)}\df \cI_0 \otimes a$ and $L_0^{(b)}\df \cI_0\otimes b$ with $a,b$ the generators of
$H^1(T^2)$, so if~\eqref{eq:I0-eigen} holds for $L_0^{(a)}$
and $L_0^{(b)}$ then it holds for all $L_0$. Thus, we require
\begin{align}
  \label{eq:I0-a-condition}
  \Phi_{a\otimes u} \ket{L_0} &= \sum_v c_v (-1)^{u\cdot v} \ket{v} = \ket{L_0} , & \Phi_{b\otimes u} \ket{L_0} &= \sum_{v\in\cI_0} c_v \ket{u+v} = \ket{L_0} ,
\end{align}
for all $u\in\cI_0$.
The first condition implies that
\begin{equation}
  c_v (-1)^{u\cdot v} = c_v\,, \qquad \forall u\in\cI_0 \,.
\end{equation}
Since $\cI_0$ is self-dual, this implies
that $c_v=0$ for $v\notin\cI_0$. The second condition then implies $c_v = c_v'$ for $v, v'\in\cI_0$, so we conclude that
\begin{equation} \label{eqn:SO30ket}
\ket{L_0} = \sum_{v \in \cI_0} \ket{v} \,,
\end{equation}
up to an overall normalization that we will not fix carefully. The partition function for $\SO(3)_0$ is therefore
\begin{equation}
  \label{eq:Z0-sum}
  Z_{\SO(3)_0}(\tau) = \sum_{v\in\cI_0} Z_v(\tau)\, .
\end{equation}

According to the results in \cite{Vafa:1994tf}, under S-duality we
have
\begin{equation} \label{eqn:K3Strans}
  Z_v(-1/\tau) = (-1)^{\frac{\chi(\Kthree)+\sigma(\Kthree)}{4}} 2^{-n}\left(\frac{\tau}{i}\right)^{-\frac{\chi(\Kthree)}{2}}\sum_{u\in H^2(\cM_4;\bZ_2)} (-1)^{u\cdot v} Z_u(\tau)\, ,
\end{equation}
with $\chi(\Kthree)=24$ and $\sigma(\Kthree) = -16$ the Euler characteristic and signature of \Kthree{}, and $n=11$ so that
$\dim(H^2(\cM_4))=2n$ as above. Likewise,\footnote{For general four-manifolds $\cM_4$ there is be an extra
  phase $e^{-2\pi i s}$, where $s=\chi(\cM_4)/12$ \cite{Vafa:1994tf},
  but we ignore it in what follows since for $\cM_4=\Kthree$ we have
  $s\equiv 0$ mod 1.}
\begin{equation}
  Z_v(\tau+1) = (-1)^{v^2/2}Z_v(\tau)\, ,
\end{equation}
in agreement with the general discussion in~\secref{sec:ninst}.

We now check how $Z_0(\tau)$ transforms under $\SL(2,\bZ)$. Invariance under $T$ is immediate, since $\cI_0$ is null. 
To derive the $S$-transformation, we use
\begin{equation}
  \label{eq:Z2-projector}
  \sum_{v\in\cI_0} (-1)^{u\cdot v} = \begin{cases}
    2^n & \text{if } u \in \cI_0\, , \\
    0 & \text{if } u \notin \cI_0\, .
  \end{cases}
\end{equation}
The case $u \in \cI_0$ is immediate. When $u \notin \cI_0$, $\cI_0$ can be divided into cosets $\cI_u$ and $\cI_u + v_u$, where $\cI_u = \{v \in \cI_0 | u\cdot v = 0\}$ and $v_u$ is any element of $\cI_0 - \cI_u$ (so that $(-1)^{u \cdot v_u} = -1$).
The cosets are of equal size $2^{n-1}$, so the positive and negative terms in the sum cancel.
Using~\eqref{eqn:K3Strans} and~\eqref{eq:Z2-projector}, we immediately conclude that
\begin{equation}
  \label{eq:Z0-S-duality}
  Z_{\SO(3)_0}(-1/\tau) = \tau^{-12} Z_{\SO(3)_0}(\tau)\, .
\end{equation}
As explained in \cite{Vafa:1994tf}, the fact that the partition
function is a modular form of non-zero weight originates from omitting higher derivative couplings to the
curvature.\footnote{It would be
  very interesting to derive these corrections from the 11d anomaly theory for type IIB
  string theory, similarly to the heuristic argument in
  \S\ref{sec:ninst}, see footnote~\ref{footnote:phases-computation}.} We obtain a fully modular invariant function by
multiplying by, e.g., $\eta(\tau)^{24}$. Something similar should
correspond to including the appropriate higher derivative terms in the calculation.

\bigskip

One can perform a similar analysis of the modular properties of the \Kthree{} partition function for $\fsu(N)$ theories. We leave the technical details to appendix~\ref{app:K3-partition-functions}, but note that the results are in perfect agreement with the type IIB viewpoint developed above as well as with~\cite{Aharony:2013hda}. For the present, we confine ourselves to the question of the existence of self-dual $\cI_0$ in the $\fsu(N)$ case (and for other algebras), as discussed below.

\subsubsection{\texorpdfstring{$\fsu(N)$}{su(N)} and other algebras}
\label{subsec:suNselfdual}

It is not difficult to extend the argument of~\secref{subsec:su2} from $\fsu(2)$ to
$\fsu(N)$ for small values of $N$ by brute force, but the
computation quickly gets unwieldy. Luckily, the mathematical problem
that we are studying has a well known general solution.\footnote{We thank Jack
Shotton for explaining the following proof to us.}  Abstractly, what
we are trying to show is that the Cartan matrix $C(E_8)$
in~\eqref{eq:C(E_8)} and $H^{\oplus 4}$ are equivalent as bilinear
forms over $\bZ_N$. This is certainly not true over $\bZ$, as $C(E_8)$ and $H^{\oplus 4}$ have different signatures, $(8,0)$ and $(4,4)$ respectively. However, the signature is not well-defined over $\bZ_N$, and so with no obvious invariant to distinguish them, it perhaps unsurprising that $C(E_8)$ and $H^{\oplus 4}$ become equivalent. In fact, we will see that any two even unimodular bilinear forms of the same dimension become equivalent over $\bZ_N$.

To show this, note that by the Chinese remainder theorem
\begin{equation}
  \bZ_N=\bZ_{p_1^{n_1}}\oplus \cdots \oplus \bZ_{p_k^{n_k}} \,,
\end{equation}
where $N=p_1^{n_1}\cdots p_k^{n_k}$ and the $p_i$ are distinct prime
numbers. Given a change of basis from $C(E_8)$ to
$H^{\oplus 4}$ modulo $N_i = p_i^{n_i}$ for each factor, we can again use the Chinese
remainder theorem to assemble them into a change of basis over $N$.
Thus, we can set $N=p^n$ for the rest of the proof without loss of generality.

We first consider the case of $p\neq 2$. We introduce the
Jacobi-Legendre symbol for $a \in \mathbb{Z}$ and prime $p$, defined as
\begin{equation}
  \left(\frac{a}{p}\right) = \begin{cases}
    1 & \text{if $a$ is a quadratic residue mod $p$ and $a\nequiv 0 \bmod p$},\\
    -1 & \text{if $a$ is not a quadratic residue mod $p$,}\\
    0 & \text{otherwise.}
  \end{cases}
\end{equation}
Then theorem 9 in section 15.7.2 of \cite{ConwaySloane}
implies\footnote{We give a simplified version that avoids
  the use of $p$-adic integers $\hat{\bZ}_p$ (distinct from $\bZ_p\df \bZ/p\bZ$). The precise statement
  in \cite{ConwaySloane} is that two $p$-adic quadratic forms
  $\hat{f}$ and $\hat{g}$ are equivalent if the conditions given in
  the text hold for every Jordan block of $\hat{f}$ and $\hat{g}$. To
  reach the statement in the text we first promote $C(E_8)$ and
  $H^{\oplus 4}$ to bilinear forms on $\hat{\bZ}_p$, then use the result
  in \cite{ConwaySloane} to prove that they are equivalent over
  $\hat{\bZ}_p$, and finally use the well-known fact that for every
  $m\geq 0$, every $\alpha\in\hat{\bZ}_p$ is congruent modulo $p^m$ to a
  unique integer $0 \leq n< p^m$ to reduce the $p$-adic
  transformations that implement the change of basis to $\bZ_{p^m}$
  transformations.} that two quadratic forms $f$ and $g$ are
equivalent over $\bZ_{p^{n}}$ if and only if they have the same
dimensions and
\begin{equation}
  \left(\frac{\det f}{p}\right) = \left(\frac{\det g}{p}\right)\, .
\end{equation}
Since we have that both $\det(H^{\oplus 4})=\det(C(E_8))=1$, it
follows that these bilinear forms (or more generally, any two
unimodular forms) are equivalent modulo $p^{n}$ for any $n>0$ and any prime $p>2$.

The case $p=2$ is addressed in
theorem 10 in section 15.7.5 of \cite{ConwaySloane}. This theorem
implies\footnote{There is also a classification theorem for odd forms;
  for simplicity we only present the statement for even forms.}  that
\emph{even} bilinear forms are equivalent over $2^{n}$ iff the same
conditions as above hold, namely that they have the same dimension and
the same Jacobi-Legendre symbol
\begin{equation}
  \left(\frac{\det f}{2}\right) = \left(\frac{\det g}{2}\right)\, .
\end{equation}
As above, this is clearly satisfied for $C(E_8)$ and $H^{\oplus 4}$,
as they are both even unimodular.

Thus, $C(E_8)$ and $H^{\oplus 4}$ are equivalent bilinear forms modulo $N$ for any integer $N$. Note, however, that per~\eqref{eq:ninst-linking} the partition function depends not just on the intersection form but also on its quadratic refinement, the Pontryagin square. Fortunately, in the torsion-free case that we are studying the quadratic refinement over $\bZ_N$ can be extracted from the intersection form over $\bZ_{2N}$, and so the above argument implies that the quadratic refinements of $C(E_8)$ and $H^{\oplus 4}$ are likewise equivalent over $\bZ_N$ for any $N$. In particular, since the above argument does not depend on the details of $C(E_8)$ and $H^{\oplus 4}$, any even unimodular bilinear form admits a basis of the form
\begin{equation} \label{eqn:specialbasis}
e_i \cdot e_j = \bar{e}_i \cdot \bar{e}_j = 0 \,, \qquad e_i \cdot \bar{e}_j = \delta_{i j} \,, \qquad e_i^2/2 = \bar{e}_i^2/2 = 0 \,,
\end{equation}
upon reduction modulo $N$. From this, we can construct a null, self-dual subspace for $\fsu(N)$ compactified on any smooth, compact, Spin $\cM_4$ without torsion,
\begin{equation} \label{eqn:I0}
\cI_0 = \Span\{e_i \} \,,
\end{equation}
just as in the case $N=2$ discussed above.

\bigskip

We briefly comment on the $(2,0)$ theories of $D$ and $E$ type. In the cases $D_{2k+1}$ and $E_k$, the defect group is cyclic, and the above analysis remains valid. For the $D_{4k+2}$ theory, the basis~\eqref{eqn:specialbasis} for $N=2$ leads to a larger basis
\begin{equation}
\alpha_i \cdot \bar{\alpha}_j = \delta_{i j} ,\qquad \beta_i \cdot \bar{\beta}_j = \delta_{i j} \,, \qquad \alpha_i^2/2 = \bar{\alpha}_i^2/2 = \beta_i^2/2 = \bar{\beta}_i^2/2 = 0\,,
\end{equation}
with $\alpha_i = e_i \otimes \alpha$, $\beta_i = e_i \otimes \beta$, etc., where $\alpha$ and $\beta$ are the generators of the $\bZ_2 \oplus \bZ_2$ defect group. Likewise, for $D_{4k}$ we obtain
\begin{equation}
\alpha_i \cdot \bar{\beta}_j = \delta_{i j} ,\qquad \beta_i \cdot \bar{\alpha}_j = \delta_{i j} \,, \qquad \alpha_i^2/2 = \bar{\alpha}_i^2/2 = \beta_i^2/2 = \bar{\beta}_i^2/2 = 0\,.
\end{equation}
In either case, the expanded basis is still of the form~\eqref{eqn:specialbasis} for $N=2$ (with twice as many generators), and so a null self-dual subspace exists as before.

While we expect that similar statements can be made for any defect group and linking pairing, we defer further consideration of this to a future work.

\subsection{2d/4d correspondences and Hecke transforms} \label{subsec:2d4d}

There is another application of the previous discussion that we will
now briefly outline. Consider a compactification of the $(2,0)$ theory
of type $\fg_\Gamma$ on $\cM_6= \cM_4 \times \Sigma$. There are two
natural limits to take: we can take the limit in which $\Sigma$ is
small, obtaining an effective four-dimensional theory $\cT_{\Sigma}$
on $\cM_4$, or alternatively we can first make $\cM_4$ small,
obtaining a two-dimensional theory $\cT_{\cM_4}$ on $\Sigma$. There
are deep relations between $\cT_{\cM_4}$ and $\cT_\Sigma$, due to
their common six-dimensional origin. It is expected that such a 2d/4d
correspondence exists for any suitable $\cM_4$/$\Sigma$ pair, as long
as we can introduce appropriate supersymmetric twists.\footnote{A case
  that has been extensively studied in the last few years is the one
  described by Alday, Gaiotto and Tachikawa \cite{Alday:2009aq}, see
  \cite{Tachikawa:2016kfc} for a clear and concise review. The best
  understood cases in this context are $\cM_4=S^4$, in which
  $\cT_{S^4}$ is Liouville theory (or perhaps more naturally, Toda
  theory \cite{Cordova:2016cmu}), and $\cM_4=S^3\times S^1$, where
  $\cT_{S^3\times S^1}$ is $q$-deformed Yang-Mills
  \cite{Gadde:2011ik}. From the point of view of this paper, the most
  interesting cases arise whenever $\cM_4$ has non-trivial one or
  two-cycles. An example of a configuration with non-trivial
  one-cycles is $\cM_4=S^3\times S^1$, and indeed in this case one can
  relate the choice of maximal isotropic subgroup in the Heisenberg
  group with the choice of the global form of the $q$-deformed
  Yang-Mills theory \cite{Tachikawa:2013hya}. The simplest example
  with non-trivial two-cycles is $\cM_4=S^2\times S^2$, which was
  studied in \cite{Bawane:2014uka}.}

A basic observable that we can compute in these
theories is the partition function. The expectation is that
\begin{equation}
  \label{eq:Z-limits}
  Z^\Gamma_{(2,0)}[\cM_4\times \Sigma; L] = Z_{\cT_{\Sigma}}[\cM_4] = Z_{\cT_{\cM_4}}[\Sigma]
\end{equation}
where the first term denotes the partition function of the
six-dimensional $(2,0)$ theory of type $\Gamma$ on
$\cM_4\times\Sigma$, with a choice of maximal isotropic subspace
$L\subset\cW_3$. Although we have chosen not to display it to avoid
cluttering the notation too much, $\cT_{\cM_4}$ and $\cT_{\Sigma}$
will depend on the choice of $L$ and $\Gamma$.

An important subtlety now arises: to obtain a genuine 4d theory, we should choose $L$ to be independent of the details of $\cM_4$. Likewise, to obtain a genuine (i.e., modular invariant) 2d theory, we should choose $L$ to be independent of the details of $\Sigma$. However, we have seen that in general $L$ cannot simultaneously be invariant under large diffeomorphisms of both $\cM_4$ and $\Sigma$! In particular, this is only seems to be possible when the corresponding $(2,0)$ theory is genuine, see~\secref{subsec:metatheories}. Thus, to make~(\ref{eq:Z-limits}) true we must either choose a genuine $(2,0)$ theory --- and the associated $\Sigma$ and $\cM_4$ reductions\footnote{In particular, the 4d theory must be modular invariant in this case.} --- or at least one of $\cT_{\cM_4}$ and $\cT_{\Sigma}$ cannot be genuine.

In this section, we will explore the consequences of making the 2d theory $\cT_{\cM_4}$ genuine (modular invariant). To do so, we choose a maximal isotropic subspace of the form~\eqref{eqn:L123}, just as above. In particular, fixing $\cM_4 = \Kthree$ and choosing a self-dual subspace $\cI_0$ of the form~\eqref{eqn:I0} leads to a modular-invariant 2d CFT $\cK_\Gamma$ on $\Sigma$.\footnote{It is possible that the particular 2d CFT obtained in this way depends on the specific choice of $\cI_0$; however, the topologically twisted Vafa-Witten partition function (essentially an elliptic genus from the 2d perspective) does not depend on this choice (cf.\ footnote~\ref{fn:toptwist}), and we so we ignore this subtlety for now.}
We have not identified this theory, but we will be able to prove some interesting
facts about its elliptic genus (as defined by the Vafa-Witten partition function~\cite{Minahan:1998vr}).

Choose, for concreteness, $\Gamma=\bZ_2$. We denote by
$\cK_2\df \cK_{\bZ_2}$ the two-dimensional modular invariant
theory that we are after. In this case~\eqref{eq:Z-limits} implies
that
\begin{equation}
  Z_{\cK_2}[T^2] = Z_{\SO(3)_0}[\Kthree]
\end{equation}
see~\eqref{eq:Z0-sum}.
 It is interesting to
compute this explicitly using the results of \cite{Vafa:1994tf}. In particular, from a general partition vector
\begin{equation}
\ket{L} = \sum_v c_v \ket{v}
\end{equation}
we obtain
\begin{equation}
Z[\Kthree] = c_0 \hat{Z}(\tau) + c_{\text{even}} Z_{\text{even}}(\tau) + c_{\text{odd}} Z_{\text{odd}}(\tau) \,, \qquad c_{\text{even}} = \sum_{\substack{v\ne 0\\v^2/2 = 0}} c_v \,, \qquad c_{\text{odd}} = \sum_{v^2/2 = 1} c_v \,,
\end{equation}
since the conformal blocks $Z_v = \{\hat{Z},Z_{\text{even}},Z_{\text{odd}}\}$ only depend on whether $v$ is zero, non-zero and even, or odd, with \cite{Vafa:1994tf}
\begin{subequations}
  \begin{align}
    \hat{Z}(\tau) & = \frac{1}{4}G(q^2) + \frac{1}{2}\left[G(q^{1/2}) + G(-q^{1/2})\right] \,, \\
    Z_{\text{even}}(\tau) & = \frac{1}{2}\left[G(q^{1/2}) + G(-q^{1/2})\right] \,,\\
    Z_{\text{odd}}(\tau) & = \frac{1}{2}\left[G(q^{1/2}) - G(-q^{1/2})\right] \,.
  \end{align}
\end{subequations}
where $q=\exp(2\pi i\tau)$ and
\begin{equation} \label{eqn:Gq}
  G(q) \df \frac{1}{\eta^{24}(q)} = \frac{1}{q\prod_{n=1}^\infty (1-q^n)^{24}}\, .
\end{equation}
Note that this is the elliptic genus for 24 left moving bosons, excluding their zero modes.

As a warmup, we describe the partition function for the genuine 4d $\fsu(2)$ theories on \Kthree{}, following the nomenclature of \cite{Aharony:2013hda}.
For $\SU(2)$, we have $c_v = (1/2) \delta_{v,0}$, where overall $1/2$ follows the conventions of~\cite{Vafa:1994tf}. By comparison, for $\SO(3)_+$ theory, we have $c_v = 1$, implying that $c_{\text{even}} = n_{\text{even}}$ and $c_{\text{odd}} = n_{\text{odd}}$ where $n_{\text{even}}=\frac{1}{2}(2^{22}+2^{11})-1$ and
$n_{\text{odd}}=\frac{1}{2}(2^{22}-2^{11})$ count the number of non-trivial $\SO(3)$ gauge bundles (see~\cite{Vafa:1994tf}). Finally, for $\SO(3)_-$ we have $c_v = (-1)^{v^2/2}$, which is the same as before except with $c_{\text{odd}} = -n_{\text{odd}}$.

We now consider the $\SO(3)_0$ theory, as defined above. From~\eqref{eqn:SO30ket}, $c_v = \delta_{v \in \cI_0}$. Since $\cI_0$ is null and self-dual, this implies $c_0 = 1$, $c_{\text{even}} = 2^{11} - 1$, and $c_{\text{odd}} = 0$, where we use $|\cI_0| = 2^{11}$. Thus, explicitly
\begin{equation}
  Z_{\cK_2}[T^2] = Z_{\SO(3)_0}[\Kthree] = \frac{1}{4}G(q^2) + 2^{10}\left[G(q^{1/2}) + G(-q^{1/2})\right]\, .
\end{equation}
It is an easy exercise, using the well-known modular transformation
properties of $\eta(q)$, to show that $Z_{\cK_2}(\tau)$ transforms as a
modular form of weight $-12$, the same as $G(q)$ itself. In fact,
the two expressions are closely connected, as
\begin{equation}
  \label{eq:Z2-Hecke}
  Z_{\cK_2}[T^2] = 2^{11}(T_2[G])(\tau)
\end{equation}
where $T_m$ is the Hecke operator (see
\cite{SerreArithmetic,1-2-3-ModularForms} for reviews, as well as~\S\ref{app:Hecke}) acting on
modular forms of weight $k$ by
\begin{equation}
  (T_m[f])(\tau) = m^{k-1}\sum_{\substack{a,d>0\\ad=m}}\frac{1}{d^k}\sum_{0 \leq b < d} f\left(\frac{a\tau+b}{d}\right)\, .
\end{equation}
In fact, the relation~\eqref{eq:Z2-Hecke} holds more generally. Let $\cK_p$ denote the theory $\cK_{\bZ_p}$ arising from the $A_{p-1}$ theory on \Kthree{}. In the case where $p$ is prime, the elliptic genus for $\cK_p$ is computed in appendix~\ref{app:su(p)-prime}, with the result
\begin{equation}
  Z_{\cK_p}[T^2] = \frac{1}{p^2} G (p
  \tau) + p^{10}  \sum_{j = 0}^{p - 1} G \left( \frac{\tau + j}{p} \right) \,,
\end{equation}
which can be easily checked to satisfy
\begin{equation} \label{eqn:KpHecke}
  Z_{\cK_p}[T^2] = p^{11}(T_p[G])(\tau)\, .
\end{equation}
By a more involved calculation, this formula can also be shown to hold for composite $N$, see~\secref{subsec:ZKN}.\footnote{To be precise, this is true when $N$ is square-free. When $N$ is divisible by a perfect square, different choices of $\cI_0$ lead to different partition functions. However, by imposing additional restrictions on $\cI_0$ this ambiguity is eliminated, and~\eqref{eqn:KpHecke} remains true.}

We see that, at least at the level of the elliptic genus,
the set of theories $\cK_N$ is in some sense generated from the theory
of 24 left-moving bosons with elliptic genus
$G(\tau)$. More precisely, there exists a family of modular-invariant
two-dimensional conformal field theories $\cK_N$, obtained by
compactification of the six-dimensional $(2,0)$ theory of type
$A_{N-1}$ on \Kthree{}, whose elliptic genera are
\begin{equation}
  \label{eq:Zp-f(tau)}
  Z_{\cK_N}[T^2](\tau) = f(\tau) (T_N[G])(\tau)\, ,
\end{equation}
where now we introduce a prefactor of $f(\tau)$ to account for
possible (unknown) curvature corrections \cite{Vafa:1994tf}. We expect
these corrections to restore modular invariance, as we ultimately have
a IIB compactification with boundary conditions invariant under large
diffeomorphisms on $\Sigma$. Modular
invariance constrains $f(\tau)$ to be a modular form of weight 12, but
it is otherwise unknown. We will conjecture a specific form below.

Note that the situation is very similar to the original discussion in
\cite{Minahan:1998vr}, where it was show that there is a similar
relation between the theory of $N$ M5 branes and the heterotic
string.\footnote{See \cite{Harvey:2018rdc,Bouchard:2018pem} for
  further recent work relating various two-dimensional CFTs via Hecke
  transforms.}  Namely, the partition function of $N$ M5 branes on
$\Kthree\times T^2$ is the same as the $N$-th Hecke transform of the
heterotic string partition function on $T^2$.\footnote{In this context
  the Hecke transform can be understood as an averaging over degree
  $N$ multi-coverings of the torus by the heterotic string. See
  \cite{Amariti:2015dxa,Amariti:2016bxv,Amariti:2016hlj} for previous
  work exploring the connection between global forms and
  multi-coverings of the torus.} While in the case of M5 branes it was
natural to expect the existence of a modular invariant theory --- as
the $\cN=4$ $U(N)$ theory is $SL(2,\bZ)$ invariant --- the existence
of the $\cK_N$ theories is a bit more surprising, and crucially
depends on the existence of the self-dual subspaces $\cI_0$
constructed in \S\ref{sec:K3-partition-function}. It would be very
interesting to learn more about this class of theories (and their
natural generalizations when we replace \Kthree{} by other
four-manifolds), particularly given their close connection to the
six-dimensional $(2,0)$ theory.

\subsubsection*{A conjecture for $f(\tau)$}

We now make a simple guess for $f(\tau)$, with interesting consequences. To motivate it, we make the following
assumptions:
\begin{enumerate}
\item $f(\tau)$ is a modular form of weight 12.
\item $f(\tau)$, coming from curvature corrections, is independent of
  $N$.
\item The form~\eqref{eq:Zp-f(tau)} holds for $N=1$, where we expect
  to have a trivial theory.
\end{enumerate}
These conditions, taken together, fix
$f(\tau)=G^{-1}(\tau)=\eta^{24}(\tau)$, up to an overall constant
which we take to be $N^{13}$ for convenience. That is, we conjecture
that the full modular invariant elliptic genus is
\begin{equation}
  \label{eq:Z-Tp-G}
  Z_{\cK_N}[T^2](\tau) = N^{13} \frac{T_N[G](\tau)}{G(\tau)}\, .
\end{equation}
Assuming that this is indeed the case, one finds the
result\footnote{The appearance of a minus sign in the constant term is
  perhaps unexpected, but since we are computing an elliptic genus
  there is no reason why this cannot occur. It would be interesting to
  better understand the significance of this term.}
\begin{equation}
  Z_{\cK_2}[T^2](\tau) = 2^{13} \eta^{24}(\tau)(T_2[\eta^{-24}])(\tau) = J(\tau) - 24 \,,
\end{equation}
where
\begin{equation}
  J(\tau) = j(\tau) - 744 = \frac{1}{q} + 196884q + 21493760q^{2} + 864299970q^{3} + \ldots
\end{equation}
is Klein's $j$-invariant without the constant term. Two dimensional
theories with partition function equal to the $j$-invariant have a
rich history, most notably the moonshine module constructed by Frenkel, Lepowsky and
Meurman \cite{frenkel1989vertex}, used by Borcherds to prove monstrous moonshine~\cite{Borcherds1992}. It is quite
enticing that the same function seems to appear (assuming that our
guess for $f(\tau)$ is correct) in trying to understand the partition
function of the $(2,0)$ $A_1$ theory on \Kthree{}.

From a purely mathematical point of view, we can understand the
appearance of the $j$-function here from the fact that
$Z_{\cK_N}[T^2](\tau)$ is a modular function that is analytic in the
upper half plane and meromorphic with a pole of order $N-1$ at $q=0$,
so it can be expressed as an order $N-1$ polynomial in $J(\tau)$ for
any $N$. The $N$ coefficients in the expression in terms of $J(\tau)$
can be determined by looking to the first $N$ terms in the Laurent
expansion of $Z_{\cK_N}[T^2](\tau)$ around $q=0$.

It is in fact possible to give concise expressions for the expansion
of $Z_{\cK_N}[T^2](\tau)$ in terms of $J(\tau)$ in the case that $N$
is prime, as a special case of the results in
\cite{Carney2012}. Define
\begin{equation}
  \cB(x;q) \df \frac{E_4^2(q)E_6(q)}{q(j(q) - x)} \,,
\end{equation}
where we have introduced the Eisenstein series
$E_4(q) = 1 + 240q + 2160q^{2} + 6720q^{3}+\ldots$ and
$E_6(q) = 1 - 504q - 16632q^{2} - 122976q^{3} + \ldots$. If we denote
by $B(m;x)$ the coefficients in the $q$-expansion of $\cB(x;q)$,
\begin{equation}
  \cB(x;q) \df \sum_{m=1}^\infty B(m;x) q^m \,,
\end{equation}
we have that
\begin{equation} \label{eqn:ZKpJexp}
  Z_{\cK_N}[T^2](\tau) = B(N-1;J(\tau)+744)\, .
\end{equation}
For reference, we find the elliptic genera for the first few primes to
be
\begin{align}
  Z_{\cK_2}[T^2](\tau) & = J(\tau)-24\, , \nonumber \\
  Z_{\cK_3}[T^2](\tau) & = J(\tau)^2 - 24J(\tau) - 393516\, , \nonumber \\
  Z_{\cK_5}[T^2](\tau) & = J(\tau)^{4} - 24 J(\tau)^{3} - 787284 J(\tau)^{2} - 71800864 J(\tau) + 75517745046\, .
\end{align}
We can also analyze directly the case of small composite $N$ by
comparing coefficients in the Laurent expansion around $q=0$. In this
way we find, for instance
\begin{align}
  Z_{\cK_4}[T^2](\tau) &= J(\tau)^3 - 24 J(\tau)^2 - 590400 J(\tau) - 55032320 \,, \nonumber \\
  Z_{\cK_6}[T^2](\tau) &= J(\tau)^5 - 24 J(\tau)^4 - 984168 J(\tau)^3 - 88569408 J(\tau)^2 + 191409608916 J(\tau) \nonumber \\ &\noeq \qquad + 19264322219040 \,.
\end{align}
Intriguingly, this agrees with~\eqref{eqn:ZKpJexp}, and it seems reasonable to conjecture that the formula is valid for all $N$.

It is also interesting to look directly to $q$ expansion in the small
$N$ cases:
\begin{align}
  Z_{\cK_2}[T^2](\tau) & = \frac{1}{q}-24 + \ldots \, , \nonumber \\
  Z_{\cK_3}[T^2](\tau) & = \frac{1}{q^2} - \frac{24}{q} + 252 + \ldots \,, \nonumber\\
  Z_{\cK_4}[T^2](\tau) & = \frac{1}{q^3} - \frac{24}{q^2} + \frac{252}{q} - 1472 + \ldots \, , \nonumber\\
  Z_{\cK_5}[T^2](\tau) & = \frac{1}{q^4} - \frac{24}{q^3} + \frac{252}{q^2} - \frac{1472}{q} + 4830 + \ldots \, . \\
  &\vdotswithin{=} \nonumber
\end{align}
These expressions have a remarkably simple structure, and the general
result is easily guessed from here
\begin{equation} \label{eqn:qexpansion}
 Z_{\cK_N}[T^2](\tau) = \frac{1}{q^N G(q)} + O(q) = \frac{1}{q^N} \sum_{n=1}^\infty q^n \uptau(n) + O(q) \,,
\end{equation}
where $\uptau(n)$ is the Ramanujan tau function, not to be confused with the modular parameter $\tau$. Note that this form is somewhat reminiscent of extremal CFTs, see, e.g.,~\cite{Witten:2007kt}, but with the vacuum character replaced by a different function.

In fact, \eqref{eqn:qexpansion} is easy to prove using the properties of the Hecke operator. From it, we obtain the exact expression,
\begin{equation}
Z_{\cK_N}[T^2](\tau) = \sum_{n=1}^{N} \uptau(n) J_{N-n}(\tau) \,,
\end{equation}
where $J_n(\tau) \df T_n[J](\tau)$ has the $q$ expansion $q^{-n} + 196884 q^n + \ldots$ for $n >0$, with $J_0(\tau) \df 1$.

\section{Conclusions} \label{sec:conclusions}

In this paper we have shown how to understand from the IIB perspective
the fact that the six-dimensional $(2,0)$ theories arising at
$\bC^2/\Gamma$ singularities typically do not have a partition
function, but rather a vector of partition functions, in which the
components mix under large diffeomorphisms. The basic observation is
that non-commutativity of RR fluxes in type IIB string theory and
non-commutativity of 2-form discrete flux operators in the $(2,0)$
theory are two sides of the same coin, and in fact they generate the same
Heisenberg group. So the problem of choosing a specific direction in
the Hilbert space of possible six-dimensional $(2,0)$ theories (that
is, the space of ``conformal blocks'') maps to the problem of choosing
boundary conditions in the non-compact IIB space. More formally, we
have shown how a theory with a non-invertible anomaly theory, the
six-dimensional $(2,0)$ theory, can arise as a subsector of type IIB
string theory, an anomaly-free theory.

One advantage of the IIB viewpoint is that it allows us to separate
the problem of determining the behaviour of the discrete 2-form
symmetries of the $(2,0)$ theory from the complicated local dynamics
of the tensionless strings. Effectively, we have geometrized this
sector of the problem into one involving free field theory for RR
forms, solved in \cite{Freed:2006ya,Freed:2006yc}. This reformulation
allows us to give simple derivations of some subtle facts in the
six-dimensional theory, in particular the structure of the commutation
relations for flux operators. While the answer for the $(2,0)$ theory
was already known by compactification on $T^2$, our derivation has the
virtue of easily generalizing to cases where the answer was not
previously known, such as the $(1,0)$ theories.

The IIB perspective also provides a first principles approach to discussing the
global structure of more exotic field theory setups, such as the
theories with duality defects studied in \S\ref{sec:polarizations},
and naturally suggests a method to compute the anomalous phases of
some strongly-coupled 4d theories under specific modular
transformations, as conjectured in \S\ref{sec:ninst}.

\bigskip

There are a number of interesting open questions that we have not
addressed. For instance, we have assumed that $\cM_6$ had no
torsion. This was done only for mathematical
simplicity, and it would be fact be physically quite interesting to
drop this assumption. The fact that K-theory plays an important role
in the IIB picture suggests that in the presence of torsion on $\cM_6$
K-theory might play a role in the classification of field theories. It
would be interesting to work this out in detail.

Along related lines, we have seen that in the absence of $\cM_6$ torsion
the IIB construction gives rise to all the global forms of the
$(2,0)$ theories known from the holographic
viewpoint~\cite{Witten:1998wy}. The presence of torsion in $\cM_6$ introduces an
interesting twist: the holographic classification of global forms for
the $(2,0)$ theory is given in terms of cohomology in
M-theory, while that for IIB is given in terms of K-theory. Comparing the
results of both descriptions would be very interesting, and is potentially somewhat analogous to but different from
\cite{Witten:1999vg,Diaconescu:2000wy}.

It would also be interesting to understand the results in this paper
from the viewpoint of the 11d anomaly theory for
IIB string theory
\cite{Belov:2004ht,Belov:2006jd,Belov:2006xj,Monnier:2011rk,Monnier:2013kna}. Our
results suggest that in some sense non-invertible
anomaly theories can be constructed by considering the behaviour of invertible anomaly
theories in non-compact spaces; it would be interesting to make this
statement precise. Such a reformulation would likely have useful
applications in the study of duality anomalies for strongly coupled
four dimensional theories, as we have sketched above.

Although 
we have focused mostly on
$\Kthree\times \bC^2/\Gamma \times T^2$, it is clear that the basic idea
will generalize to more involved geometries. For instance, one can
study the global structure in the case of Vafa-Witten
topologically-twisted compactifications of $\cN=4$ on other
four-manifolds $\cM_4$, such as $\bP^2$, where we also know the answer
for the partition function \cite{Vafa:1994tf,Minahan:1998vr}, by
considering cases in which $\cM_4$ appears as a submanifold of
threefolds \cite{Minahan:1998vr} or $G_2$ manifolds
\cite{Bershadsky:1995qy}. More generally, we might consider
different theories in four dimensions, such as those coming from other
choices for the 4d topological twist, or alternatively
the $\Omega$-deformed backgrounds that lead to the 4d/2d
correspondence found by Alday, Gaiotto and Tachikawa
\cite{Alday:2009aq}. (See \cite{Hellerman:2011mv,Hellerman:2012zf} for
discussion on how to realize these backgrounds in string theory.)

Along similar lines, we could also consider compactifications
on singular Calabi-Yau theefolds in the context of geometric
engineering (starting with
\cite{Klemm:1996bj,Katz:1997eq,Gukov:1999ya,Shapere:1999xr}, and more
recently
\cite{Cecotti:2010fi,Cecotti:2011gu,DelZotto:2011an,DelZotto:2015rca,Xie:2015rpa,Chen:2016bzh,Wang:2016yha,Chen:2017wkw}). The
global structure of any theory that can be engineered in terms of a
singular threefold or fourfold with isolated singularities can in
principle be obtained via an extension of the methods described
here. It would be very interesting to do this in detail.

Another assumption that would be interesting to drop is that $\cM_6$
is compact. We might, for instance, consider IIB on spacetimes of the
form $\bC^2/\Gamma_1\times \bC^2/\Gamma_2\times T^2$. As argued in
\cite{Dijkgraaf:2007sw}, such a configuration leads to a chiral WZW
model (with algebra determined by $\Gamma_1$ and $\Gamma_2$) living on
$T^2$. It is natural to conjecture that a careful analysis of the
boundary conditions of IIB in this background should reproduce the
structure of conformal blocks of the chiral WZW model, and in
particular give a geometric picture for the Verlinde formula for these
theories
\cite{Verlinde:1988sn,Dijkgraaf:1988tf,Witten:1988hf,Moore:1988qv,Blau:1993tv}
(and relatedly, a direct string theory interpretation of the work by
Nakajima \cite{nakajima1994}). Additionally, considering such
non-compact geometries would be the starting point for understanding
the behaviour of the global structure under gluing,
along the lines of \cite{Gadde:2013sca,Gukov:2018iiq}.

The inclusion of non-simply laced algebras in lower dimensions is another
important open problem. This would require the analysis of IIB
backgrounds with orientifold actions, changing the type of K-theory
that we need to consider.

Finally, we could also ask what happens if one replaces $\bC^2/\Gamma$
by a multi-centered Taub-NUT space (for simplicity we refer to the
$\Gamma=\bZ_N$ case here). The local dynamics are unaffected, but
there is an additional normalizable mode, corresponding to a centre of
mass degree of freedom. In the six-dimensional $(2,0)$ theory this
mode leads to a free tensor multiplet, which is expected to reduce in
the $\cN=4$ theory to the $U(1)$ factor of $U(N)$. In light of the
results of this paper, it is natural to ask whether this is the only
possibility, or one can obtain the other gauge groups with algebra
$\fu(1)\oplus\fsu(N)$ by choosing boundary conditions appropriately in
the same IIB background. It would be interesting to work this out in
detail; we expect the work of Belov and Moore
\cite{Belov:2004ht,Belov:2006jd,Belov:2006xj} to be relevant here.

\acknowledgments

We would like to thank Ofer Aharony, Philip Argyres, David Berman,
Ralph Blumenhagen, Matt Buican, Mat Bullimore, Federico Carta,
Andr\'{e}s Collinucci, Michele del Zotto, Jacques Distler, Michael
Fuchs, Davide Gaiotto, Amadeo Jim\'enez-Alba, Mario Martone, Mark
Powell, Diego Rodr\'iguez-G\'omez, Carlos Shahbazi, Jack Shotton,
Douglas Smith, Yuji Tachikawa, Timo Weigand and Edward Witten for
illuminating discussions. This research was supported in part by
Perimeter Institute for Theoretical Physics. Research at Perimeter
Institute is supported by the Government of Canada through Industry
Canada and by the Province of Ontario through the Ministry of Economic
Development and Innovation.  The authors would like to thank the Aspen
Center for Physics, which is supported by National Science Foundation
grant PHY-1066293, for hospitality while this work was being
initiated. I.G.-E. would also like to thank the organizers of the 2017
Pollica Summer Workshop on Dualities in Superconformal Field Theories,
for providing an ideal working environment during the early stages of
the development of this paper. B.H. is supported by National Science
Foundation grant PHY-1914934.

\appendix

\section{K-theory groups of \texorpdfstring{$S^3/\Gamma$}{S3/Gamma}}

\label{app:homology-sphere-K-theory}

In this appendix we compute the (complex) K-theory
groups of $S^3/\Gamma$, with $\Gamma$ an arbitrary discrete subgroup
of $U(2)$ acting freely on $S^3$. These spaces are orientable, with
cohomology groups\footnote{The result for $H^1$ follows from the
  universal coefficient theorem (see theorem 3.2 in
  \cite{Hatcher:478079})
  \begin{equation}
    0\to \Ext(H_{n-1}(X), \bZ) \to H^n(X) \to \Hom(H_n(X), \bZ) \to 0\, .
  \end{equation}
  Since $H_0(X)$ is free, this implies that
  $H^1(X)=\Hom(H_1(X), \bZ)$, which vanishes since
  $H_1(X)=\Gamma^{\text{ab}}$ is torsion.}
\begin{equation}
  \label{eq:spherical-space-cohomology}
  H^i(S^3/\Gamma) = \begin{cases}
    \bZ & \text{for } i\in\{0,3\}\\
    \Gamma^{\text{ab}} & \text{for } i = 2\\
    0 & \text{for } i =1
  \end{cases} \,,
\end{equation}
where $\Gamma^{\text{ab}}\coloneqq\Gamma/[\Gamma,\Gamma]$ denotes the
abelianization of $\Gamma$, as given by the Hurewicz
homomorphism. Note that $\Gamma^{\text{ab}}$ is pure torsion, since
$\Gamma$ itself is.

\subsubsection*{The Atiyah-Hirzebruch spectral sequence}

\begin{figure}
  \centering
  \includegraphics{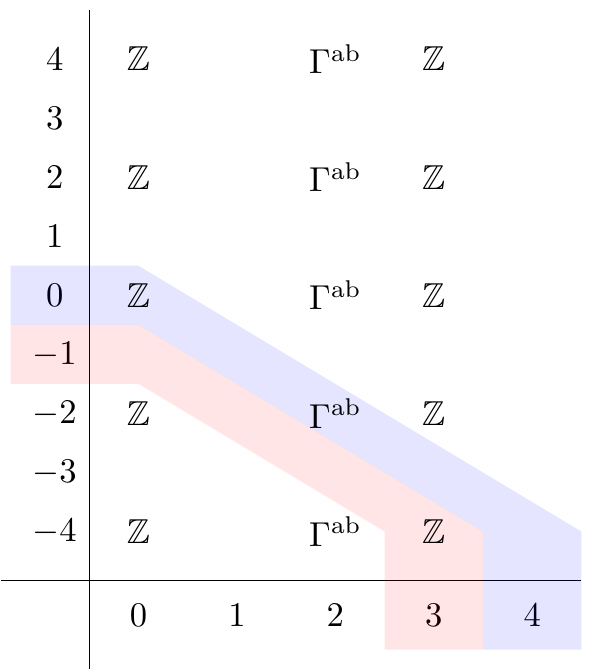}
  \caption{Contributions of the second page $E_2=E_\infty$ to
    $K^0(S^3/\Gamma)$ (blue, on top) and
    $K^{-1}(S^3/\Gamma)=K^1(S^3/\Gamma)$ (pink, below).}
  \label{fig:AHSS-S3/Gamma}
\end{figure}

Say that we are interested in computing the (complex) K-theory groups
$K^p(X)$ of some manifold $X$, where $p$ is the degree. Given that
$K^p$ is a generalized cohomology theory (see for example \S13.90 of
\cite{Switzer} for a description of the associated spectrum), we can
compute the groups of interest using the Atiyah-Hirzebruch spectral
sequence (we refer the reader to \cite{McCleary} for background on the
computation of spectral sequences)
\begin{equation}
  \label{eq:AHSS-K}
  E_2^{p,q} = H^p(X; K^q(\pt)) \Rightarrow E_\infty^{p+q}(X)
\end{equation}
associated to the fibration $0\to\pt\to X \to X\to 0$. Using
$K^*(\pt) = \bZ[x,x^{-1}]$
and the cohomology groups~\eqref{eq:spherical-space-cohomology} we
can immediately write the $E_2$ terms in the Atiyah-Hirzebruch
spectral sequence. We show the result in
figure~\ref{fig:AHSS-S3/Gamma}.

There are two small complications one encounters in going from $E_2$
to $K^i(S^3/\Gamma)$. First, there might be non-vanishing
differentials acting on the modules in the spectral sequence. By
dimensional reasons, the only potentially non-vanishing differential
is
\begin{equation}
  d_3\colon E_2^{0,p} \to E_2^{3,p-2}
\end{equation}
for $p$ even.  This differential necessarily vanishes, though, because
otherwise we would find $K^0(S^3/\Gamma)=\Gamma^{\text{ab}}$, which is
incompatible with the Chern
homomorphism~\eqref{eq:Chern-isomorphism}. The other issue to deal
with is that of potential non-trivial extensions. Recall that we have
\begin{equation}
  E_\infty^{p,q} = \frac{F^pK^{p+q}}{F^{p+1}K^{p+q}}
\end{equation}
for the filtration
\begin{equation}
  K^n(S^3/\Gamma) = F^0K^n \supset \ldots F^iK^n \supset F^{i+1}K^n \supset \ldots
\end{equation}
The only potentially non-trivial step is
\begin{equation}
  E_\infty^{0,0} = \bZ = \frac{K^n(S^3/\Gamma)}{\Gamma^{\text{ab}}} \,,
\end{equation}
or in other words that
$0 \to \Gamma^{\text{ab}} \to K^n(S^3/\Gamma) \to \bZ \to 0$
is exact. But this sequence splits, since $\bZ$ is free, and thus we
find
\begin{align}
  K^0(S^3/\Gamma) &= \bZ \oplus \Gamma^{\text{ab}} \,, & K^1(S^3/\Gamma) & = \bZ \,.
\end{align}
So in this case K-theory is completely determined by
cohomology:
\begin{equation}
  K^i(S^3/\Gamma) \cong \bigoplus_{n \equiv i \text{ mod } 2} H^n(S^3/\Gamma, \bZ)\, .
\end{equation}

\section{\texorpdfstring{\Kthree{}}{K3} partition functions for \texorpdfstring{$\cN=4$}{N=4} theories with algebra \texorpdfstring{$\fsu(N)$}{su(N)}}

\label{app:K3-partition-functions}

In this appendix we perform two related calculations of the $\cN=4$ Vafa-Witten partition function on \Kthree~\cite{Vafa:1994tf, Minahan:1998vr} for gauge algebra $\fsu(N)$. Firstly, we compute the partition function for all the genuine theories classified by \cite{Aharony:2013hda}, and verify that their duality relations are as expected. Secondly, we compute the elliptic genus of the modular-invariant $\cK_N$ theory described in~\secref{subsec:2d4d}.

\subsection{The Hecke operator}

\label{app:Hecke}

Before diving into the physics, we briefly review the Hecke operator and some associated identities.

Recall that a modular form $f$ of weight $w \in 2\bZ$ is a holomorphic function satisfying
\begin{equation} \label{eqn:modularform}
  f \biggl( \frac{a \tau + b}{c \tau + d} \biggr) = (c \tau + d)^w f (\tau)
\end{equation}
for any integers $a, b, c, d$ satisfying $a d - b c = 1$. More generally, a non-holomorphic modular form of weight $(w,\tilde{w})$ satisfies
\begin{equation}
  f \biggl( \frac{a \tau + b}{c \tau + d}, \frac{a \bar{\tau} + b}{c \bar{\tau} + d} \biggr) = (c \tau + d)^w (c \bar{\tau} + d)^{\tilde{w}} f (\tau, \bar{\tau}) \,.
\end{equation}
where $w - \tilde{w} \in 2 \bZ$.

For any integer $m\ge 1$, we define the Hecke operator $T_m$ (see
\cite{SerreArithmetic,1-2-3-ModularForms} for reviews) by its action on
a modular form of weight $w$
\begin{equation}
  (T_m[f])(\tau) \df m^{w-1}\sum_{\substack{a,d>0\\ad=m}}\frac{1}{d^w}\sum_{0 \leq b < d} f\biggl(\frac{a\tau+b}{d}\biggr)\, .
\end{equation}
For a non-holomorphic modular form we replace $w \to w+\tilde{w}$. 

We now show that $T_m[f]$ is itself a modular form of weight $w$. It is easy to see that $(T_m[f])(\tau+1) = (T_m[f])(\tau)$. To compute the $S$ transformation, we first derive a useful identity. Consider~\eqref{eqn:modularform} with $c \ne 0$, and 
define $t \equiv c \tau + d$, so that $\tau = \frac{t - d}{c}$ and
\begin{equation}
  f \biggl( - \frac{1}{c t} + \frac{a}{c} \biggr) = t^w f \biggl( \frac{t}{c} -
  \frac{d}{c} \biggr) \,.
\end{equation}
Thus, for coprime $k, N$:
\begin{equation} \label{eqn:coprimeSidentity}
  f \biggl( - \frac{1}{N \tau} + \frac{k}{N} \biggr) = \tau^w f \biggl(
  \frac{\tau}{N} + \frac{k'}{N} \biggr) \,,
\end{equation}
where $k k' \equiv - 1 \pmod{N}$. More generally:
\begin{equation} \label{eqn:generalSidentity}
  f \biggl( - \frac{1}{\tau} + \frac{k}{N} \biggr) = \biggl( \frac{\gcd (k, N)
  \tau}{N} \biggr)^w f \biggl( \frac{\gcd (k, N)^2}{N^2} \tau + \frac{k' \gcd
  (k, N)}{N} \biggr) \,,
\end{equation}
where $k'$ is the solution to the equation:
\begin{equation}
  k k' \equiv - \gcd (k, N) \quad (\tmop{mod} N) \,,
\end{equation}
and we take the convention $\gcd (0, n) = n$ for $n > 0$.

Thus,
\begin{align} \label{eqn:HeckeSdual}
  (T_m[f])(-1/\tau) &= m^{w-1} \sum_{\substack{a,d>0\\ad=m}}\frac{1}{d^w}\sum_{0 \leq b < d} f\biggl(-\frac{a}{d \tau} + \frac{b}{d}\biggr) \nonumber \\
  &= m^{w-1} \tau^w \sum_{\substack{a,d>0\\ad=m}} \sum_{0 \leq b < d} \biggl(\frac{\gcd(b,d)}{m}\biggr)^w f\biggl(\frac{\gcd(b,d)^2}{m} \tau + \frac{b' \gcd(b,d)}{d}\biggr) \,,
\end{align}
where $b b' \equiv -\gcd(b,d) \pmod d$. Define $\tilde{a} \df \gcd(b,d)$, $\tilde{d} \df m/\gcd(b,d)$, and $\tilde{b} \df b' a$, and note that there is a bijective relationship between $(a,d,b)$ and $(\tilde{a},\tilde{d},\tilde{b})$ (with $0 \le b < d$ and $0 \le \tilde{b} \le \tilde{d}$); in particular $a = \gcd(\tilde{b},\tilde{d})$, $d = m/\gcd(\tilde{b},\tilde{d})$, and $b \tilde{b} \equiv - a \tilde{a} \pmod m$, so the map is its own inverse. Thus,
\begin{equation}
(T_m[f])(-1/\tau) = m^{w-1} \tau^w \sum_{\substack{\tilde{a},\tilde{d}>0\\\tilde{a}\tilde{d}=m}} \frac{1}{\tilde{d}^w} \sum_{0 \leq \tilde{b} < \tilde{d}} f\biggl(\frac{\tilde{a} \tau+\tilde{b}}{\tilde{d}}\biggr) = \tau^w (T_m[f])(\tau) \,,
\end{equation}
so $T_m[f]$ is indeed a modular form of weight $w$.

\subsection{\texorpdfstring{$\fsu(p)$}{su(p)} with \texorpdfstring{$p$}{p} prime}

\label{app:su(p)-prime}

As a warmup, we first consider the case where $N = p$ is prime, taking a somewhat ad hoc approach. Later, we return to the general $N$ case and proceed more systematically.

We start with
the partition function of $\SU(p)$ on \Kthree{} for prime $p$~\cite{Vafa:1994tf,Minahan:1998vr}:
\begin{equation} \label{eqn:zeroblock}
  Z_{\SU (p)}(\tau) = \frac{1}{p}  \hat{Z}(\tau) \qquad, \qquad \hat{Z}(\tau) =
  \frac{1}{p^2} G (p \tau) + \frac{1}{p}  \sum_{j = 0}^{p - 1} G \left(
  \frac{\tau + j}{p} \right) \,,
\end{equation}
where $G(\tau) = 1/\eta(\tau)^{24}$ is a modular form of weight $-12$, $\hat{Z}(\tau)$ is the conformal block associated to the trivial gauge bundle, and we follow the normalization conventions of~\cite{Vafa:1994tf}.
Under $S$ ($\tau \to - 1/\tau$) we obtain:
\begin{subequations}
\begin{align}
  G (p \tau) &\rightarrow G \left( - \frac{p}{\tau} \right) = (\tau / p)^{-
  12} G \left( \frac{\tau}{p} \right) \,, \\
  G \left( \frac{\tau}{p} \right) & \rightarrow G \left( - \frac{1}{\tau p}
  \right) = (\tau p)^{- 12} G (\tau p) \,, \\
  G \left( \frac{\tau + j}{p} \right) & \rightarrow G \left( - \frac{1}{\tau
  p} + \frac{j}{p} \right) = \tau^{- 12} G \left( \frac{\tau + j'}{p} \right) \,,
  \qquad \left( j j' \equiv - 1 \quad \tmop{mod} p \right) \,,
\end{align}
\end{subequations}
using~\eqref{eqn:coprimeSidentity}. Therefore,
\begin{equation}
  Z_{\SU (p)}\biggl(-\frac{1}{\tau}\biggr) = \frac{1}{\tau^{12}}  \Biggl[ p^9 G \biggl(
  \frac{\tau}{p} \biggr) + \frac{1}{p^{14}} G (p \tau) + \frac{1}{p^2} 
  \sum_{j = 1}^{p - 1} G \biggl( \frac{\tau + j}{p} \biggr) \Biggr] = \frac{1}{(p
  \tau)^{12}} Z_{(\SU (p) /\mathbb{Z}_p)_0}(\tau) \,,
\end{equation}
where we define:
\begin{equation} \label{eqn:SUp0}
  Z_{(\SU (p) /\mathbb{Z}_p)_0}(\tau) \df \frac{1}{p^2} G (p \tau) + p^{21} G
  \left( \frac{\tau}{p} \right) + p^{10}  \sum_{j = 1}^{p - 1} G \left(
  \frac{\tau + j}{p} \right) \,.
\end{equation}
Applying $T^k$ ($\tau \to \tau+k$), we obtain:
\begin{equation} \label{eqn:SUpk}
  Z_{(\SU (p) /\mathbb{Z}_p)_k}(\tau) \df Z_{(\SU (p) /\mathbb{Z}_p)_0}(\tau+k) = \frac{1}{p^2} G (p \tau) + p^{21} G
  \left( \frac{\tau + k}{p} \right) + p^{10}  \sum_{0 \leqslant j < p}^{j \neq
  k} G \left( \frac{\tau + j}{p} \right) \,.
\end{equation}
Applying $S$ again, we find:
\begin{subequations}
\begin{align}
  Z_{(\SU (p) /\mathbb{Z}_p)_0}\!\biggl(-\frac{1}{\tau}\biggr) &= \frac{1}{\tau^{12}} \Biggl[ p^{10}
  G \biggl( \frac{\tau}{p} \biggr) + p^9 G (p \tau) + p^{10}  \sum_{j = 1}^{p -
  1} G \biggl( \frac{\tau + j}{p} \biggr) \Biggr] = \frac{p^{12}}{\tau^{12}}
  Z_{\SU (p)}(\tau) \,, \\
  Z_{(\SU (p) /\mathbb{Z}_p)_k}\!\biggl(-\frac{1}{\tau}\biggr) &= \frac{1}{\tau^{12}} \Biggl[ p^{10}
  G \biggl( \frac{\tau}{p} \biggr) + p^{21} G \biggl( \frac{\tau + k'}{p}
  \biggr) + \frac{1}{p^2} G (p \tau) + p^{10}  \sum_{0 < j < p}^{j \neq k} G
  \biggl( \frac{\tau + j'}{p} \biggr) \Biggr] \nonumber \\
  & = \frac{1}{\tau^{12}} Z_{(\SU (p) /\mathbb{Z}_p)_{k'}}(\tau) \,.
\end{align}
\end{subequations}
where $k k' \equiv - 1 \pmod p$.
We have deliberately chosen notation in~\eqref{eqn:SUp0} and~\eqref{eqn:SUpk} in line with \cite{Aharony:2013hda}. Indeed, as we will show later, these are the partition functions for the $\SU(p)/\bZ_p$ theory with different discrete theta angles. As a simple check, note that
$S \colon (\SU (p) /\mathbb{Z}_p)_k \leftrightarrow (\SU (p)
/\mathbb{Z}_p)_{k'}$ for $k k' \equiv - 1 \pmod p$ is a
special case of the rules given in \cite{Aharony:2013hda}.

These partition functions can be decomposed into conformal blocks
\begin{equation}
  Z_{(\SU (p) /\mathbb{Z}_p)_k}(\tau) = \hat{Z}(\tau) + \sum_{a = 0}^{p - 1} n_a
  \omega_p^{k a} Z^{(a)}(\tau) \,,
\end{equation}
where $\hat{Z}(\tau)$ is the $v=0$ block, as above, and $Z^{(a)}(\tau)$ is the $v\ne 0$ block with $v^2/2 \equiv a \pmod p$, and $n_a$ denotes the gauge bundle multiplicities within $H^2(\Kthree; \bZ_p)$ in each category. We can determine these multiplicities by choosing a basis of the form~\eqref{eqn:specialbasis}, so that $v = v^i e_i + \bar{v}^i \bar{e}_i$ for $v^i, \bar{v}^i = 0, \ldots, (p-1)$, and
\begin{equation}
v^2/2 \equiv \delta_{i j} v^i \bar{v}^j \pmod p \,.
\end{equation}
For fixed $v^2/2 \equiv a \pmod p$ with $a \ne 0$, this equation can be solved to eliminate one of the $\bar{v}^i$s provided that $v^i \ne 0$ for some $i$. Therefore,
\begin{align}
n_a &= (p^{11} - 1) p^{10} = \frac{p^{22} - p^{11}}{p} \,,\qquad a\not\equiv 0 \pmod p \,, \\
n_0 &= \frac{p^{22}-p^{11}}{p} +p^{11}- 1 \,,
\end{align}
where the remaining case $n_0$ ($v^2/2 \equiv 0 \pmod p$ with $v \ne 0$) is fixed by the requirement $p^{22} = 1+\sum_a n_a$. Equivalently, the extra $p^{11} - 1$ gauge bundles correspond to the non-zero elements of the self-dual subspace $\cI_0$.

The conformal blocks $Z^{(a)}(\tau)$ remain to be determined. These are not given explicitly in~\cite{Vafa:1994tf,Minahan:1998vr}, but fortunately we can reverse engineer them from the partition functions that we have derived using the modular properties of the $\SU(p)$ and $\SU(p)/\bZ_p$ theories.
We find
\begin{align}
  \frac{1}{p}  \sum_{k = 0}^{p - 1} Z_{(\SU (p) /\mathbb{Z}_p)_k}(\tau) &=
  \hat{Z} + n_0 Z^{(0)} = \frac{1}{p^2} G (p \tau) + \frac{p^{22} + (p - 1)
  p^{11}}{p} \cdot \frac{1}{p} \sum_{j = 0}^{p - 1} G \biggl( \frac{\tau +
  j}{p} \biggr) \,, \nonumber \\
  \frac{1}{p}  \sum_{k = 0}^{p - 1} \omega_p^{- a k} Z_{(\SU (p)
  /\mathbb{Z}_p)_k}(\tau) &= n_a Z^{(a)} = \frac{p^{22} - p^{11}}{p} \cdot
  \frac{1}{p}  \sum_{j = 0}^{p - 1} \omega_p^{- a j} G \biggl( \frac{\tau +
  j}{p} \biggr)\,, \quad a\not\equiv 0 \pmod p \,.
\end{align}
Comparing with~\eqref{eqn:zeroblock}, we obtain the simple result
\begin{equation} \label{eqn:ablock}
  Z^{(a)}(\tau) = \frac{1}{p} \sum_{j = 0}^{p - 1} \omega_p^{- a j} G \biggl( \frac{\tau + j}{p} \biggr) \,.
\end{equation}
We verify that these are the correct conformal blocks by independent means below.

With the conformal blocks in hand, we can compute the partition function for the modular-invariant theory $\cK_p$:
\begin{equation}
Z_{\cK_p}(\tau) = \sum_{v \in \cI_0} Z_v(\tau) = \hat{Z}(\tau) + \sum_a c_a Z^{(a)}(\tau) \,,
\end{equation}
where $c_0 = p^{11} - 1$ and $c_a = 0$ for $a \not\equiv 0 \pmod p$ since $\cI_0$ is null and $|\cI_0| = p^{11}$.
We obtain
\begin{equation}
  \label{eq:su(p)-self-dual-polarization}
  Z_{\cK_p}(\tau) \df \frac{1}{p^2} G (p
  \tau) + p^{10}  \sum_{j = 0}^{p - 1} G \biggl( \frac{\tau + j}{p} \biggr) \,.
\end{equation}
This is nothing but the Hecke transform of $G(\tau)$, up to normalization:
\begin{equation} \label{eqn:ZKp}
Z_{\cK_p}(\tau) = p^{11} T_p[G](\tau) \,,
\end{equation}
from which it is evident that $Z_{\cK_p}(\tau)$ is a modular form of weight $-12$, in line with the expectation the $\cK_p$ is a genuine 2d CFT. Below, we show that~\eqref{eqn:ZKp} generalizes to arbitrary square-free $N$, and to any composite $N$ once an ambiguity in the definition of the $\cK_N$ theory is appropriately resolved.

\subsection{Conformal blocks for general \texorpdfstring{$\fsu(N)$}{su(N)}}

We now consider $\fsu(N)$ for general $N$. The $\U(N)$ partition function was obtained by~\cite{Minahan:1998vr}:
\begin{gather} \label{eqn:UNpartition}
  Z_{\U(N)}(\tau, \bar{\tau}) = T_N[Z_{\U(1)}](\tau, \bar{\tau}) = \frac{1}{N^2} \sum_{\substack{a,d>0\\ad=m}} d \sum_{0 \leq b < d} Z_{U (1)}\! \biggl( \frac{a \tau + b}{d}, \frac{a \bar{\tau} + b}{d} \biggr)\,, \\
  \intertext{where}
  Z_{U (1)}(\tau,\bar{\tau}) = G (\tau) \theta_{\Gamma^{19, 3}} (\tau, \bar{\tau})\,, \qquad
  \theta_{\Gamma^{19, 3}} (\tau, \bar{\tau}) \df \sum_{V \in \Gamma^{19,
  3}} q^{\frac{1}{2} V_L^2}  \bar{q}^{\frac{1}{2} V_R^2}\,.
\end{gather}
Here $\Gamma^{19,3}$ is the even self-dual lattice of signature $(19,3)$, $\theta_{\Gamma^{19, 3}}(\tau,\bar{\tau})$ is the associated theta function, and $Z_{\U(N)}$ has weights $(-5/2,3/2)$, with $(-12,0)$ coming from $G(\tau)$ and $(19/2,3/2)$ coming from $\theta_{\Gamma^{19, 3}}$. Note that, since \Kthree{} has negative signature $\sigma = -16$, the geometric intersection form on \Kthree{} is
\begin{equation}
U \circ V = U_R\cdot V_R - U_L\cdot V_L \,, \qquad \text{so that} \qquad \frac{1}{2} V^2 = \frac{1}{2} V_R^2 - \frac{1}{2} V_L^2 \,.
\end{equation}
This is the reverse of the usual worldsheet convention, but we will stick to it to maintain consistency with the rest of the paper.

The $\U(N)$ partition function can be decomposed as
\begin{equation}
Z_{\U(N)}(\tau, \bar{\tau}) = \sum_{V \in \Gamma^{19, 3}} (q^{1/N})^{\frac{1}{2} V_L^2} (\bar{q}^{1/N})^{\frac{1}{2} V_R^2} Z_V(\tau) \,,
\end{equation}
where $q \df \e^{2 \pi i \tau}$ and $Z_V(\tau)$ is holomorphic and satisfies the periodicity condition $Z_V(\tau) = Z_{V+N U}(\tau)$ for any $U \in \Gamma^{19, 3}$. In particular, $Z_v(\tau)$ are the conformal blocks, indexed by gauge bundles $v$ valued in
\begin{equation}
H^2(\Kthree; \bZ_N) \cong \frac{H^2(\Kthree)}{N\cdot H^2(\Kthree)} = \frac{\Gamma^{19, 3}}{N\cdot \Gamma^{19, 3}}\,,
\end{equation}
since \Kthree{} is torsionless.

Our goal is now to extract the conformal blocks $Z_v(\tau)$ from the $\U(N)$ partition function $Z_{\U(N)}(\tau, \bar{\tau})$. First, noting that there are not $N^{22}$ distinct conformal blocks but rather many of them equal due to the symmetries of \Kthree{}, we list the data on which $Z_v(\tau)$ can depend. Clearly, these include the order $k$ of $v$ within $\bZ_N^{22}$, where $k$ divides $N$ by elementary group theory. Likewise, the blocks can depend on the Pontryagin square $v^2/2$, which is single-valued on $\bZ_N^{22}$ modulo $N$ because
\begin{equation}
\frac{(v+N u)^2}{2} = \frac{v^2}{2} + N v \circ u + N^2 \frac{u^2}{2} \,.
\end{equation}
However, if $v$ has order $k<N$ then it lies within $\frac{N}{k} \Gamma^{19, 3}$, and therefore $v \circ u$ is a multiple of $N/k$, implying that $\frac{1}{2} v^2$ is single-valued modulo $N^2 / k$, or equivalently that the refined Pontryagin square $\frac{1}{2} (k v/N)^2$ is single-valued modulo $k$. This is naturally interpreted as the Pontryagin square on the subgroup $H^2(\Kthree; \bZ_k) = \frac{N}{k} H^2(\Kthree; \bZ_N)$ of $\SU(N)/\bZ_k$ gauge bundles modulo $\SU(N)$ gauge bundles. Note that $\frac{1}{2} v^2 \pmod N$ is fixed by $\frac{1}{2} (k v/N)^2 \pmod k$, but the converse is not true when $\gcd(k, N/k) > 1$, which holds for some $k$ whenever $N$ is divisible by a perfect square.

Thus, we can categorize the conformal blocks $Z_v(\tau)$ by the order $k$ of $v$, as well as $a \equiv \frac{1}{2} (k v/N)^2 \pmod k$, or more physically as $\SU(N)/\bZ_k$ gauge bundles that do not lift to any covering group, where the associated Stiefel-Whitney class $v = w_2$ has Pontryagin square $a$ modulo $k$. In fact, this is all the data on which the conformal blocks can depend, because the $\U(N)$ partition function only depends on $\frac{1}{2} V^2$, and we have extracted all data from this that is invariant under shifts $V \to V+N U$. Thus, we denote the $\fsu(N)$ conformal blocks as $Z_{N ; k}^{(a)}(\tau)$.

To compute these blocks, define the sublattice $\Gamma_{N ; k} = \frac{N}{k} \Gamma^{19, 3}$ for each $k|N$, let $\Gamma_{N ; k}^{(a)}$ be the subset of $\Gamma_{N ; k}$ with $\frac{(v / (N / k))^2}{2} \equiv a \pmod k$, and let $\hat{\Gamma}^{(a)}_{N ; k}$ denote the subset of $\Gamma_{N; k}^{(a)}$ that is not in $\Gamma_{N; \tilde{k}}$ for any $\tilde{k} < k$.\footnote{Since the intersection of $\Gamma_{N; k}$ and $\Gamma_{N; \tilde{k}}$ is $\Gamma_{N; \gcd(k,\tilde{k})}$, we can restrict to $\tilde{k} | k$.}
Correspondingly, we have theta functions
\begin{subequations}
\begin{align}
  \theta_{N ; k} \df & \sum_{P \in \Gamma_{N ; k}} (q^{1 /
  N})^{\frac{1}{2} P_L^2}  (\bar{q}^{1 / N})^{\frac{1}{2} P_R^2} \,, \\
  \theta_{N ; k}^{(a)} \df & \sum_{P \in \Gamma_{N ; k}^{(a)}} (q^{1 /
  N})^{\frac{1}{2} P_L^2}  (\bar{q}^{1 / N})^{\frac{1}{2} P_R^2} \,, \\
  \hat{\theta}_{N ; k}^{(a)} \df & \sum_{P \in \hat{\Gamma}_{N ; k}^{(a)}}
  (q^{1 / N})^{\frac{1}{2} P_L^2}  (\bar{q}^{1 / N})^{\frac{1}{2} P_R^2} \,,
\end{align}
\end{subequations}
so that (cf.~\cite{Minahan:1998vr})
\begin{equation}
  Z_{U (N)} = \sum_{k|N} \sum_{a = 0}^{k - 1} \hat{\theta}_{N ; k}^{(a)} Z_{N; k}^{(a)} \,.
\end{equation}
Explicitly, we find:
\begin{align} \label{eqn:thetadecompose1}
  \theta_{N ; k}(\tau) & = \theta_{\Gamma^{19, 3}}\! \biggl( \frac{N}{k^2} \tau
  \biggr)\,, &
  \theta_{N ; k}^{(a)}(\tau) & = \frac{1}{k}  \sum_{j = 0}^{k - 1} \omega_k^{a j}
  \theta_{\Gamma^{19, 3}}\! \biggl( \frac{N}{k^2} \tau + \frac{j}{k} \biggr) \,,
\end{align}
where $\sum_{a = 0}^{k - 1} \theta_{N ; k}^{(a)} = \theta_{N ; k}$ as
required.
The functions $\hat{\theta}_{N ; k}^{(a)}$ are determined implicitly by
\begin{equation}
  \theta_{N ; k}^{(a)} = \sum_{\ell |k} \sum_{\substack{0 \leqslant b < k /
  \ell \\ \ell^2 b \equiv a \bmod k}} \hat{\theta}_{N ; k /
  \ell}^{(b)} \,. \label{eqn:thetaincl}
\end{equation}
This can be inverted using the M\"obius inversion formula, but this is unnecessary.

To extract the conformal blocks from $Z_{U (N)}$, we apply
\begin{equation}
  \theta_{\Gamma^{19, 3}} \! \biggl( \frac{N}{k^2} \tau + \frac{j}{k} \biggr) =
  \sum_{a = 0}^{k - 1} \omega_k^{- a j} \theta_{N ; k}^{(a)}(\tau) = \sum_{\ell |k}
  \sum_{a = 0}^{k / \ell - 1} \omega_{k / \ell}^{- a j \ell}  \hat{\theta}_{N; k / \ell}^{(a)}(\tau) \,,
\end{equation}
to the Hecke transform formula~\eqref{eqn:UNpartition} to obtain
\begin{equation}
  Z_{U (N)}(\tau) = \sum_{k|N} \frac{k}{N^2}  \sum_{j = 0}^{k - 1} G \left(
  \frac{N}{k^2} \tau + \frac{j}{k} \right)  \sum_{\ell |k} \sum_{a = 0}^{k /
  \ell - 1} \omega_{k / \ell}^{- a j \ell}  \hat{\theta}_{N ; k / \ell}^{(a)} \,.
\end{equation}
Resumming, we obtain:
\begin{equation}
  Z_{U (N)}(\tau) =
  \sum_{k|N} \sum_{a = 0}^{k - 1} \hat{\theta}_{N ; k}^{(a)}  \sum_{\ell |
  \frac{N}{k}} \frac{k \ell}{N^2} \sum_{j = 0}^{k \ell - 1} \omega_k^{- a j
  \ell} G \biggl( \frac{N}{k^2 \ell^2} \tau + \frac{j}{k \ell} \biggr) = \sum_{k|N} \sum_{a = 0}^{k - 1} \hat{\theta}_{N ; k}^{(a)} Z_{N ; k}^{(a)} \,,
\end{equation}
from which we read off:
\begin{equation}
  Z_{N ; k}^{(a)} = \sum_{\ell | \frac{N}{k}} \frac{k \ell}{N^2} \sum_{j =
  0}^{k \ell - 1} \omega_k^{- a j \ell} G \biggl( \frac{N}{k^2 \ell^2} \tau +
  \frac{j}{k \ell} \biggr) \,,
\end{equation}
since the $\hat{\theta}_{N ; k}^{(a)}$ are linearly independent. This is
nothing but a modified Hecke transform:
\begin{equation}
  Z_{N ; k}^{(\alpha)} = \frac{1}{N^2}  \sum_{\substack{a,d>0\\ad=N \\ k|d}} d \sum_{0 \leq b < d} \omega_{k}^{-\alpha b d/k} G \biggl( \frac{a \tau +
  b}{d} \biggr) \,,
\end{equation}
with the added constraint that $d$ is a multiple of $k$ and a phase factor in the sum.

In the special case where $N$ is prime, we recover:
\begin{equation}
\begin{aligned}
  \hat{Z} = Z_{N ; 1}^{(0)} & = \frac{1}{N^2} G (N \tau) + \frac{1}{N} \sum_{j
  = 0}^{N - 1} G \biggl( \frac{\tau + j}{N} \biggr) \,, \\
  Z^{(a)} = Z_{N ; N}^{(a)} & = \frac{1}{N} \sum_{j = 0}^{N - 1} \omega_N^{- a
  j} G \biggl( \frac{\tau + j}{N} \biggr) \,,
\end{aligned}
\end{equation}
which matches~\eqref{eqn:zeroblock}, \eqref{eqn:ablock}.

\subsection{Partition functions for genuine theories}

The partition function for the $(\SU (N) /\mathbb{Z}_k)_{\ell}$ theory
as defined by \cite{Aharony:2013hda} is then
\begin{equation} \label{eqn:Zkm}
  Z_{(\SU (N) /\mathbb{Z}_k)_{m}} = \frac{k}{N}  \sum_{d|k} \sum_{a =
  0}^{d - 1} \omega_d^{a m (k / d)}  \hat{n}_{N ; d}^{(a)} Z_{N ; d}^{(a)}
\end{equation}
where we normalize using the conventions of~\cite{Vafa:1994tf}---$N / k$ being the volume of the $\mathbb{Z}_{N / k}$ center---and
$\hat{n}_{N ; d}^{(a)}$ is the number of gauge bundles in $H^2(\Kthree; \bZ_N)$ of the indicated type.

The multiplicity factor $\hat{n}_{N ; d}^{(a)}$ can be computed using the modular properties of the
theta function $\theta_{\Gamma^{19, 3}}$, as follow. Consider for instance,
\begin{equation}
 n_k \df \lim_{q \rightarrow 1} \frac{\theta_{N ; k}}{\theta_{N ; 1}} \,.
\end{equation}
This counts the index of the sublattice $\Gamma_{N ; 1} \subseteq \Gamma_{N ;
k}$. Since $q \rightarrow 1$ corresponds to $\tau \rightarrow 0$, we can
relate it to $q \rightarrow 0$ ($\tau \rightarrow i \infty$) by $\tau
\rightarrow - 1 / \tau$. We have:
\begin{equation}
  \theta_{\Gamma^{19, 3}} (- 1 / \tau, - 1 / \bar{\tau}) = \tau^8  | \tau |^3
  \theta_{\Gamma^{19, 3}} (\tau, \bar{\tau}) \,,
\end{equation}
since $\theta_{\Gamma^{19, 3}}$ is  modular form of weight $(19/2,3/2)$. Thus,
\begin{equation}
  n_k = \lim_{\tau
  \rightarrow 0} \frac{\theta_{\Gamma^{19, 3}} \left( \frac{N}{k^2} \tau
  \right)^{}}{\theta_{\Gamma^{19, 3}} (N \tau)^{}} = \lim_{\tau \rightarrow i
  \infty} \frac{(k^2 / N)^{11}}{(1 / N)^{11}} \cdot \frac{\theta_{\Gamma^{19,
  3}} \left( \frac{k^2}{N} \tau \right)^{}}{\theta_{\Gamma^{19, 3}} \left(
  \frac{1}{N} \tau \right)^{}} = k^{22} \,.
\end{equation}
While this result is obvious, we can derive similar, less obvious results in
the same way. In particular:
\begin{align}
  n_k^{(a)} & \df \lim_{q \rightarrow 1} \frac{\theta_{N ;
  k}^{(a)}}{\theta_{N ; 1}} = \frac{1}{k}  \sum_{j = 0}^{k - 1} \omega_k^{a j}
  \lim_{\tau \rightarrow 0}  \frac{\theta_{\Gamma^{19, 3}} \Bigl(
  \frac{N}{k^2} \tau + \frac{j}{k} \Bigr)}{\theta_{\Gamma^{19, 3}} (N
  \tau)}\nonumber\\
  & = \frac{1}{k}  \sum_{j = 0}^{k - 1} \omega_k^{a j} \lim_{\tau \rightarrow
  i \infty}  \frac{\Bigl( \frac{k \gcd (j, k)}{N} \Bigr)^{11}
  \theta_{\Gamma^{19, 3}} \Bigl( \frac{\gcd (j, k)^2}{N} \tau + \frac{j' \gcd
  (j, k)}{k} \Bigr)}{(1 / N)^{11} \theta_{\Gamma^{19, 3}} \bigl( \frac{1}{N}
  \tau \bigr)}
   = k^{11} \cdot \frac{1}{k}  \sum_{j = 0}^{k - 1} \omega_k^{a j} \gcd (j,
  k)^{11}  \label{eqn:nkformula}
\end{align}
using~\eqref{eqn:generalSidentity}, where $j'$ is the
solution to $j' j \equiv - \gcd (j, k) \; \tmop{mod} k$. Note that
 $n_k = \sum_{a = 0}^{k - 1} n_k^{(a)}$,
as required. Now $\hat{n}^{(a)}_k = \lim_{q \rightarrow 1}
\frac{\hat{\theta}_{N ; k}^{(a)}}{\theta_{N ; 1}}$, is given implicitly by the formula:
\begin{equation}
  n_k^{(a)} = \sum_{\ell |k} \sum_b^{\ell^2 b \equiv a \; (\tmop{mod} k)}
  \hat{n}_{k / \ell}^{(b)} \label{eqn:nincl} \,.
\end{equation}
As with (\ref{eqn:thetaincl}), it is not necessary to solve this explicitly.

Use~\eqref{eqn:nkformula}, we obtain
\begin{gather}
  k^{11} \gcd (j, k)^{11} = \sum_{a = 0}^{k - 1} \omega_k^{- a j} n_k^{(a)} =
  \sum_{\ell |k} \sum_{a = 0}^{k / \ell - 1} \omega_{k / \ell}^{- a j \ell} 
  \hat{n}_{k / \ell}^{(a)} \,. \\
\intertext{Applying this to~\eqref{eqn:Zkm} gives:}
\begin{aligned}
  Z_{(\SU (N) /\mathbb{Z}_k)_m} & =
  \sum_{\ell |N} \sum_{j = 0}^{\ell - 1} \frac{k \ell}{N^3} G \biggl(
  \frac{N}{\ell^2} \tau + \frac{j}{\ell} \biggr)  \sum_{d| \gcd (k, \ell)}
  \sum_{a = 0}^{d - 1} \omega_d^{a \frac{m k - j \ell}{d}}  \hat{n}_d^{(a)} \\
  & = \sum_{\ell |N} \sum_{j = 0}^{\ell - 1} \frac{k \ell}{N^3} \gcd (k,
  \ell)^{11} \gcd \biggl( \frac{j \ell - m k}{\gcd (k, \ell)}, \gcd (k, \ell)
  \biggr)^{11} G \biggl( \frac{N}{\ell^2} \tau + \frac{j}{\ell} \biggr) \,,
\end{aligned}
\intertext{which can be rewritten somewhat more concisely as:}
  Z_{(\SU (N) /\mathbb{Z}_k)_m} = \sum_{\ell |N} \frac{k \ell}{N^3} 
  \sum_{j = 0}^{\ell - 1} \gcd [j \ell - m k, k^2, \ell^2]^{11} G \biggl(
  \frac{N}{\ell^2} \tau + \frac{j}{\ell} \biggr) \,.
\end{gather}
Thus, we get a Hecke transform modified by the weight $\gcd [\ldots]^{11}$.

We now verify that this has the modular properties predicted by \cite{Aharony:2013hda}. Under $T : \tau
\rightarrow \tau + 1$, we find:
\begin{equation}
  T : Z_{(\SU (N) /\mathbb{Z}_k)_m} \rightarrow Z_{(\SU (N)
  /\mathbb{Z}_k)_{m + N / k}}
\end{equation}
in agreement with \cite{Aharony:2013hda}. Under $S : \tau \rightarrow - 1 / \tau$, we find:
\begin{multline}
  Z_{(\SU (N) /\mathbb{Z}_k)_m}(-1/\tau) =
  \sum_{\ell |N} \frac{k \ell}{N^3}  \sum_{j = 0}^{\ell - 1} \gcd [j \ell
  - m k, k^2, \ell^2]^{11}  \left( \frac{\ell \gcd (j, \ell)}{N} \tau
  \right)^{- 12} \\ 
    \times G \Biggl( \frac{\gcd (j, \ell)^2}{N} \tau + \frac{\hat{j}'
  \gcd (j, \ell)}{\ell} \Biggr) \,,
  \qquad \left( j j' \equiv - \gcd (j, \ell) \; \tmop{mod} \ell
  \right) \,.
\end{multline}
We resum this in terms of the variables:
\begin{equation}
  \tilde{\ell} = \frac{N}{\gcd (j, \ell)} \qquad, \qquad \tilde{j} = \frac{j' N}{\ell} \,,
\end{equation}
which are ``dual'' to the original variables $j, \ell$, in that there is bijective involution between them, i.e., $\gcd(j', \ell') = N/\ell$ and $j \tilde{j} \equiv - \frac{N^2}{\ell \tilde{\ell}} \pmod N$, just as discussed below~\eqref{eqn:HeckeSdual}.

Let $(\tilde{k}, \tilde{m})$ be related to $(k, m)$ by the same bijective involution.
Observe that:
\begin{equation}
  \frac{\gcd [j \ell - m k, k^2, \ell^2]}{k \ell} = \frac{\gcd [\tilde{j} \tilde{\ell} - \tilde{m}
  \tilde{k}, \tilde{k}^2, \tilde{\ell}^2]}{\tilde{k}\tilde{\ell}} \,. \label{formulaToProve}
\end{equation}
Using this formula, we obtain:
\begin{equation}
Z_{(\SU (N) /\mathbb{Z}_k)_m}(-1/\tau) = (\tilde{k} \tau / k)^{- 12} Z_{(\SU (N) /\mathbb{Z}_{\tilde{k}})_{\tilde{m}}}(\tau)
\end{equation}
which reproduces the predictions of \cite{Aharony:2013hda}.

\subsubsection*{Proof of~\eqref{formulaToProve}}

To prove (\ref{formulaToProve}), it is convenient to generalize it slightly to
\begin{equation}
  \frac{\gcd [j \ell X - m k Y, k^2, \ell^2]}{k \ell} = \frac{\gcd [j' \ell' Y
  - m' k' X, k^{\prime 2}, \ell^{\prime 2}]}{k' \ell'}
  \label{inductiveFormula}
\end{equation}
for any $X, Y \in \mathbb{Z}$ such that $\gcd (X, N) = \gcd (Y, N) = 1$. We
proceed inductively in the prime factors of $N$.

Consider first the case where $N = p^n$ is a prime power. We then have
\begin{equation}
  k = p^{\kappa} \;, \; m = m_0 p^{\mu} \;, \; \ell = p^{\lambda} \;, \; j =
  j_0 p^{\theta}\,, \qquad \left( \gcd (k, m) = p^{\mu}, \quad \gcd (\ell, j) =
  p^{\theta} \right)\,,
\end{equation}
for integers $\kappa, \mu, \lambda, \theta, m_0, j_0$ satisfying $0 \leqslant
\mu \leqslant \kappa \leqslant n$ and $0 \leqslant \theta \leqslant \lambda
\leqslant n$. We can chose $m$ in the range $0 < m \leqslant k$ and $j$ in the
range $0 < j \leqslant \ell$, in which case $\gcd (m_0, p) = \gcd (j_0, p) =
1$. Therefore:
\begin{subequations}
\begin{align}
  k' = p^{n - \mu} \;, & \; m' = m_0' p^{n - \kappa}\,, \qquad \left( m_0 m_0'
  \equiv - 1 \;  (\tmop{mod} p^{\kappa - \mu}) \right) \,,\\
  \ell' = p^{n - \theta} \;, & \; j' = j_0' p^{n - \lambda}\,, \qquad \left( j_0
  j_0' \equiv - 1 \;  (\tmop{mod} p^{\lambda - \theta}) \right) \,.
\end{align}
\end{subequations}

We have
\begin{subequations}
\begin{align}
  \frac{\gcd [j \ell X - m k Y, k^2, \ell^2]}{k \ell} & = \frac{\gcd [j_0
  p^{\lambda + \theta} X - m_0 p^{\kappa + \mu} Y, p^{2 \lambda}, p^{2
  \kappa}]}{p^{\kappa + \lambda}} \,,\\
  \frac{\gcd [j' \ell' Y - m' k' X, k^{\prime 2}, \ell^{\prime 2}]}{k' \ell'}
  & = \frac{\gcd [j_0' p^{2 n - \lambda - \theta} Y - m_0' p^{2 n - \kappa -
  \mu} X, p^{2 n - 2 \mu}, p^{2 n - 2 \theta}]}{p^{2 n - \mu - \theta}} \,.
\end{align}
\end{subequations}
We can assume $\lambda + \theta \geqslant \kappa + \mu$ without loss
of generality due to the symmetry of the identity to be proven under
$(\ell, j, X) \leftrightarrow (k, m, Y)$. We then obtain:
\begin{subequations}
\begin{align}
  \frac{\gcd [j \ell X - m k Y, k^2, \ell^2]}{k \ell} & = p^{\mu - \lambda}
  \gcd [j_0 p^{\lambda + \theta - \kappa - \mu} X - m_0 Y, p^{2 \lambda -
  \kappa - \mu}, p^{\kappa - \mu}] \,, \\
  \frac{\gcd [j' \ell' Y - m' k' X, k^{\prime 2}, \ell^{\prime 2}]}{k' \ell'}
  & = p^{\mu - \lambda} \gcd [j_0' Y - m_0' p^{\lambda + \theta - \kappa -
  \mu} X, p^{\lambda + \theta - 2 \mu}, p^{\lambda - \theta}] \,.
\end{align}
\end{subequations}
If $\lambda + \theta - \kappa - \mu > 0$, then the first term is not divisible
by $p$ since $\gcd (m_0 Y, p) = \gcd (j_0' Y, p) = 1$, so we obtain $p^{\mu -
\lambda}$ in both cases. Conversely, if $\lambda + \theta = \kappa + \mu$, we
find:
\begin{subequations}
\begin{align}
  \frac{\gcd [j \ell X - m k Y, k^2, \ell^2]}{k \ell} & = p^{\mu - \lambda}
  \gcd [j_0 X - m_0 Y, p^{\lambda - \theta}, p^{\kappa - \mu}] \,, \\
  \frac{\gcd [j' \ell' Y - m' k' X, k^{\prime 2}, \ell^{\prime 2}]}{k' \ell'}
  & = p^{\mu - \lambda} \gcd [j_0' Y - m_0' X, p^{\kappa - \mu}, p^{\lambda -
  \theta}]  \,.
\end{align}
\end{subequations}
We have:
\begin{equation}
  m_0 m_0' = a p^{\kappa - \mu} - 1 \qquad, \qquad j_0 j_0' = b p^{\lambda -
  \theta} - 1 \,,
\end{equation}
so that
\begin{equation}
  (j_0 X - m_0 Y) j_0' m_0' = b X m_0' p^{\lambda - \theta} - a Y j_0'
  p^{\kappa - \mu} + j_0' Y - m_0' X \,.
\end{equation}
Let $r = \min (\lambda - \theta, \kappa - \mu) \geqslant 0$. We find:
\begin{equation}
  \gcd [j_0 X - m_0 Y, p^r] = \gcd [(j_0 X - m_0 Y) j_0' m_0', p^r] = \gcd
  [j_0' Y - m_0' X, p^r] \,,
\end{equation}
and the identity is proven for $N$ a prime power.

Next, consider the case where $N = N_1 N_2$ with $\gcd (N_1, N_2) = 1$. We can
split $\ell = \ell_1 \ell_2$ and $\ell' = \ell_1' \ell_2'$ such that $\ell_i
|N_i$ and $\ell_i' |N_i$. Observe that $\gcd (\ell, j) = \gcd (\ell_1, j) \gcd
(\ell_2, j)$, so that
\begin{equation}
  \ell_i' = \frac{N_i}{\gcd (\ell_i, j)} \qquad, \qquad \ell_i =
  \frac{N_i}{\gcd (\ell_i', j')} \,,
\end{equation}
We define $j_i \equiv j / \gcd (j, \ell_{i \pm 1}) = j \ell_{i \pm 1}' / N_{i
\pm 1}$, and observe that $\gcd (\ell_i, j) = \gcd (\ell_i, j_i)$, so that
\begin{equation}
  \ell_i' = \frac{N_i}{\gcd (\ell_i, j_i)} \qquad, \qquad \ell_i =
  \frac{N_i}{\gcd (\ell_i', j_i')} \,.
\end{equation}
Moreover,
\begin{equation}
  j j' = a N_1 N_2 - \frac{N_1^2 N_2^2}{\ell_1 \ell'_1 \ell_2 \ell_2'} \,,
\end{equation}
for some $a \in \mathbb{Z}$, which implies
\begin{equation}
  j_1 j_1' = a \left( \frac{\ell_2 \ell_2'}{N_2} \right) N_1 -
  \frac{N_1^2}{\ell_1 \ell'_1} \,.
\end{equation}
Since $\ell_2 \ell_2' / N_2 = \ell_2 / \gcd (\ell_2, j_2) \in \mathbb{Z}$, we
find
\begin{equation}
  j_1 j_1' \equiv - \frac{N_1^2}{\ell_1 \ell'_1} \quad (\tmop{mod} N_1) \,,
\end{equation}
and likewise for $j_2 j_2'$. Assuming that (\ref{inductiveFormula}) is true
for $N_1$, we obtain
\begin{align}
  \frac{\gcd [j \ell X - m k Y, k_1^2, \ell_1^2]}{k_1 \ell_1} & = \frac{\gcd
  \left[ j_1 \ell_1 \left( \frac{\ell_2}{\ell_2'} N_2 \right) X - m_1 k_1
  \left( \frac{k_2}{k_2'} N_2 \right) Y, k_1^2, \ell_1^2 \right]}{k_1
  \ell_1}\nonumber\\
  & = \frac{\gcd \left[ j_1' \ell_1'  \left( \frac{k_2}{k_2'} N_2 \right) Y -
  m_1' k_1'  \left( \frac{\ell_2}{\ell_2'} N_2 \right) X, (k'_1)^2, (\ell'_1)^2
  \right]}{k_1' \ell_1'}\nonumber\\
  & = \frac{\gcd \left[ j_1' \ell_1'  \left( \frac{\ell_2'}{\ell_2} N_2
  \right) Y - m_1' k_1'  \left( \frac{k_2'}{k_2} N_2 \right) X, (k'_1)^2,
  (\ell'_1)^2 \right]}{k_1' \ell_1'}\nonumber\\
  & = \frac{\gcd [j' \ell' Y - m' k' X, (k'_1)^2, (\ell'_1)^2]}{k_1' \ell_1'} \,,
\end{align}
where in the penultimate step we make use of our freedom to exchange $\gcd (x,
y) \leftrightarrow \gcd (z x, y)$ provided that $\gcd (y, z) = 1$. Thus, if
(\ref{inductiveFormula}) is also true for $N_2$, we find:
\begin{align}
  \frac{\gcd [j \ell X - m k Y, k^2, \ell^2]}{k \ell} & = \frac{\gcd [j \ell X
  - m k Y, k_1^2, \ell_1^2]}{k_1 \ell_1} \cdot \frac{\gcd [j \ell X - m k Y,
  k_2^2, \ell_2^2]}{k_2 \ell_2}\nonumber\\
  & = \frac{\gcd [j' \ell' Y - m' k' X, (k'_1)^2, (\ell'_1)^2]}{k'_1 \ell'_1}
  \cdot \frac{\gcd [j' \ell' Y - m' k' X, (k'_2)^2, (\ell'_2)^2]}{k'_2 \ell'_2}\nonumber\\
  & = \frac{\gcd [j' \ell' Y - m' k' X, k^{\prime 2}, \ell^{\prime 2}]}{k'
  \ell'}  \,,
\end{align}
and the formula holds for $N = N_1 N_2$. Thus, (\ref{inductiveFormula}) is
proven by induction, and (\ref{formulaToProve}) follows as a corollary.

\subsection{The $\cK_N$ partition function} \label{subsec:ZKN}

We now consider the $\cK_N$ theory.
\begin{equation}
Z_{\cK_N}(\tau) = \sum_{v \in \cI_0} Z_v(\tau) = \sum_{k|N} \sum_{a=0}^{k-1} \hat{C}_{k}^{(a)}[\cI_0] Z_{N ; k}^{(a)}(\tau) \,,
\end{equation}
where $\hat{C}_{k}^{(a)}[\cI_0]$ counts the multiplicity of the indicated type of gauge bundle on the null self-dual subspace $\cI_0$.

Above, we have indicated the possibility that the multiplicities $\hat{C}_{k}^{(a)}[\cI_0]$ depend on the choice of null self-dual subspace $\cI_0$. In fact, unless $N$ is square-free, this is true. For instance, for $N=4$, the $\bZ_2^{22}$ subspace of $H^2(\Kthree,\bZ_4) = \bZ_4^{22}$ is null self-dual, but there are also null self-dual $\bZ_4^{11}$ subspaces of the form \eqref{eqn:I0}, where $\hat{C}_{4}^{(a)} = 0$ for all $a$ in the former case but not in the latter.

However, there is a special type of null self-dual subspace that unique determines $\hat{C}_{k}^{(a)}$ for all $N$. We say that $\cI_0$ is ``completely null'' if, for any $v \in \cI_0$, $k v = 0$ (modulo $N$) implies $\frac{1}{2}(kv/N)^2 \equiv 0 \pmod k$. In other words, for all $k|N$, the intersection of $\cI_0$ with $H^2(\Kthree; \bZ_k) = \frac{N}{k} H^2(\Kthree; \bZ_N)$ is null with respect to the Pontryagin square on $H^2(\Kthree; \bZ_k)$. Note that a completely null self-dual subspace $\cI_0$ exists for any $N$, because we can apply the method of~\secref{subsec:suNselfdual} first to $N' = 2 N^2$ and then reduce by modular congruence to determine the Pontryagin square on each subgroup $H^2(\Kthree; \bZ_k)$.

Because a null subspace of $H^2(\Kthree; \bZ_k)$ can have at most $k^{11}$ elements (where the bound is saturated in the self-dual case), we conclude that completely null $\cI_0$ has at most $k^{11}$ elements whose order divides $k$ for each $k|N$, that is $C_k \le k^{11}$, where
\begin{equation}
C_k = \sum_{a = 0}^{k-1} C_k^{(a)} \,, \qquad  C_k^{(a)} = \sum_{\ell |k} \sum_{\substack{b \\ \ell^2 b \equiv a \bmod k}}
  \hat{C}_{k / \ell}^{(b)}  \,,
\end{equation}
are the inclusive counts. Suppose for instance that $N = p^n$ is a prime power. Then, as a finite abelian group, $\cI_0$ can be decomposed into the direct sum of cyclic groups of prime power order:
\begin{equation}
\cI_0 = \bZ_{p^{n_1}} \oplus \ldots \oplus \bZ_{p^{n_\ell}} \,,\qquad n_i \le n \,.
\end{equation}
The number of elements with order dividing $p$ is then $p^\ell$, so $\ell \le 11$ by the above argument. However, if $\cI_0$ is self-dual, then $|\cI_0| = N^{11} = p^{11n}$, so that
\begin{equation}
\sum_i n_i = 11 n \,.
\end{equation}
Taken together, these constraints have only one solution, $\cI_0 = \bZ_N^{11}$, so that the bound $C_k \le k^{11}$ is saturated for all $k|N$. The same result generalizes to any $N$ using the Chinese remainder theorem. Thus, if $\cI_0$ is self-dual and completely null, then its intersection with each subspace $H^2(\Kthree; \bZ_k)$ is null self-dual.

In particular, this implies the counts
\begin{equation}
C_k^{(0)} = k^{11} = \sum_{\ell |k} \hat{C}_{k / \ell}^{(0)} \,, \qquad C_k^{(a)}=\hat{C}_k^{(a)}=0 \,, \qquad (a \not\equiv 0 \pmod k) \,.
\end{equation}
For general $N$, we define the IIB boundary conditions for the $\cK_N$ theory by a choice of $\cI_0$ that is self-dual and completely null. The partition function is then:
\begin{align}
Z_{\cK_N}(\tau) &= \sum_{k|N} \hat{C}_k^{(0)} Z_{N ; k}^{(0)}(\tau) = \sum_{k|N} \hat{C}_k^{(0)} \sum_{\ell | \frac{N}{k}} \frac{k \ell}{N^2} \sum_{j =
  0}^{k \ell - 1} G \biggl( \frac{N}{k^2 \ell^2} \tau +
  \frac{j}{k \ell} \biggr) \nonumber \\
  &=  \frac{1}{N^2} \sum_{d|N} d^{12} \sum_{j =
  0}^{d - 1} G \biggl( \frac{N}{d^2} \tau +
  \frac{j}{d} \biggr) = N^{11} T_N[G](\tau) \,. \label{eqn:ZKN}
\end{align}
This remarkably simple result (generalizing~\eqref{eqn:ZKp}) makes it manifest that $Z_{\cK_N}(\tau)$ is a modular form of weight $-12$.

\bibliographystyle{JHEP}
\bibliography{refs}

\end{document}